\documentclass[12pt]{article}
\usepackage{a4wide,amssymb,cite}
\pdfoutput=1

\usepackage{a4wide,amssymb,graphicx}
\usepackage{epsfig}
\usepackage[usenames,dvipsnames]{color}
\usepackage{slashed}
\usepackage{amssymb,cite,graphicx}
\usepackage{slashed}
\usepackage{amsmath,bm,bbm}
\usepackage{amsfonts}
\usepackage[titletoc,title]{appendix}
\usepackage[small]{caption}
\usepackage[margin=1in]{geometry}
\usepackage[multiple]{footmisc}
\usepackage{mathtools}
\usepackage{slashed}
\usepackage[nottoc]{tocbibind}
\usepackage{xcolor}

\newcommand{\be}{\begin{equation}}
\newcommand{\ee}{\end{equation}}
\newcommand{\bea}{\begin{eqnarray}}
\newcommand{\eea}{\end{eqnarray}}

\def\circa#1{\,\raise.3ex\hbox{$#1$\kern-.75em\lower1ex\hbox{$\sim$}}\,}

\begin{document}

\begin{titlepage}
%
%


%

\begin{centering}
\vspace{1cm}
{\Large {\bf Self-resonant dark matter with $Z_4$ gauged symmetry }} \\

\vspace{1.5cm}

\begin{centering}
{\bf  Lucca Radicce Justino$^{1,\star}$, Seong-Sik Kim$^{1,2,\sharp}$, Hyun Min Lee$^{1,\dagger}$, \\ and Jun-Ho Song$^{1,\ddagger}$}
\end{centering}
\\
\vspace{.5cm}

{\it $^1$Department of Physics, Chung-Ang University, Seoul 06974, Korea.} \\
{\it $^2$Department of Physics and Institute of Quantum Systems (IQS), Chungnam National University, Daejeon 34134, Korea.}

\vspace{.5cm}


\end{centering}
\vspace{2cm}

\begin{abstract}
\noindent
We present a new model for two-component scalar dark matter (DM), consisting of two complex scalar fields. In this model, both the DM components are stable due to the remaining $Z_4$ gauge symmetry, which is the remnant of the $U(1)^\prime$ local symmetry. When the resonance condition for DM masses is fulfilled, we show that the elastic co-scattering processes between two components of dark matter ($u$-channel processes) are enhanced due to the Yukawa potential with a small effective mass for the lighter DM mediator, so we can use such co-scattering processes for dark matter to explain the small-scale problems at galaxies. Moreover, there are also semi-annihilation processes that two components of dark matter annihilate into one dark matter particle and a dark photon/Higgs, which are enhanced by the $u$-channel Sommerfeld factor. Focusing on some benchmark models for two-component dark matter satisfying the observed relic density, we obtain the bounds for the dark photon portal couplings from the direct detection for boosted dark matter, which is produced from the semi-annihilation processes at the galactic center.

\end{abstract}

\vspace{2.5cm}

\begin{flushleft} 
$^\star$Email: luccarj2024@cau.ac.kr \\
${}^{\sharp}$Email: sskim.working@gmail.com  \\
$^\dagger$Email: hminlee@cau.ac.kr  \\
$^\ddagger$Email: thdwnsgh1003@gmail.com
\end{flushleft}

\end{titlepage}

\section{Introduction}

Dark matter (DM) is ubiquitous in the universe, from galaxies, large scale structures to cosmological scales, and it is measured precisely by Cosmic Microwave Background (CMB) data from Planck \cite{Planck:2018vyg}. However, there is no clue for identifying the nature of dark matter,  although there have been a plenty of models for dark matter and a lot of efforts to test them by direct and indirect experiments. Weakly Interacting Massive Particles (WIMP) paradigm has been one of the interesting avenues in searching for dark matter, but the bounds from the direct detection experiments such as  XENONnT \cite{xenonnt}, PandaX-4T \cite{PandaX:2024qfu}, LZ \cite{LZCollaboration:2024lux}, etc, have ruled out most of the parameter space for weak-scale dark matter.

Dark matter has been assumed to be composed of a single species of particles in the extension of the Standard Model (SM), so there has been an interesting interplay between the DM annihilation, the DM-nucleon scattering and the DM production, providing a guideline for constructing a consistent model for dark matter. However, dark matter can be composed of more than one species of particles in the so called multi-component dark matter scenarios, so new interactions between dark matter particles can play a crucial role not only in the evolution of dark matter abundance, large scale structures and galaxies, but also in the detection prospects for dark matter.

The extended dark sector is welcoming to make the N-body DM simulation compatible with the observed rotation velocities in galaxies at small scales. For instance, light mediators for dark matter ($t$-channel mechanism)\cite{SIDM} or a multi-component dark matter with resonant mass conditions ($u$-channel mechanism)  \cite{SRDM,review} can make the dark matter self-interactions enhanced and velocity-dependent \cite{velocitydep}. Then, self-interacting dark matter can provide a solution to the small-scale problems such as core-cusp problem, too-big-to-fail problem, diversity problem, etc \cite{smallscale,SIDM}, although we need a more precise understanding of the baryonic feedback on the DM simulation \cite{baryon}. In particular, in the latter $u$-channel mechanism where there are more than one species for dark matter candidates with comparable masses,  the dark matter self-scattering can be enhanced to solve the small-scale problems for a small velocity of dark matter at dwarf galaxies without a light mediator, and there are interesting signals for direct and indirect detections, depending on the relative fractions of dark matter abundances \cite{SRDM}.

In this article, we consider a model with $Z_4$ discrete gauge symmetry for two-component dark matter. 
The $U(1)'$ local symmetry is spontaneously broken to $Z_4$ such that a complex scalar field $\phi_1$ and the lighter component of another complex scalar field $\phi_2$ can be both candidates for stable dark matter. In comparison, in the $Z_2$ model  considered in Ref.~\cite{SRDM},  $\phi_1$ was taken as being a complex singlet scalar field charged under the $U(1)'$, while $\phi_2$ was considered as being a neutral singlet scalar under the $U(1)'$. In this case, $\phi_2$ would not be stable due to  the  loop-induced decays with $Z'$ interactions unless a twin partner of $\phi_1$ with a similar direct coupling to $\phi_2$  is introduced \cite{review}. In the current paper,  the $Z_4$ symmetry provides a consistent framework for two-component dark matter because it makes both $\phi_2$ (or the lighter of scalar fields belong to $\phi_2$) and $\phi_1$ absolutely stable. 

Focusing on the case with degenerate masses for two real components of $\phi_2$, we just call the second dark matter $\phi_2$. The $Z_4$-invariant self-interactions between $\phi_1$ and $\phi_2$ are responsible for the $u$-channel mechanism for enhancing the DM self-scattering cross sections under the resonant mass condition as well as determining the relic density for multi-component dark matter.  We call  the dark matter in our model ``self-resonant dark matter''. 
We remark that the qualitative discussion of the current work is still valid even for the general case with split masses for $\phi_2$, after the resonant mass condition is introduced between $\phi_1$ and the lighter component of $\phi_2$.

In the presence of the resonant mass condition for $\phi_1$ and $\phi_2$, namely, $m_2\simeq 2m_1$, the tree level co-scattering processes between $\phi_1$ and $\phi_2$ (or $\phi^\dagger_2$)  become divergent in the limit of a small momentum transfer for dark matter. Thus, we develop the Bethe-Salpeter formalism for getting the enhanced co-scattering cross sections following closely the method taken by some of the authors in Ref.~\cite{SRDM}. Moreover, we show how the semi-annihilation processes, $\phi_1\phi^{(\dagger)}_2\to \phi^\dagger_1 Y$, with $Y=X, h_X$, and their complex conjugates, where $X$ is the dark photon and $h_X$ is the dark Higgs, can be also enhanced by the $u$-channel Sommerfeld factor. The boosted dark matter $\phi_1$ produced from the semi-annihilation processes carries a sizable kinetic energy such that it can be probed by direct detection experiments above the detector threshold. We discuss the compatibility of the effective self-scattering cross section for dark matter with the observed rotation velocities in galaxies and galaxy clusters. We also choose some benchmark models for dark matter with weak-scale or sub-GeV scale masses and show how the parameter space for dark photon portal couplings is constrained by the relic density, indirect detection such as CMB bound, and the direct detection for boosted dark matter coming from the galactic center. 

The paper is organized as follows.
We begin with the model setup consisting of two complex scalar fields for dark matter and a complex scalar field for the dark Higgs mechanism and identify the dark sector interactions as well as the DM interactions with the SM. We apply the Bethe-Salpeter formalism for the non-relativistic co-scattering processes between $\phi_1$ and $\phi_2$  and obtain the effective Yukawa-type potential only for one linear combination of the initial states, $| \phi_1\phi_2\rangle$ and $| \phi_1\phi^\dagger_2\rangle$.
Then, we obtain the velocity-averaged cross section for the co-scattering between two different species for dark matter and the one for the self-scattering for $\phi_1$ only and compare between them.
Next we provide the Boltzmann equations for the relic abundances for two-component dark matter and present the direct detection cross sections for halo dark matter, the bound from CMB recombination, direct detection for boosted dark matter.
Finally conclusions are drawn. There are three appendices containing the DM annihilation cross sections in the model, the direct detection cross sections for Higgs and dark photon portal couplings as well as the kinematics for the semi-annihilation processes for multi-component dark matter.

\section{The setup}

For a working model for self-resonant dark matter \cite{SRDM,review}, we introduce a complex scalar $\chi$ for the symmetry breaking of $U(1)'$ to $Z_4$, two additional complex scalars, and $\phi_1, \phi_2$, for multi-component dark matter. The extra gauge boson associated with $U(1)'$ is denoted by $X_\mu$. We assume that the SM particles including the Higgs doublet are neutral under the $U(1)'$.
We specify the charges for a local $U(1)'$ symmetry and the $Z_4$ transformations for the dark scalars, as summarized in Table 1.

\begin{table}[!ht]
    \centering
    \begin{tabular}{|c|c|c|c|}
    \hline
         & $\phi_1$ & $\phi_2$ & $\chi$  \\
        \hline\hline
        $U(1)'$ & $+1$ & $+2$ & $+4$ \\
        \hline
        $U(1)'\supset Z_4$  &  $i$  &  $-1$ & $+1$  \\
        \hline
    \end{tabular}
    \caption{The $U(1)'$ charges and $Z_4$  transformations for dark scalars. }
    \label{table:1}
\end{table}

The $U(1)'$ invariant Lagrangian for the SM Higgs and dark scalars is given by
\bea
{\cal L}=|D_\mu\chi|^2+ |D_\mu \phi_1|^2 +|D_\mu \phi_2|^2 -\frac{1}{4} X_{\mu\nu} X^{\mu\nu} -\frac{1}{2} \sin\xi\, X_{\mu\nu}B^{\mu\nu}-V(\chi,\phi_1,\phi_2)+{\cal L}_{H,{\rm portal}} \label{Lag}
\eea
where $X_{\mu\nu}=\partial_\mu X_\nu -\partial_\nu X_\mu$ is the field strength tensor for the extra gauge boson, the covariant derivatives are $D_\mu\chi=(\partial_\mu-4i g_X X_\mu)\chi$, $D_\mu \phi_1=(\partial_\mu-ig_XX_\mu)\phi_1$, $D_\mu\phi_2=(\partial_\mu -2 ig_X X_\mu)\phi_2$, and the scalar potential is
\bea
V(\chi,\phi_1,\phi_2)&=&m_\chi^2 |\chi|^2 +\lambda_\chi |\chi|^4 +m_1^2 |\phi_1|^2 +\lambda_1 |\phi_1|^4+m_2^2 |\phi_2|^2 +\lambda_2 |\phi_2|^4  \nonumber \\
&& + \sum_{i=1,2}\lambda_{\chi i}|\chi|^2 |\phi_i|^2 + \lambda_{12} |\phi_1|^2 |\phi_2|^2 + (g_1 m_1 \phi^\dagger_2 \phi_1^2  + {\rm h.c.}) \nonumber \\
&&+\Big(\kappa_1 \chi^\dagger \phi^2_2 + \kappa_2 \chi^\dagger  \phi_2 \phi_1^2+{\rm h.c.}\Big).
\eea
Moreover,  $\sin\xi$ is a gauge kinetic mixing term between the extra gauge boson and the $U(1)_Y$ gauge boson. We can also introduce  Higgs-portal couplings like $ |H|^2{\cal O}_{\rm dark}$ with ${\cal O}_{\rm dark}=|\chi|^2, |\phi_1|^2, |\phi_2|^2$.
That is, the Lagrangian for the SM Higgs doublet is extended with the Higgs-portal interactions to the dark scalars,
\bea
{\cal L}_{H,{\rm portal}}=|D_\mu H|^2-\lambda_H |H|^4-m^2_H |H|^2-\lambda_{\chi H}|H|^2|\chi|^2-\lambda_{H1}|H|^2|\phi_1|^2-\lambda_{H2}|H|^2|\phi_2|^2.
\eea

\subsection{$U(1)'$ symmetry breaking vacuum}

We assume that the SM Higgs $H$ and the dark Higgs $\chi$ get nonzero VEVs by $H=(0,v_H)^T/\sqrt{2}$ and $\langle\chi\rangle= \frac{1}{\sqrt{2}}\, v_\chi$, respectively, but $\phi_1, \phi_2$ do not get VEVs.
By setting the VEVs for $\phi_1$ and  $\phi_2$ to zero, we find that the minimization conditions for the potential with respect to  the dark Higgs and the SM Higgs are given by
\bea
0&=&m^2_\chi +\lambda_\chi v_\chi^2 + \frac{1}{2}\lambda_{\chi H} v^2_H, \\ 
0&=&m^2_H +\lambda_H v_H^2 +\frac{1}{2}\lambda_{\chi H} v^2_\chi,
\eea
or the VEVs are determined by
\bea
v^2_H&=& \frac{2(\lambda_{\chi H} m^2_\chi-2\lambda_\chi m^2_H)}{4\lambda_H \lambda_\chi-\lambda^2_{\chi H}}, \nonumber \\
v^2_\chi &=& \frac{2(\lambda_{\chi H} m^2_H-2\lambda_H m^2_\chi)}{4\lambda_H \lambda_\chi-\lambda^2_{\chi H}}.
\eea
Furthermore, from the mass matrix for the SM Higgs and the dark Higgs, it turns out that the conditions for a local minimum are given \cite{Z4IDM} by
\bea
&&4\lambda_H \lambda_\chi-\lambda^2_{\chi H}>0, \\ 
&&\lambda_{\chi H} m^2_\chi -2\lambda_\chi m^2_H>0, \\
&&\lambda_{\chi H} m^2_H -2\lambda_H m^2_\chi>0.
\eea
We note that there are other important constraints on the quartic couplings in our model, such as unitarity, perturbativity, vacuum stability \cite{Z4IDM}, when the fields are far away from the local minimum or the energy increases far above the VEVs \cite{U1RG}. However, we don't consider the extra constraints on the quartic couplings explicitly in the following discussion, as energies or field ranges for the typical processes are near the VEVs.

Then, after $\chi$ gets a VEV, the $U(1)'$ gauge symmetry is broken spontaneously to $Z_4$ and the $U(1)'$ gauge boson gets a nonzero mass as $m_X=4 g_X v_\chi$. 
Moreover, the effective mass terms and the dimensionful interaction terms for $\phi_1, \phi_2$, which are relevant for the $u$-channel processes,  are shifted, as follows, 
\bea
{\cal L}_{\rm eff} = -m^2_{1,{\rm eff}} |\phi_1|^2 - m^2_{2,{\rm eff}} |\phi_2|^2  - (m^2_3\phi_2^2  + g_1 m_1 \phi^\dagger_2 \phi_1^2  +  g_2 m_1 \phi_2 \phi_1^2+{\rm h.c.})  \label{DMself}
\eea
with
\bea
m^2_{1,{\rm eff}} &\equiv & m^2_1 + \frac{1}{2} \lambda_{\chi 1} v^2_\chi+\frac{1}{2}\lambda_{H1}v^2_H, \\
m^2_{2,{\rm eff}} &\equiv & m^2_2 + \frac{1}{2} \lambda_{\chi 2} v^2_\chi +\frac{1}{2}\lambda_{H2}v^2_H, \\
m^2_3 &\equiv & \frac{1}{\sqrt{2}} \kappa_1 v_\chi, \\
g_2&\equiv & \frac{\kappa_2 v_\chi}{\sqrt{2}m_1}.
\eea
In particular, the scalars in $\phi_2$, namely, $\phi_2= \frac{1}{\sqrt{2}}(s+ia)$, have split masses, due to the effective $\phi_2^2$ term, as
\bea
m^2_s &=&m^2_{2,{\rm eff}} +2m^2_3, \\
m^2_a &=&m^2_{2,{\rm eff}} -2m^2_3.
\eea
As a result, the direction for $\phi_1$ and the scalars in $\phi_2$ are stable as far as $m^2_{1,{\rm eff}}>0$,  $m^2_s>0$ and $m^2_a>0$, which constrain the bare mass parameters and the quartic couplings for the DM scalars, as follows,
\bea
&&m^2_1+ \frac{1}{2} \lambda_{\chi 1} v^2_\chi+\frac{1}{2}\lambda_{H1}v^2_H>0, \\
&& m^2_2 + \frac{1}{2} \lambda_{\chi 2} v^2_\chi+\frac{1}{2}\lambda_{H2}v^2_H\pm \sqrt{2} \kappa_1 v_\chi>0.
\eea

We note that two real scalar fields of $\phi_2$ are degenerate in mass, i.e. $m_s=m_a=m_{2,{\rm eff}}$, for $m_3=0$ or $\kappa_1=0$. 
In this case, for the later discussion, we use the notations, $m_2$ for $m_{2,{\rm eff}}$, for simplicity. 
If $m_3\neq 0$ or $\kappa_1\neq 0$, $m_s\neq m_a$: $a(s)$ is the second dark matter candidate for $m^2_3>0(m^2_3<0)$.
In the latter case, the resonant mass condition changes from $m_2=2m_1$ (for $m_3=0$) to $m_{a(s)}=2m_1$ (for $m_3\neq 0$).

A nonzero $\kappa_1$ splits the masses between $s$ and $a$, but it is sufficient to take the mass relation between $\phi_1$ and one of $s$ and $a$ for the $u$-channel enhancement. So, the existence of the $u$-channel enhancement does not depend on the mass splitting between $s$ and $a$.
So, for simplicity, we can take the same masses for $s$ and $a$.

In general, the dark matter abundance at present depends on the mass splitting between $s$ and $a$, $|m_s-m_a|$,  and the lifetime of the heavier one. As far as $|m_s-m_a|\lesssim T_{\rm f.o.}$, where $T_{\rm f.o.}$ is the freeze-out temperature of about $0.2\,{\rm min}(m_s,m_a) $,  the relic density calculations would depend little on the mass splitting. Moreover, the $u$-channel condition would be satisfied  for both $s$ and $a$, if $|m_s-m_a|\lesssim M=m_{s(a)}\sqrt{2-\frac{m_{s(a)}}{m_1}}$ where $M$ is the effective mass for the $u$-channel resonance, as will be shown in the later section. 
Henceforth, for simplicity, we focus on the former case with degenerate masses for $\phi_2$, namely,  $m_s=m_a$.

\subsection{DM self-interactions}

First, from eq.~(\ref{DMself}), we collect the self-interaction terms for dark scalars, $\phi_1, s, a$, as follows,
\bea
{\cal L}_{\phi_1-s,a} &=&-\frac{1}{\sqrt{2}}(g_1+g_2) m_1 s (\phi^2_1+\phi^{\dagger 2}_1)+\frac{1}{\sqrt{2}}i (g_1-g_2)m_1 a (\phi^2_1-\phi^{\dagger 2}_1) \nonumber\\
&&-\frac{1}{2} \lambda_{12} (s^2+a^2) |\phi_1|^2 - \frac{1}{4}\lambda_2 (s^2+a^2)^2-\lambda_1 |\phi_1|^4. \label{scalars1}
\eea
Here, we assumed that all the couplings for dark scalars are real.  After the $U(1)'$ is broken, the stability for $\phi_1$ and the lighter particle among $s$ and $a$ are guaranteed because of the remaining $Z_4$ symmetry:
\bea
Z_4:\,\, \phi_1\to i\phi_1, \quad s\to -s, \quad a\to -a,
\eea
as far as $m_s, m_a<2m_1$. As a result, the multi-component dark matter can be realized in our model.

\subsection{DM interactions with the dark sector}

Expanding the dark Higgs $\chi$ around the VEV by $\chi=\frac{1}{\sqrt{2}}(v_\chi+h_X)$ in unitary gauge, we can also get the interaction terms for the dark Higgs $h_X$ and the dark scalars, from eq.~(\ref{Lag}), as follows,
\bea
{\cal L}_{h_X-\phi_1,s,a}&=&-\lambda_\chi v_\chi h^3_\chi-\frac{1}{4}\lambda_\chi h^4_\chi- \frac{1}{2} \lambda_{\chi 1}(2v_\chi h_X + h^2_X) |\phi_1|^2 - \frac{1}{4} \lambda_{\chi 2}(2v_\chi h_X + h^2_X)(s^2+a^2) \nonumber \\
&&-\frac{1}{\sqrt{2}}\kappa_1 h_X(s^2-a^2) - \frac{1}{2}\kappa_2 h_X s(\phi^2_1+\phi^{\dagger 2}_1)-  \frac{1}{2}i\kappa_2 h_X a(\phi^2_1-\phi^{\dagger 2}_1). \label{scalars2}
\eea
We note that the Higgs-portal interaction for $\chi$, namely, $|H|^2|\chi|^2$, gives rise to a Higgs mixing after electroweak symmetry is also broken.

Moreover, the $U(1)'$ gauge interactions for the dark Higgs $h_X$  and dark scalars, $\phi_1, s, a$, are given by
\bea
{\cal L}_{X} &=&\frac{m^2_X}{v_\chi}\, h_X X_\mu X^\mu+ 8g^2_X h^2_X X_\mu X^\mu  \nonumber \\
 &&+ig_X X_\mu (\phi^\dagger_1 \partial^\mu\phi_1-\phi_1 \partial^\mu\phi^\dagger_1)+ g^2_X X_\mu X^\mu \phi^\dagger_1\phi_1 \nonumber \\
&& +2 ig_X X_\mu (\phi^\dagger_2 \partial^\mu\phi_2-\phi_2 \partial^\mu\phi^\dagger_2)+4 g^2_XX_\mu X^\mu \phi^\dagger_2\phi_2  \nonumber \\
&=&\frac{m^2_X}{v_\chi}\, h_X X_\mu X^\mu+ 8g^2_X h^2_X X_\mu X^\mu \nonumber \\
&&+ ig_X X_\mu (\phi^\dagger_1 \partial^\mu\phi_1-\phi_1 \partial^\mu\phi^\dagger_1)+ g^2_X X_\mu X^\mu \phi^\dagger_1\phi_1 \nonumber \\
&&- 2g_X X_\mu (s\partial^\mu a -a\partial^\mu s)+2 g^2_X X_\mu X^\mu (s^2+a^2). \label{gauge}
\eea
Due to the gauge interactions in eq.~(\ref{gauge}), $s$ decays into $a+X$ for $m_s>m_a+m_X$ while $a$ decays into $s+X$ for $m_a>m_s+m_X$. Even if the $X$ boson produced from the decays of $s$ or $a$ is off-shell, either $s$ or $a$ is unstable, as $X$ can decay into the light fermions in the SM through the gauge kinetic mixing, so we can regard only the lighter particle among $s$ and $a$ as being a stable dark matter, whereas the relic abundance of the heavier particle depends on the mass splitting, $|m_s-m_a|$, and the lifetime of the heavier particle.

\subsection{DM interactions with the SM sector}

For $|\xi|\ll 1$, there is a small mixing between the dark gauge boson and the $Z$ boson, for which the neutral current interactions \cite{gmix1,gmix2} become
\bea
{\cal L}_{\rm NC}\simeq A_\mu J^\mu_{\rm EM} +Z_\mu \bigg(J^\mu_Z +\varepsilon \, t_W J^\mu_X \bigg)+X_\mu \bigg(-\varepsilon J^\mu_{\rm EM}+ J^\mu_X \bigg), \label{gaugeportal}
\eea
with $\varepsilon\equiv c_W\xi$, and $J^\mu_{\rm EM}, J^\mu_Z$ and $J^\mu_X$ are electromagnetic, neutral and dark currents, given by
\bea
J^\mu_{\rm EM} &=& e {\bar f} \gamma^\mu Q_f f, \\
 J^\mu_Z &=& \frac{e}{2s_W c_W}\, {\bar f}\gamma^\mu(\tau^3-2s^2_W Q_f)f, \\
 J^\mu_X&=&ig_X (\phi^\dagger_1\partial^\mu\phi_1-\phi_1\partial^\mu\phi^\dagger_1)+2i g_X (\phi^\dagger_2 \partial^\mu\phi_2-\phi_2 \partial^\mu\phi^\dagger_2)  \nonumber \\
 &=&ig_X (\phi^\dagger_1\partial^\mu\phi_1-\phi_1\partial^\mu\phi^\dagger_1)-2g_X (s\partial^\mu a-a\partial^\mu s).
\eea
Here, we note that  the gauge kinetic mixing has been included for the linear interactions of the extra gauge boson in addition to those in  eq.~(\ref{gauge}).

Moreover, the SM Higgs $h$, (from $H^T=\frac{1}{\sqrt{2}}(0,v+h)$), mixes with the dark Higgs $h_X$ by
\bea
h_1 =\cos\theta\, h_X -\sin\theta\, h, \quad h_2=\sin\theta\, h_X + \cos\theta\, h,
\eea
with 
\bea
\tan2\theta =\frac{\lambda_{\chi H} v_\chi v}{\lambda_H v^2-\lambda_\chi v^2_\chi}.
\eea
Then, we also obtain the effective couplings for the dark Higgs-like scalar $h_1$ and the SM Higgs-like scalar $h_2$ to multi-component dark matter and the SM fermions, as follows,
\bea
{\cal L}_{h_1,h_2}&=&- y_{h_1\phi^\dagger_1\phi_1} h_1 |\phi_1|^2 - y_{h_2 \phi^\dagger_1\phi_1} h_2  |\phi_1|^2-\frac{1}{2}y_{h_1 ss} h_1 \, s^2-\frac{1}{2}y_{h_1 aa} h_1\, a^2 \nonumber \\
&&-\frac{1}{2}y_{h_2 ss} h_2 \,s^2-\frac{1}{2}y_{h_2 aa} h_2\, a^2- (\lambda_{f1}h_1+\lambda_{f2} h_2) {\bar f}f, \label{Higgsportal}
\eea
with
\bea
y_{h_1\phi^\dagger_1\phi_1} &=&\lambda_{\chi 1}v_\chi\,\cos\theta-\lambda_{H1}v\sin\theta, \\
y_{h_2\phi^\dagger_1\phi_1} &=&\lambda_{\chi 1}v_\chi\,\sin\theta+\lambda_{H1}v\cos\theta, \\
y_{h_1 ss} &=&(\lambda_{\chi 2}v_\chi +\sqrt{2}\kappa_1)\cos\theta-\lambda_{H2}v\sin\theta, \\
y_{h_2 ss} &=&(\lambda_{\chi 2}v_\chi +\sqrt{2}\kappa_1)\sin\theta+\lambda_{H2}v\cos\theta, \\
y_{h_1 aa} &=&(\lambda_{\chi 2}v_\chi -\sqrt{2}\kappa_1)\cos\theta-\lambda_{H2}v\sin\theta, \\
y_{h_2 aa} &=&(\lambda_{\chi 2}v_\chi -\sqrt{2}\kappa_1)\sin\theta+\lambda_{H2}v\cos\theta, \\
\lambda_{f 1} &=&-\frac{m_f}{v}\,\sin\theta, \\
\lambda_{f 2} &=&\frac{m_f}{v}\,\cos\theta.
\eea
For $m_s=m_a=m_2$, namely, for $\kappa_1=0$, we can identify the effective couplings for the dark matter $\phi_2$ by $y_{h_i\phi^\dagger_2\phi_2}=y_{h_i ss}$, with $i=1,2,$.

In the later discussion, the $u$-channel resonances appear for multi-component dark matter with $m_s\simeq 2m_1$ and/or $m_a\simeq 2m_1$, but we focus on  the benchmark models with $m_s=m_a\simeq 2m_1$ for the DM self-scattering, the relic density and the direct and indirect detections,  for which the scalar self-interactions in eqs.~(\ref{scalars1}) and (\ref{scalars2}) and the gauge interactions in eq.~(\ref{gauge}) as well as gauge-portal and Higgs-portal interactions in eqs.~(\ref{gaugeportal}) and (\ref{Higgsportal}) are relevant.

\section{DM self-scattering with $u$-channel resonances}

We discuss the dark matter co-scattering processes with $u$-channel dominance at tree level and present the Bethe-Salpeter formalism for the non-perturbative results for them.
In the following discussion, we assume the case $m_s=m_a$, without loss of generality.

\subsection{Dark matter co-scattering at tree level}

We consider the elastic scattering, $\phi_1 s\rightarrow \phi_1s$ and its complex conjugate, which can have a non-perturbative enhancement near $m_s\sim 2m_1$ \cite{SRDM}. The dark Higgs also contributes to the elastic scattering in the $t$-channel but we don't consider it in the following discussion by assuming that the dark Higgs is not so light as compared to the DM mass.

For the elastic scattering process, $\phi_1(q) s(p)\rightarrow \phi_1(q') s(p')$, 
the tree-level scattering amplitude ${\widetilde \Gamma}_{s,u}(p,q;p,q')$ is given by
\bea
 {\widetilde \Gamma}_{s,u}(p,q;p',q') =-\frac{4g^2_s m^2_1}{(p-q')^2-m^2_1}=\frac{4g^2_s m^2_1}{|{\vec p}-{\vec q}'|^2+m^2_1-\omega^2}
\eea
with $g_s\equiv \frac{1}{\sqrt{2}}(g_1+g_2)$ and $\omega=p_0-q'_0$ being the energy exchange. 
Then, in the non-relativistic limit for initial particles, we get 
\bea
m^2_1-\omega^2&=&m^2_1-\Big(\sqrt{m^2_s+{\vec p}^2}-\sqrt{m^2_1+{\vec q}^{\prime 2}}\Big)^2 \nonumber \\
&\approx & m_s(2m_1-m_s)+\Big(-1+\frac{m_1}{m_s}\Big){\vec p}^2+\Big(-1+\frac{m_s}{m_1} \Big){\vec q}^{\prime 2}. \label{energytransfer}
\eea
We note that the energy exchange $\omega$ does not vanish in the non-relativistic limit, unlike the case for the elastic scattering between the same particles, so the process is not instantaneous.

As a result, the tree-level scattering amplitude ${\widetilde \Gamma}_{s,u}(p,q;p',q')$ becomes
\bea
 {\widetilde \Gamma}_{s,u}(p,q;p',q') \approx \frac{4g^2_s m^2_1}{\Big(\sqrt{\frac{m_1}{m_s}} {\vec p}-\sqrt{\frac{m_s}{m_1}}{\vec q}'\Big)^2+m_s(2m_1-m_s)}. \label{4point-tree}
\eea
Then, we find that the effective squared mass of the $u$-channel mediator is semi-positive definite for $m_s\leq 2m_1$.  For ${\vec p}={\vec q}'=0$, the above tree-level scattering amplitude diverges at $m_s=2m_1$.

Similarly, for the elastic scattering process, $\phi_1(q) a(p)\rightarrow \phi_1(q') a(p')$, 
the tree-level scattering amplitude ${\widetilde \Gamma}_{a,u}(p,q;p,q')$ is given by
\bea
 {\widetilde \Gamma}_{a,u}(p,q;p',q') =-\frac{4g^2_a m^2_1}{(p-q')^2-m^2_1}=\frac{4g^2_a m^2_1}{|{\vec p}-{\vec q}'|^2+m^2_1-\omega^2}
\eea
with $g_a\equiv \frac{1}{\sqrt{2}}(g_1-g_2)$. Following a similar step for the non-relativistic limit for the initial particles, we can approximate the above result to
\bea
 {\widetilde \Gamma}_{a,u}(p,q;p',q') \approx \frac{4g^2_a m^2_1}{\Big(\sqrt{\frac{m_1}{m_a}} {\vec p}-\sqrt{\frac{m_a}{m_1}}{\vec q}'\Big)^2+m_a(2m_1-m_a)}. 
\eea
So, the above tree-level scattering amplitude diverges at $m_a=2m_1$ for ${\vec p}={\vec q}'=0$.

\begin{figure}[t]
\centering
\includegraphics[width=0.20\textwidth,clip]{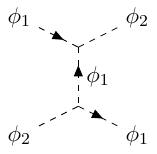} \,\,\,\,\,\,\,\,
\includegraphics[width=0.20\textwidth,clip]{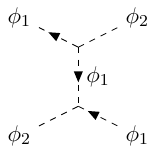} 
\caption{Tree-level Feynman diagrams for the DM co-scattering processes with $u$-channel, $\phi_1\phi_2\to \phi_1\phi_2$ and $\phi^\dagger_1\phi_2\to \phi^\dagger_1\phi_2$; $\phi_1\phi^\dagger_2\to \phi_1\phi_2$ and $\phi^\dagger_1\phi^\dagger_2\to \phi^\dagger_1\phi_2$.
} 
\label{Fig:uch-diag}
\end{figure}

For $m_s=m_a=m_2$, for the DM co-scattering processes of our focus in the later discussion, $\phi_1(q) \phi_2(p)\to \phi_1(q')\phi_2(p')$ and $\phi_1(q) \phi^\dagger_2(p)\to \phi_1(q')\phi_2(p')$ and their complex conjugates, the corresponding $u$-channel amplitudes at tree level shown in Fig.~\ref{Fig:uch-diag} are given by
\bea
 {\widetilde \Gamma}_{\phi_1\phi_2\to \phi_1\phi_2}(p,q;p',q') &=&-\frac{4g^2_1 m^2_1}{(p-q')^2-m^2_1},  \label{ph2a} \\
  {\widetilde \Gamma}_{\phi_1\phi^\dagger_2\to \phi_1\phi_2}(p,q;p',q') &=&-\frac{4g_1 g_2 m^2_1}{(p-q')^2-m^2_1}.  \label{ph2b}
\eea
Similarly, for $\phi_1(q) \phi^\dagger_2(p)\to \phi_1(q')\phi^\dagger_2(p')$ and $\phi_1(q) \phi_2(p)\to \phi_1(q')\phi^\dagger_2(p')$ and their complex conjugates, as shown in Fig.~\ref{Fig:uch-diag}, we also obtain
\bea
 {\widetilde \Gamma}_{\phi_1\phi^\dagger_2\to \phi_1\phi^\dagger_2}(p,q;p',q') &=&-\frac{4g^2_2 m^2_1}{(p-q')^2-m^2_1},    \label{ph2c} \\
 {\widetilde \Gamma}_{\phi_1\phi_2\to \phi_1\phi^\dagger_2}(p,q;p',q') &=&-\frac{4g_1 g_2 m^2_1}{(p-q')^2-m^2_1}.   \label{ph2d}
\eea
Then, the tree-level $u$-channel scattering amplitudes with a complex scalar $\phi_2$ in eqs.~(\ref{ph2a})-(\ref{ph2d}) diverge at $m_2=2m_1$ for ${\vec p}={\vec q}'=0$.
Other DM co-scattering processes with $u$-channels are complex conjugates of the above processes.

In the non-relativistic limit for two-component dark matter, the DM co-scattering cross section for $\phi_1 C \to \phi_1 C$, with $C=s, a$ or $\phi_2$, is, at tree level, given by 
\bea
\sigma_{\phi_1 C\to \phi_1 C} = \frac{1}{16\pi (m_1+m_C)^2}\,  \Big|{\widetilde \Gamma}_{u,\phi_1 C\to \phi_1 C}\Big|^2.
\eea
However, as will be discussed shortly, we need to take into account the non-perturbative effects from the $u$-channels for the DM annihilation.

For $m_s=m_a=m_2$, the non-perturbative co-scattering cross section can be written with $s$-wave dominance \cite{SRDM}, as 
\bea
\sigma^{\rm total}_{\psi_1 \to \phi_1\phi_2}=\frac{4\pi}{k^2}\,\sin^2\delta_0,  \label{coscatt}
\eea
and its complex conjugate, with $\psi_1$ being a mixture of two DM pairs, $(\phi_1\phi_2)$ and $(\phi_1\phi^\dagger_2)$, $k=\mu v$, $\mu$ being the reduced mass for $\phi_1-\phi^{(\dagger)}_2$, and $\delta_0$ being the phase shift for the elastic co-scattering, which is determined by solving the Bethe-Salpeter equation for $\psi_1$ or its complex conjugate, as will be shown in the next subsection.

\subsection{Bethe-Salpeter equations for DM co-scattering}

In this section, we use the Bethe-Salpeter(BS) formalism for the enhanced self-scattering cross section as well as the DM annihilation cross section, based on the $u$-channel dominance.

For simplicity, we consider the case that the real and imaginary parts of the complex scalar $\phi_2$ have degenerate masses, namely, for $m_s=m_a=m_2$. However, the results of the following discussion are applicable to the general case with split masses for $\phi_2$, as far as the resonant mass condition is satisfied between $\phi_1$ and the lighter component of $\phi_2$. 
In this case, we can work in the basis with two complex scalars, $(\phi_1,\phi_2)$. 

Then, we only have to consider four pairs of combinations of fields for self-scattering with $u$-channel dominance, namely, $(\phi_2, \phi^\dagger_2)\phi_1\to \phi_2\phi_1$, $(\phi_2,\phi^\dagger_2)\phi_1\to  \phi^\dagger_2\phi_1$, $(\phi^\dagger_2,\phi_2)\phi^\dagger_1\to \phi_2 \phi^\dagger_1$ and $(\phi^\dagger_2,\phi_2)\phi^\dagger_1\to \phi^\dagger_2\phi^\dagger_1$.

First, we consider a pair of self-scattering processes, $\phi_2(p)\phi_1(q)\to \phi_2(p')\phi_1(q')$ and $\phi^\dagger_2(p)\phi_1(q)\to \phi_2(p')\phi_1(q')$. Then, the corresponding non-perturbative scattering amplitudes satisfy the following relations,
\bea
i\chi_A(p,q) &\simeq& -G_2(p)G_1(q)\int \frac{d^4k}{(2\pi)^4} \bigg[{\widetilde \Gamma}_{\phi_2\phi_1\to\phi_2\phi_1} (p,q;p+q-k,k) \chi_A(p+q-k,k)  \nonumber \\
&&+{\widetilde \Gamma}_{\phi_2\phi_1\to\phi^\dagger_2\phi_1} (p,q;p+q-k,k) \chi_B(p+q-k,k) \bigg], \label{BS1} \\
i\chi_B(p,q) &\simeq& -G_2(p)G_1(q)\int \frac{d^4k}{(2\pi)^4} \bigg[{\widetilde \Gamma}_{\phi^*_2\phi_1\to\phi_2\phi_1} (p,q;p+q-k,k) \chi_A(p+q-k,k)  \nonumber \\
&&+{\widetilde \Gamma}_{\phi^\dagger_2\phi_1\to\phi^\dagger_2\phi_1} (p,q;p+q-k,k) \chi_B(p+q-k,k) \bigg] \label{BS2}
\eea
where $G_{1,2}(p)=\frac{i}{p^2-m^2_{1,2}}$ are the propagators for dark scalars, $\widetilde \Gamma$'s are the tree-level scattering amplitudes, and 
\bea
\chi_A(p,q) &\equiv & G_2(p)G_1(q) \Gamma_{\phi_2\phi_1\to\phi_2\phi_1}(p,q;p',q'), \\
\chi_B(p,q) &\equiv & G_2(p)G_1(q) \Gamma_{\phi^\dagger_2\phi_1\to\phi_2\phi_1}(p,q;p',q').
\eea
By using the relations between the tree-level scattering amplitudes,
\bea
{\widetilde \Gamma}_{\phi_2\phi_1\to\phi^\dagger_2\phi_1} &=&{\widetilde \Gamma}_{\phi^\dagger_2\phi_1\to\phi_2\phi_1}=\frac{g_2}{g_1}\, {\widetilde \Gamma}_{\phi_2\phi_1\to\phi_2\phi_1}, \\
{\widetilde \Gamma}_{\phi^\dagger_2\phi_1\to\phi^\dagger_2\phi_1}&=& \bigg(\frac{g_2}{g_1}\bigg)^2 {\widetilde \Gamma}_{\phi_2\phi_1\to\phi_2\phi_1},
\eea
we can rewrite the BS equations in eqs.~(\ref{BS1}) and (\ref{BS2}) as
\bea
i\chi_A(p,q) &\simeq& -G_2(p)G_1(q)\int \frac{d^4k}{(2\pi)^4} {\widetilde \Gamma}_{\phi_2\phi_1\to\phi_2\phi_1} (p,q;p+q-k,k)\times\nonumber\\  \nonumber \\
&&\times \bigg[\chi_A(p+q-k,k) +\frac{g_2}{g_1}\chi_B(p+q-k,k) \bigg],  \label{BS1}\\
i\chi_B(p,q) &\simeq& -\frac{g_2}{g_1}G_2(p)G_1(q)\int \frac{d^4k}{(2\pi)^4}{\widetilde \Gamma}_{\phi_2\phi_1\to\phi_2\phi_1} (p,q;p+q-k,k) \times   \nonumber \\
&&\times\bigg[\chi_A(p+q-k,k) +\frac{g_2}{g_1} \chi_B(p+q-k,k) \bigg]  \label{BS2}
\eea
Thus, from eqs.~(\ref{BS1}) and (\ref{BS2}), we find that 
\bea
\chi_B(p,q)\simeq \frac{g_2}{g_1}\chi_A(p,q). \label{relation}
\eea
Therefore, plugging eq.~(\ref{relation}) into eqs. (\ref{BS1}) or (\ref{BS2}), both $\chi_B(p,q)$ and $\chi_A(p,q)$ satisfy the same BS equations,
\bea
i\chi_{A,B}(p,q) &\simeq& -\Big(1+\frac{g^2_2}{g^2_1}\Big)G_2(p)G_1(q)\int \frac{d^4k}{(2\pi)^4} {\widetilde \Gamma}_{\phi_2\phi_1\to\phi_2\phi_1} (p,q;p+q-k,k)\times\nonumber\\  \nonumber \\
&&\times \chi_{A,B}(p+q-k,k).
\eea
Then, redefining $\chi_{A,B}(p,q)={\widetilde\chi}_{A,B}(P,Q)$ with 
\bea
P=\frac{1}{2}(p+q), \quad Q=\mu\bigg(\frac{p}{m_2}-\frac{q}{m_1}\bigg),
\eea
with $\mu=m_1 m_2/(m_1+m_2)$ and using the BS wave functions in momentum space, 
\bea
{\widetilde\psi}_{A,B}({\vec Q})&=& \int \frac{dQ_0}{2\pi}\, {\widetilde\chi}_{A,B}(P,Q), 
\eea
we obtain the following coupled BS equations,
\bea
\bigg( \frac{{\vec Q}^2}{2\mu} -E\bigg){\widetilde\psi}_{A,B}({\vec Q}) =-\Big(1+\frac{g^2_2}{g^2_1}\Big)\int \frac{d^3k'}{(2\pi)^3} {\widetilde V}\bigg(\sqrt{\frac{m_1}{m_2}}{\vec Q}+\sqrt{\frac{m_2}{m_1}} {\vec k}' \bigg) {\widetilde\psi}_{A,B}({\vec k}'), \label{BSa}
\eea
with
\bea
 {\widetilde V}\bigg(\sqrt{\frac{m_1}{m_2}}{\vec Q}+\sqrt{\frac{m_2}{m_1}} {\vec k}' \bigg)\equiv -\frac{1}{4m_1m_2} \cdot\, \frac{4g^2_1 m^2_1}{\big(\sqrt{\frac{m_1}{m_2}}{\vec Q}+\sqrt{\frac{m_2}{m_1}} {\vec k}'\big)^2+m_2(2m_1-m_2)}.
\eea

As a result, in terms of the BS wave functions in position space,
\bea
\psi_{A,B}({\vec x})= \int \frac{d^3{\vec Q}}{(2\pi)^3}\, e^{i{\vec Q}\cdot {\vec x}}\, {\widetilde \psi}_{A,B}({\vec Q}),
\eea
the coupled BS equations in eq.~(\ref{BSa}) can be recast into the following,
\bea
-\frac{1}{2\mu}\nabla^2\psi_{A,B}({\vec x})+ V_{\rm eff}({\vec x}) \psi_{A,B}\Big(-\frac{m_2}{m_1}{\vec x}\Big)  = E \psi_A({\vec x}),
\label{Schroeq} 
\eea
where
\bea
V_{\rm eff}({\vec x}) =-\Big(1+\frac{g^2_2}{g^2_1}\Big)\frac{\alpha}{r}\, e^{-M r}, \label{effpot}
\eea
with $\alpha\equiv \frac{g^2_1}{4\pi}$ and $M\equiv m_2\sqrt{2-\frac{m_2}{m_1}}$. Here, from eq.~(\ref{relation}), the BS wavefunctions in position space are related by
\bea
\psi_{B}({\vec x})\simeq \frac{g_2}{g_1}\psi_{A}({\vec x}). \label{relation2}
\eea

It is illuminating to consider the BS equations before inserting eq.~(\ref{relation}) into  eqs. (\ref{BS1}) or (\ref{BS2}), in the form of the following coupled BS equations,
\bea
\bigg( \frac{{\vec Q}^2}{2\mu} -E_A\bigg){\widetilde\psi}_{A}({\vec Q}) &=&-\int \frac{d^3k'}{(2\pi)^3} {\widetilde V}\bigg(\sqrt{\frac{m_1}{m_2}}{\vec Q}+\sqrt{\frac{m_2}{m_1}} {\vec k}' \bigg)\bigg( {\widetilde\psi}_{A}({\vec k}') +\frac{g_2}{g_1}{\widetilde\psi}_{B}({\vec k}')   \bigg),  \label{BSb} \\
\bigg( \frac{{\vec Q}^2}{2\mu} -E_B\bigg){\widetilde\psi}_{A}({\vec Q}) &=&-\frac{g_2}{g_1}\int \frac{d^3k'}{(2\pi)^3} {\widetilde V}\bigg(\sqrt{\frac{m_1}{m_2}}{\vec Q}+\sqrt{\frac{m_2}{m_1}} {\vec k}' \bigg)\bigg( {\widetilde\psi}_{A}({\vec k}') +\frac{g_2}{g_1}{\widetilde\psi}_{B}({\vec k}')   \bigg), \label{BSc}
\eea
which can be recast into the following,
\bea
-\frac{1}{2\mu}\nabla^2 \left(\begin{array}{c} \psi_A({\vec x}) \\ \psi_B({\vec x}) \end{array}\right) +V({\vec x})\left(\begin{array}{cc} 1 & \frac{g_2}{g_1} \\  \frac{g_2}{g_1} & \big(\frac{g_2}{g_1}\big)^2 \end{array} \right)  \left(\begin{array}{c} \psi_A\big(-\frac{m_2}{m_1}{\vec x}\big) \vspace{0.1cm}\\ \psi_B\big(-\frac{m_2}{m_1}{\vec x}\big) \end{array}\right) = \left(\begin{array}{cc} E_A & 0 \\ 0 & E_B  \end{array}\right) \left(\begin{array}{c} \psi_A({\vec x}) \\ \psi_B({\vec x}) \end{array}\right).
\eea
Then, for the BS wave functions with the same energy eigenvalues, $E_A=E_B=E$, we find that only one of the linear combinations of the BS wave functions, $\psi_1=\frac{1}{\sqrt{N}}\big(\psi_A+\frac{g_2}{g_1} \psi_B\big)$, with $N=1+\frac{g^2_2}{g^2_1}$, has a nonzero effective Yukawa-type  potential given by eq.~(\ref{effpot}),
while the orthogonal combination, $\psi_2\equiv \frac{1}{\sqrt{N}}\big(\psi_B-\frac{g_2}{g_1} \psi_A\big)$, is free from the Yukawa potential.
Therefore, for the analysis of the enhanced cross sections with the $u$-channel resonances, we can set $\psi_B=\frac{g_2}{g_1} \psi_A$, which is the same as  eq.~(\ref{relation2}). 

We note that in our model with the remaining $Z_4$ symmetry, the result for a toy model for self-resonant dark matter in Ref.~\cite{SRDM} is modified by the overall factor, $\big(1+\frac{g^2_2}{g^2_1}\big)$, in the effective Yukawa potential in the BS equation as in eq.~(\ref{effpot}).

In addition to $(\phi_2,\phi^\dagger_2)\phi_1\to \phi_2\phi_1$, which was considered above, there are three other pairs of coupled self-scattering processes with $u$-channel enhancement which preserve the $Z_4$ symmetry: $(\phi_2,\phi^\dagger_2)\phi_1\to  \phi^\dagger_2\phi_1$, $(\phi^\dagger_2,\phi_2)\phi^\dagger_1\to \phi_2 \phi^\dagger_1$ and $(\phi^\dagger_2,\phi_2)\phi^\dagger_1\to \phi^\dagger_2\phi^\dagger_1$. Then, we find that there is an universal $u$-channel enhancement  in all the DM co-scattering processes, because the linear combinations of the DM pairs for the other channels \footnote{$A$ and $B$ stand for the ordered pair of the initial states in the DM co-scattering.}, $\psi_1=\frac{1}{\sqrt{N}}\big(\psi_A+\frac{g_2}{g_1} \psi_B\big)$, satisfy the same BS equations as for the case with $(\phi_2,\phi^\dagger_2)\phi_1\to \phi_2\phi_1$.  Actually, $(\phi^\dagger_2,\phi_2)\phi^\dagger_1\to \phi^\dagger_2\phi^\dagger_1$ is the CP conjugate of $(\phi_2,\phi^\dagger_2)\phi_1\to \phi_2\phi_1$, whereas $(\phi_2,\phi^\dagger_2)\phi_1\to  \phi^\dagger_2\phi_1$  is the CP conjugate of $(\phi^\dagger_2,\phi_2)\phi^\dagger_1\to \phi_2 \phi^\dagger_1$.

For the DM co-scattering, we can use the numerical solution for the DM co-scattering cross section, $\sigma^{\rm total}_{\psi_1\to \phi_1\phi_2}$, and perform the velocity average for dark matter in the halo. Then, we can define the effective $u$-channel scattering rate for dark matter as
\bea
\Gamma_{u,{\rm scatt}}=\frac{n_1n_2}{n^2_{\rm DM}}\,\langle  n_{\rm DM}\,\sigma^{\rm total}_{\psi_1\to \phi_1\phi_2} \,v\rangle \equiv \rho_{\rm DM}\, \frac{\langle\sigma v\rangle_{u,{\rm eff}}}{m_1} \label{scattrate}
\eea
where $\langle \sigma v\rangle_{u,{\rm eff}}=\langle\sigma^{\rm total}_{\psi_1\to\phi_1\phi_2}v\rangle\, r_1(1-r_1)/(\frac{m_2}{m_1}\,r_1+1-r_1)$, with $r_1$ being the fraction of $\phi_1$ in the DM relic abundance, namely, $r_1=\Omega_1/\Omega_{\rm DM}$. Here, we note that $n_{\rm DM}=n_1+n_2$ and $\rho_{\rm DM}=m_1 n_1+m_2 n_2$.

\begin{figure}[t]
\centering
\includegraphics[width=0.45\textwidth,clip]{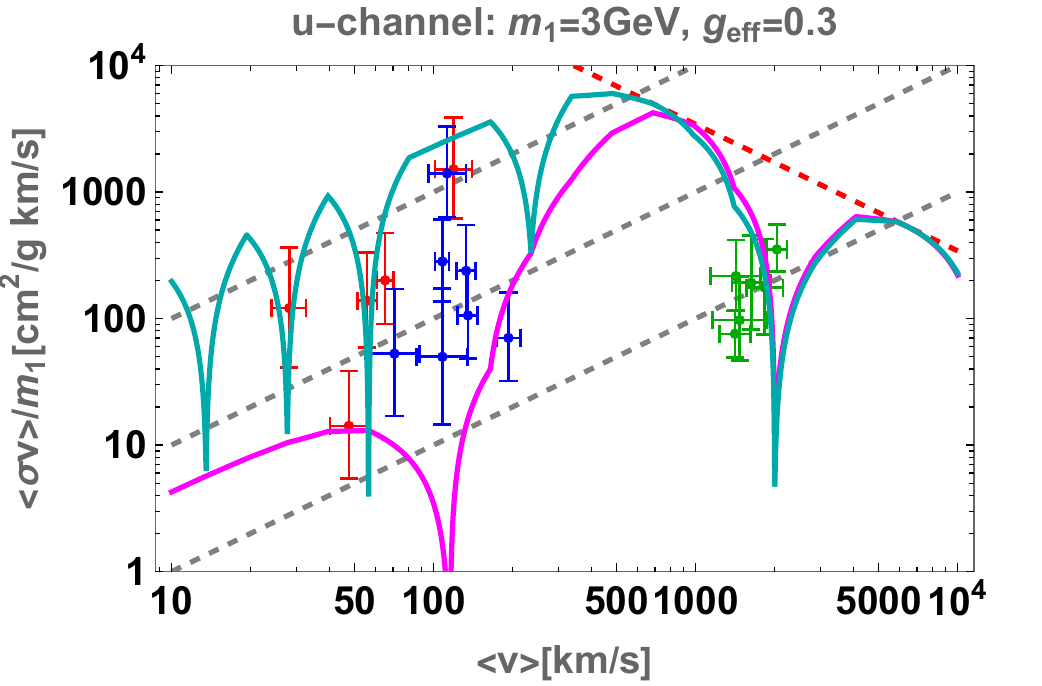} \,\,\,\,\
\includegraphics[width=0.45\textwidth,clip]{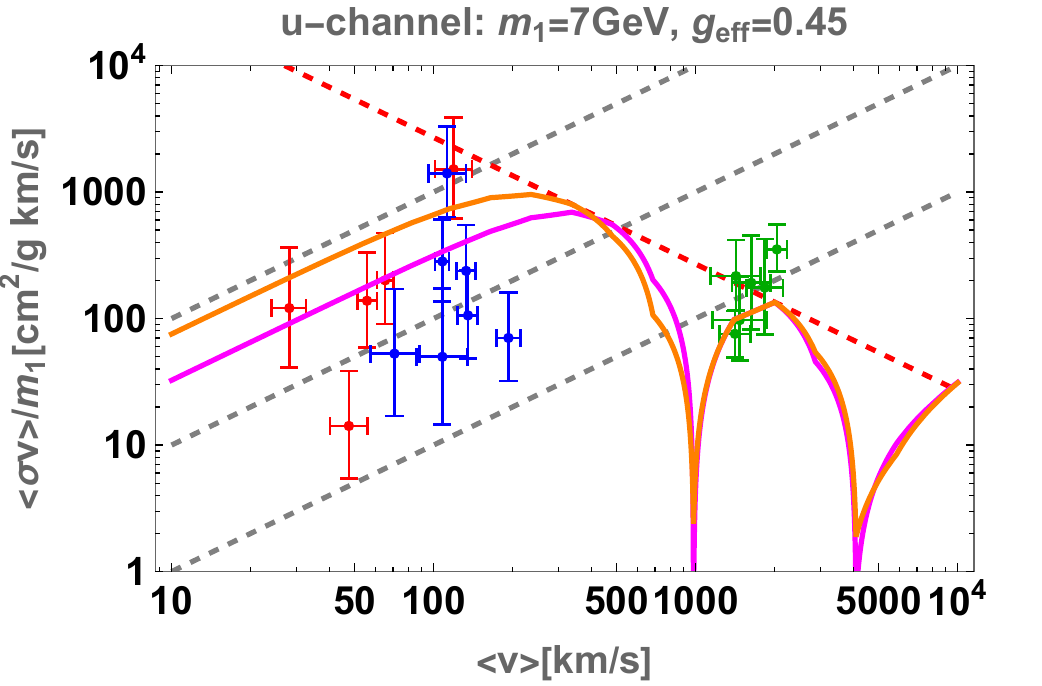} 
\caption{The effective $u$-channel self-scattering cross section for $\phi_1\phi_2\to \phi_1\phi_2$ as a function of the averaged velocity $\langle v\rangle$.  We chose the DM self-couplings as $g_{\rm eff}\equiv \sqrt{g^2_1+g^2_2}=0.3, 0.45$ on the left and right plots, respectively, and $r_1=\frac{1}{2}$ for both plots. The black dashed lines correspond to $\sigma_{\rm self}/m_1=10, 1, 0.1\,{\rm cm^2/g}$ from top to bottom, and the regions above the red dashed lines exceed the unitarity bound.
} 
\label{Fig:uch}
\end{figure}

In Fig.~\ref{Fig:uch}, we depict the effective $u$-channel self-scattering cross section  for $\phi_1\phi_2\to \phi_1\phi_2$  as a  function of the averaged velocity $\langle v\rangle$, for $m_1=3, 7\, {\rm GeV}$ on left and right plots, respectively.  We chose  $g_{\rm eff}\equiv \sqrt{g^2_1+g^2_2}=0.3, 0.45$ on the left and right plots, respectively, and $r_1=\frac{1}{2}$ for both plots.  For  $\Delta\equiv 1-m_2/(2m_1)$, we also took $\Delta=10^{-6}, 3\times 10^{-6}$ on left, and $\Delta= 10^{-6}, 2\times 10^{-6}$ on right, from top to bottom. In Fig.~\ref{Fig:uch}, we showed the fitted data used from Ref.~\cite{velocitydep} for comparison, but we don't pursue fitting the $u$-channel self-scattering cross section. Instead we just illustrate how  the self-scattering cross section can be made velocity-dependent due to the $u$-channel resonance such that the self-scattering gets suppressed at large DM velocities for galaxy clusters for perturbative DM couplings \cite{SRDM}.

\subsection{Some limiting cases of the Bethe-Salpeter equation}

By taking the BS wave function for the DM self-scattering with the $u$-channel resonances \cite{SRDM} as
\bea
\psi_{\rm BS}({\vec x})=\frac{u_l(x)}{x}\, Y^m_l(\theta,\phi),
\eea
with $x\equiv \frac{1}{2}\mu \alpha' r$ and $\alpha'\equiv(1+g^2_2/g^2_1)\alpha$,  we can rewrite the BS equation in eq.~(\ref{Schroeq}) with $E=\frac{1}{2}\mu v^2$ as
\bea
\bigg[\frac{d^2}{dx^2}-\frac{l(l+1)}{x^2}\bigg]u_l(x) +\frac{4}{b\,x}\, e^{-cx}(-1)^l u_l(bx)+a^2 u_l(x)=0 \label{Schroeq2}
\eea
where
\bea
a=\frac{2v}{\alpha'}, \quad b=\frac{m_2}{m_1}, \quad c=\frac{2M}{\mu\alpha'}.
\eea

Then, for $cx\lesssim 1$, we can approximate $e^{-cx}\simeq 1-cx$, so  the $s$-wave solution, $u_0(x)$, satisfies the following approximate equation,
\bea
\frac{d^2 u_0(x)}{dx^2}+\frac{4}{b\,x} u_0(bx) + \bigg(a^2 u_0(x)-\frac{4c}{b} u_0(bx)\bigg)=0. \label{Schroeq-swave}
\eea
As a result, for $a^2\gtrsim 4c/b$, namely,
\bea
v\gtrsim \sqrt{\frac{2\alpha' M m_1}{m_2\mu}}\simeq \sqrt{\frac{3\alpha' M}{2m_1}}
\eea
where $m_2\simeq 2m_1$ is taken on the right-hand side,
the BS equation for $s$-wave in eq.~(\ref{Schroeq-swave}) becomes Coulomb-like or weakly-coupled (Coulomb limit), as far as the non-linear wave function is comparable to the linear counterpart, $u_0(bx)\sim u_0(x)$, for $cx\lesssim 1$. In this case, we can apply the standard results for the DM self-scattering or the Sommerfeld factor for the case with the Coulomb potential \cite{self-scattering}. 
We note that the lower bound on the DM velocity in the Coulomb limit can be rewritten as the upper bound on $\Delta\equiv 1-m_2/(2m_1)$, as follows,
\bea
\Delta \lesssim \frac{v^4}{18\alpha^{\prime 2}}.
\eea
Thus, for a given averaged DM velocity in the dwarf galaxies, the amount of detuning from $m_2=2m_1$ is constrained.

On the other hand, in the opposite limit with $a^2\lesssim 4c/b$ or $v\lesssim  \sqrt{\frac{3\alpha' M}{2m_1}}$,  the BS equation in  eq.~(\ref{Schroeq-swave}) becomes the radial equation for the hydrogen atom for $u_0(bx)\sim u_0(x)$, so the DM bound states  between $\phi_1$ and $\phi_2$ form only for $4c/b\sim \frac{1}{n^2}$ with $n$ being integer or
\bea
\frac{8M m_1}{\alpha' \mu m_2}\sim \frac{1}{n^2}.
\eea 
Then, for $m_2\simeq 2m_1$, the above resonant condition for the DM bound states with the $u$-channel becomes
\bea
m_1\sim \frac{12M}{\alpha'}\, n^2 \label{rescond}
\eea
or
\bea
\Delta\sim \frac{\alpha^{\prime 2}}{1152 n^4}.
\eea
We note that the combination of the above resonant condition with the non-perturbative limit, namely, $v\lesssim  \sqrt{\frac{3\alpha' M}{2m_1}}$, gives rise to the upper bound on the DM velocity by $v\lesssim \frac{1}{2\sqrt{2}n} \alpha'$.

Even if the DM bound states do not form, the self-scattering cross section \cite{boundstate1,self-scattering} and the DM annihilation cross section \cite{boundstate2,self-scattering} can be enhanced significantly near the resonant conditions.
In particular, the resonant condition in eq.~(\ref{rescond}) leads to the phase shift for the DM co-scattering with the $u$-channel being $\delta_0\simeq \frac{\pi}{2}$ \cite{boundstate1,self-scattering}, so the corresponding DM co-scattering cross section in eq.~(\ref{coscatt}) and the  co-scattering rate in eq.~(\ref{scattrate}) becomes maximized, because $\sigma^{\rm total}_{\psi_1\to \phi_1\phi_2}\simeq \frac{4\pi}{k^2}$.

\subsection{Comparison with $t$-channel case and enhancement factors for $u$-channel case}

We consider the analytic results for the Yukawa type potential, $V(r)=-\frac{\alpha'}{r}\,e^{-Mr}$, in the case of the $t$-channel resonances for the DM self-scattering and interpret them for the DM co-scattering cross section and the Sommerfeld factor for the DM annihilation in the case of the $u$-channel resonances in our work.

There exist analytic solutions to the BS equation in the case for the DM self-scattering with the $t$-channel resonances, when the Yukawa potential is replaced by the Hulth\'en potential,
\bea
V_H(r)=  -\frac{\alpha'\, \delta \,e^{-\delta r}}{1-e^{-\delta r}},
\eea
with $\delta=\frac{\pi^2}{6}\, M$. Then, the phase shift $\delta_0$ for the $s$-wave self-scattering for dark matter \cite{boundstate1,self-scattering} is given by
\bea
\delta_0={\rm arg}\bigg(\frac{i\Gamma(\lambda_++\lambda_--2)}{\Gamma(\lambda_+)\Gamma(\lambda_-)} \bigg) \label{phaseshift}
\eea
where
\bea
\lambda_\pm =1+iw \pm w\sqrt{\frac{y}{w}-1},
\eea
with $w=\frac{k}{\delta}$ and $y=\frac{\alpha'}{v}$. Moreover, the Sommerfeld factor for the $s$-wave annihilation for dark matter is also given \cite{boundstate2,self-scattering} by
\bea
S_0&=&\bigg|\frac{\Gamma(\lambda_+)\Gamma(\lambda_-)}{\Gamma(\lambda_++\lambda_--1)}\bigg| \nonumber \\
&=& \frac{\pi y \sinh(2\pi w)}{\cosh(2\pi w)-\cos\big(2\pi w\sqrt{\frac{y}{w}-1}\big)}. \label{Sommerfeld}
\eea
We note that the Sommerfeld factor can be recast into another form, which is more convenient when the Coulomb limit with  $y/w\lesssim 1$ is taken, in the following,
\bea
S_0=\frac{\pi y \sinh(2\pi w)}{2\sinh[\pi w(1-\sqrt{1-\frac{y}{w}})] \sinh[\pi w(1+\sqrt{1-\frac{y}{w}})]}.
\eea

Using the above results from the $t$-channel resonances, we interpret the conditions for the large self-scattering and the Sommerfeld factor for dark matter in our case.  We first note that $w\gtrsim 1$ is the classical regime and $w\lesssim 1$ is the quantum regime; $y/w\lesssim 1$ is the Coulomb limit (or weakly-coupled regime) and $y/w\gtrsim 1$ is the strongly-coupled regime \cite{Colquhoun:2020adl}. 

First, in the Coulomb limit with $y/w\lesssim 1$, which corresponds to $\alpha' \delta/(kv)\lesssim 1$, the Sommerfeld factor becomes
\bea
S_0\simeq \frac{\pi \alpha'}{v}\, \frac{1}{1-e^{- \frac{\pi \alpha'}{v}}},
\eea
and the phase shift becomes 
\bea
\delta_0&=&{\rm arg}\bigg(\frac{\Gamma(\lambda_++\lambda_--1)}{\Gamma(\lambda_+)\Gamma(\lambda_-)} \bigg) \nonumber \\
&\simeq & {\rm arg}\bigg(\Gamma\Big(1-i\frac{y}{2}\Big)\bigg) +\frac{1}{2}y\ln (2w) \label{phaseC}
\eea
where we used $\Gamma(\lambda_++\lambda_--1)=(\lambda_++\lambda_--2)\Gamma(\lambda_++\lambda_--2)=2iw \Gamma(\lambda_++\lambda_--2)$ in the first line and we took $w=k/\delta\gtrsim 1$ (classical) in the second line.
Then, the first term in eq.~(\ref{phaseC}) is the $l$-dependent phase shift and the second term corresponds to the universal phase shift \cite{Landau}, which diverges in the limit of $\delta\to 0$.
As a result, for $y=\alpha'/v\gtrsim 1$ (non-relativistic), we obtain 
\bea
\delta_0 \simeq \frac{\alpha'}{2v}\ln \bigg(\frac{k v}{\delta \alpha'}\bigg)+\frac{\alpha'}{2v} -\frac{\pi}{4}.
\eea
Therefore, for $\alpha'/v\gtrsim 1$, the Sommerfeld factor becomes larger than unity and the phase shift varies between $0$ and $\frac{\pi}{2}$.

Second, from eq.~(\ref{phaseshift}),  the resonance condition for the DM self-scattering, $\delta_0=\frac{\pi}{2}$, is achieved for $\lambda_-\simeq -n$ with $n$ being zero or negative integer, namely,
\bea
w\sqrt{\frac{y}{w}-1}\simeq n+1, \label{exactres}
\eea
for $w\lesssim 1$ (quantum).
Thus,  for  $y/w\gtrsim 1$ (strongly-coupled), the resonance condition becomes 
\bea
M=\frac{4\alpha' m_1}{\pi^2} \,\frac{1}{(n+1)^2}, \quad n=0,-1,-2,-3,\cdots, \label{resmass}
\eea
which corresponds to eq.~(\ref{rescond}) in the previous subsection. Here, we took $\mu\simeq \frac{2}{3} m_1$ for $m_2\simeq 2m_1$.
We note that under the condition given by eq.~(\ref{exactres}), the Sommerfeld factor in eq.~(\ref{Sommerfeld}) becomes
\bea
S_0 &\simeq& \frac{\pi y\sinh(2\pi w)}{\cosh(2\pi w)-1} \nonumber \\
&\simeq & \frac{y}{w}\simeq \frac{\alpha^{\prime 2}}{\pi^2 v^2}
\eea
where we took $w\lesssim 1$ and eq.~(\ref{resmass}) in the second line. Therefore, for $w\lesssim 1$, the phase shift can be made close to $\frac{\pi}{2}$ by the resonance condition and the Sommerfeld factor becomes larger than unity for $\alpha'/(\pi v)\gtrsim 1$.

We also remark that away from the poles of the gamma function, a low momentum expansion of the phase shift can be taken in the effective range theory, as follows,
\bea
k\cot\delta_0 =-\frac{1}{a} +\frac{1}{2} r_0 k^2,
\eea 
with 
\bea
a&=& \frac{1}{\delta}\Big(\psi^{(0)}(1+\eta)+\psi^{(0)}(1-\eta)+2\gamma \Big), \\
r_0 &=& \frac{2}{3}a -\frac{1}{3\delta \eta} \bigg[ \psi^{(0)}(1+\eta)+ \psi^{(0)}(1-\eta)+2\gamma \bigg]^{-2} \nonumber \\
&\times& \bigg[3\Big(  \psi^{(1)}(1+\eta)- \psi^{(1)}(1-\eta)\Big)+\eta \Big( \psi^{(2)}(1+\eta)+ \psi^{(2)}(1-\eta)+16\zeta(3)\Big) \bigg].
\eea
Here, $\eta\equiv yw=\frac{\alpha' \mu}{\delta}$ is a velocity-dependent quantity,  $\gamma\simeq 0.5772$ is the Euler-Mascheroni constant, $\psi^{(n)}(z)$  are the polygamma functions of order $n$, and $\zeta(3)$ is the Riemann zeta function. 
As a result, applying to the DM co-scattering cross section in eq.~(\ref{coscatt}), we get
\bea
\sigma^{\rm total}_{\psi_1\to\phi_1\phi_2} \simeq \frac{4\pi a^2}{1+k^2 (a^2-a r_0)+\frac{1}{4}a^2 r^2_0 k^4}. 
\eea
We note that the effective range term can be ignored for $k\ll 1/a$ in the Born approximation, so $\sigma^{\rm total}_{\psi_1\to\phi_1\phi_2}\simeq 4\pi a^2$, which is velocity-independent.

\section{DM self-scattering with $s$-channel resonances}

We discuss the implications of the resonant condition taken in the $u$-channel co-scattering processes for the $s$-channel processes for self-annihilation processes.  We perform the velocity averaged self-scattering cross section for the lighter component of dark matter and compare it with the $u$-channel co-scattering case.

\subsection{Dark matter self-scattering at tree level}

We also consider the elastic scattering, $\phi_1 \phi_1\rightarrow \phi^{(\dagger)}_1\phi^{(\dagger)}_1$ and its complex conjugate, for which the dark scalars, $s, a$, contributes to the $s$-channel. In this case, the contribution from the dark scalars $s,a$ can be important because they are absolutely stable for $m_s, m_a<2m_1$. On the other hand, in the case of $\phi^*_1\phi_1\rightarrow \phi^*_1\phi_1$, $ss\to ss, aa\to aa$,  and $sa\to sa$ (or $\phi_2\phi_2\to  \phi^{(\dagger)}_2\phi^{(\dagger)}_2$ and $\phi^\dagger_2\phi_2\to\phi^\dagger_2\phi_2$ for degenerate masses for $s$ and $a$), there are also $s$-channel contributions from the dark Higgs $h_X$ or the extra gauge boson $X$, but they are limited by nonzero decay widths. Thus, we focus on $\phi_1 \phi_1\rightarrow \phi^{(\dagger)}_1\phi^{(\dagger)}_1$ with the DM $s$-channel resonances.

For the DM scattering process, $\phi_1(q) \phi_1(p)\rightarrow \phi_1(q') \phi_1(p')$, with $s,a$ in the $s$-channels,
the corresponding tree-level scattering amplitude ${\widetilde \Gamma}_{s,1}(p,q;p,q')$ is given by
\bea
 {\widetilde \Gamma}_{s,1}(p,q;p',q') &=&-\frac{4g^2_s m^2_1}{(p+q)^2-m^2_s}-\frac{4g^2_a m^2_1}{(p+q)^2-m^2_a} \nonumber \\
 &=&\frac{4g^2_s m^2_1}{|{\vec p}+{\vec q}|^2+m^2_s-E^2_{\rm cm}}+\frac{4g^2_a m^2_1}{|{\vec p}+{\vec q}|^2+m^2_a-E^2_{\rm cm}}, \label{schannel1}
\eea
with $E_{\rm cm}=p_0+q_0$ being the COM energy.
For the non-relativistic limit for dark matter, $E_{\rm cm}\approx 2m_1+\frac{1}{2m_1}({\vec p}^2+{\vec q}^2)$, so we can approximate eq.~(\ref{schannel1}) in the COM frame to
\bea
 {\widetilde \Gamma}_{s,1}(p,q;p',q')\approx \frac{4g^2_s m^2_1}{m^2_s-4m^2_1-4{\vec p}^2}+\frac{4g^2_a m^2_1}{m^2_a-4m^2_1-4{\vec p}^2}. 
\eea
Thus, for ${\vec p}=0$,  the above tree-level scattering amplitude diverges at $m_s=2m_1$ or $m_a=2m_1$, similarly to the case with the $u$-channel resonances. In general, we can take $m_s, m_a<2m_1$ such that the lighter one between $s$ and $a$ is stable, so there is no decay width appearing in the propagator for either $s$ or $a$.

Similarly, for the DM scattering process, $\phi_1(q) \phi_1(p)\rightarrow \phi^\dagger_1(q') \phi^\dagger_1(p')$, with $s,a$ in the $s$-channels,
the corresponding tree-level scattering amplitude ${\widetilde \Gamma}_{s,2}(p,q;p,q')$ is given by
\bea
 {\widetilde \Gamma}_{s,2}(p,q;p',q') &=&-\frac{4g^2_s m^2_1}{(p+q)^2-m^2_s}+\frac{4g^2_a m^2_1}{(p+q)^2-m^2_a} \nonumber \\
 &=&\frac{4g^2_s m^2_1}{|{\vec p}+{\vec q}|^2+m^2_s-E^2_{\rm cm}}-\frac{4g^2_a m^2_1}{|{\vec p}+{\vec q}|^2+m^2_a-E^2_{\rm cm}}.
 \label{schannel2}
\eea
Then, in the non-relativistic limit for dark matter, we can approximate eq.~(\ref{schannel2}) in the COM frame to
\bea
 {\widetilde \Gamma}_{s,2}(p,q;p',q')\approx \frac{4g^2_s m^2_1}{m^2_s-4m^2_1-4{\vec p}^2}-\frac{4g^2_a m^2_1}{m^2_a-4m^2_1-4{\vec p}^2}. 
\eea
Thus, for ${\vec p}=0$,  the above tree-level scattering amplitude diverges at $m_s=2m_1$ or $m_a=2m_1$, similarly to the case with the $u$-channel resonances. As compared to $ {\widetilde \Gamma}_{s,1}$, there is a deconstructive interference between $s$ and $a$ contributions, so $ {\widetilde \Gamma}_{s,2}$ vanishes in the limit of an unbroken $U(1)'$, namely, for $g_s=g_a$ and $m_s=m_a$.

\begin{figure}[t]
\centering
\includegraphics[width=0.20\textwidth,clip]{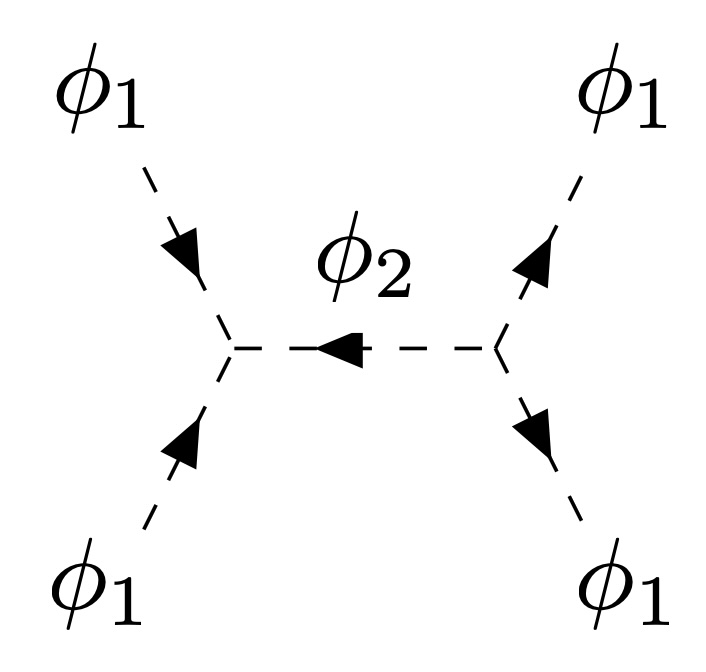} \,\,\,\,\,\,\,\,
\includegraphics[width=0.20\textwidth,clip]{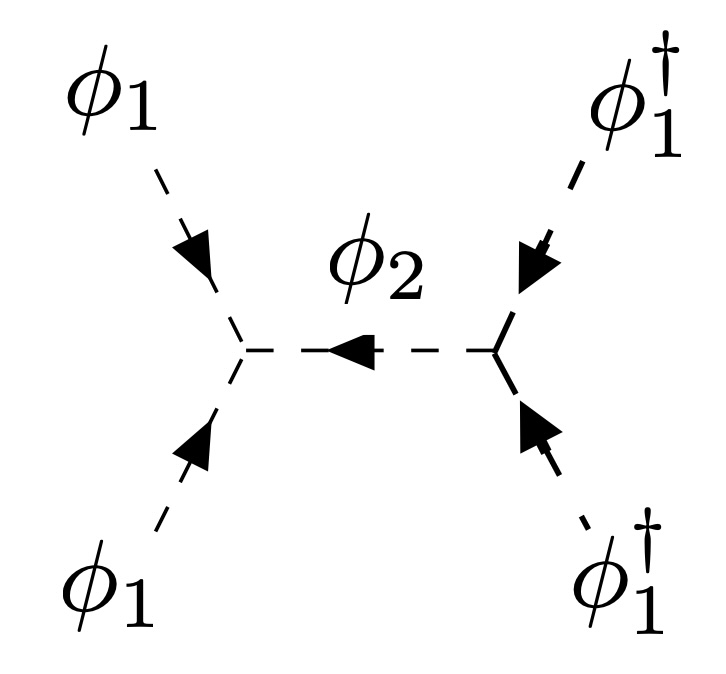} 
\caption{Tree-level Feynman diagrams for the DM self-scattering processes with $s$-channel, $\phi_1\phi_1\to \phi_1\phi_1$ and $\phi_1\phi_1\to \phi^\dagger_1\phi_1^\dagger$.
} 
\label{Fig:sch-diag}
\end{figure}

In the following discussion, as in the $u$-channel case, we focus the case that the real and imaginary parts of the complex scalar $\phi_2$ have degenerate masses, namely, for $m_s=m_a=m_2$. In this case, from the tree-level Feynman diagrams shown in Fig.~\ref{Fig:sch-diag}, the $s$-channel self-scattering amplitudes become
\bea
 {\widetilde \Gamma}_{s,1}(p,q;p',q') &=&-\frac{4(g^2_1+g^2_2) m^2_1}{(p+q)^2-m^2_2}, \\
  {\widetilde \Gamma}_{s,2}(p,q;p',q') &=&-\frac{8g_1g_2 m^2_1}{(p+q)^2-m^2_2}.
\eea
Then, we can take $m_2<2m_1$ such that $\phi_2$ is absolutely stable, so there is no decay width appearing in the propagator for $\phi_2$. For the general case with split masses for $\phi_2$, as far as the resonant mass condition is satisfied between $\phi_1$ and the lighter component of $\phi_2$, the current discussion on the $s$-channel case  is still valid.

The DM self-scattering cross section with $s$-channel dominance, $\phi_1\phi_1\to \phi^{(\dagger)}_1 \phi^{(\dagger)}_1$, is, at tree level, given by
\bea
\sigma_{\phi_1\phi_1\to \phi^{(\dagger)}_1 \phi^{(\dagger)}_1}=\frac{1}{16\pi s} \Big| {\widetilde \Gamma}_{s,\phi_1\phi_1\to \phi^{(\dagger)}_1 \phi^{(\dagger)}_1}\Big|^2.
\eea
Then,  again considering the case with $m_s=m_a=m_2$ and taking the non-relativistic limit for $\phi_1$,  the above DM self-scattering cross section  becomes
\bea
\sigma_{\phi_1\phi_1\to \phi^{(\dagger)}_1 \phi^{(\dagger)}_1}\simeq \frac{{\tilde g}^4}{8\pi m^2_1}\, \frac{1}{(v^2_c+v^2)^2}, \label{selfscatt}
\eea
with $v_c^2\equiv 4\Delta(2-\Delta)$, $\Delta\equiv 1-m_2/(2m_1)\geq 0$, and 
\bea
{\tilde g}\equiv \left\{ \begin{array}{c}\sqrt{g^2_1+g^2_2}, \quad \phi_1\phi_1\to \phi_1\phi_1, \\  \sqrt{2g_1 g_2},  \quad \phi_1\phi_1\to \phi^\dagger_1\phi^\dagger_1. \end{array} \right.
\eea

We comment on the $s$-channel self-scattering in comparison to the $u$-channel co-scattering discussed in the previous section. Comparing the  $u$-channel co-scattering cross section in eq.~(\ref{coscatt}) and the $s$-channel self-scattering cross section for $\phi_1\phi_1\to \phi_1\phi_1$ in eq.~(\ref{selfscatt}), the former is dominant if
\bea
\frac{4\pi}{k^2}\,\sin^2\delta_0\gtrsim \frac{2\pi \alpha^{\prime 2}}{m^2_1(v^2_c+v^2)^2}.
\eea
For $v\gtrsim v_c\simeq \sqrt{8\Delta}$, we need $v^2\gtrsim \frac{2}{9}{\alpha^{\prime 2}}/\sin^2\delta_0$, for which the enhancement factor for the $u$-channel is mild for $\delta_0\simeq \frac{\pi}{2}$, as discussed in the previous section.  On the other hand, for $v\lesssim  v_c\simeq \sqrt{8\Delta}$ and $\delta_0\simeq \frac{\pi}{2}$,  we need $v^2\lesssim 288 (\Delta/\alpha')^2$. Therefore, taking $v^2\sim 10^{-8}$ at dwarf galaxies, we find that the $u$-channel co-scattering is dominant over the $s$-channel self-scattering, as long as $\alpha'\lesssim 0.17(\Delta/10^{-6})\,|\sin\delta_0|$. However, for a small $\delta_0$, namely, off the resonance, we also need to take into account the $s$-channel self-scattering.
A similar conclusion can be drawn for the cross section for another $s$-channel self-scattering process, $\phi_1\phi_1\to\phi^\dagger_1\phi^\dagger_1$, with $g_1=g_2$.

\subsection{Velocity-averaged self-scattering}

There are several definitions for the averaged self-scattering cross section, such as the velocity-averaged cross section $\langle \sigma v\rangle$, the momentum-transfer cross section $\langle \sigma v^2\rangle$, the energy-transfer cross section, $\langle \sigma v^3\rangle$ \cite{Colquhoun:2020adl}, etc.
In this work, we consider the velocity-averaged self-scattering cross section, as follows,
\bea
\langle\sigma v\rangle=\int^{v_{\rm esp}}_0 f(v,v_0) \,(\sigma v) dv, \label{velocityave}
\eea
where the Maxwell-Boltzmann distribution for the dark matter velocity is given by
\bea
f(v,v_0)=\frac{4v^2}{v^3_0 \sqrt{\pi}}\, e^{-v^2/v^2_0},
\eea
with $v_0$ being related to the averaged velocity by $\langle v\rangle\simeq 2v_0/\sqrt{\pi}$,
and $v_{\rm esp}$ is the escape velocity in the dark matter halo. 
Here, $\sigma $ can be $\sigma^{\rm total}_{\phi_1 A\to \phi_1 A}$ in eq.~(\ref{coscatt}) for the DM co-scattering or $\sigma_{\phi_1\phi_1\to \phi^{(\dagger)}_1 \phi^{(\dagger)}_1}$ in eq.~(\ref{selfscatt}) for the DM self-scattering.

For the DM self-scattering, we obtain the velocity-averaged cross section as
\bea
\langle\sigma v\rangle_{\phi_1\phi_1\to \phi^{(\dagger)}_1 \phi^{(\dagger)}_1}=\frac{{\tilde g}^4}{2\pi^{3/2} m^2_1 v^3_0}\, \cdot F\Big(\frac{v^2_c}{v^2_0}\Big),
\eea
with
\bea
F(a^2)&=& \int^{\frac{v_{\rm esp}}{v_0}}_0\, \frac{t^3\,  e^{-t^2}}{(a^2+t^2)^2}\,dt \nonumber \\
&\simeq & \frac{1}{2} \Big((1+a)\,e^a \Gamma(0,a)-1 \Big).
\eea
Here,  $\Gamma(0,a)$ is the incomplete gamma function given by
\bea
\Gamma(0,a)=\int^\infty_a \,\frac{e^{-t}}{t}\, dt.
\eea

Therefore, from the effective $s$-channel scattering rate for dark matter, 
\bea
\Gamma_{s,{\rm scatt}}=\frac{n^2_1}{n^2_{\rm DM}}\, \langle n_1\sigma^{\rm total}_s v\rangle \equiv \rho_{\rm DM}\, \frac{\langle\sigma v\rangle_{s,{\rm eff}}}{m_1},
\eea
with $\sigma^{\rm total}_s=\sigma_{\phi_1\phi_1\to \phi_1 \phi_1}+\sigma_{\phi_1\phi_1\to \phi^\dagger_1 \phi^\dagger_1}$,
we define the effective $s$-channel self-scattering cross section as $\langle\sigma v\rangle_{s,{\rm eff}}= \langle \sigma^{\rm total}_s v\rangle\, r_1^3/(\frac{m_2}{m_1}\,r_1+1-r_1)^2$.

\begin{figure}[t]
\centering
\includegraphics[width=0.45\textwidth,clip]{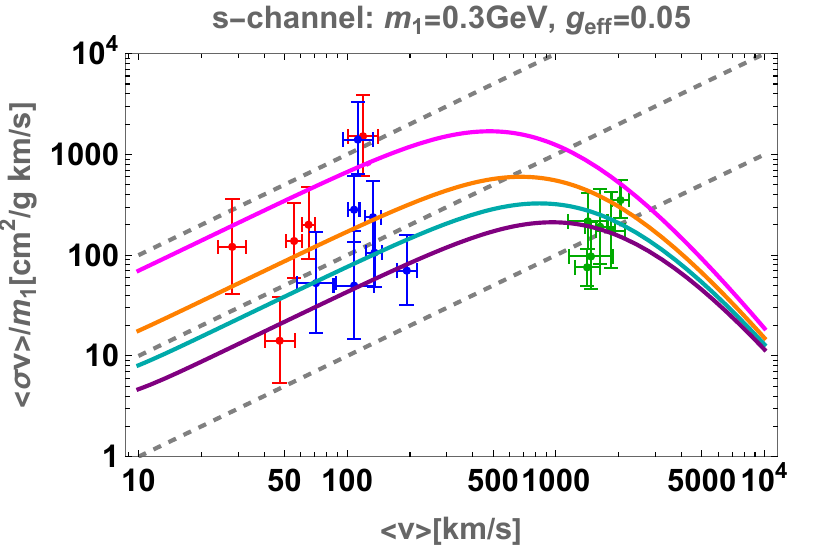} \,\,\,\,\
\includegraphics[width=0.45\textwidth,clip]{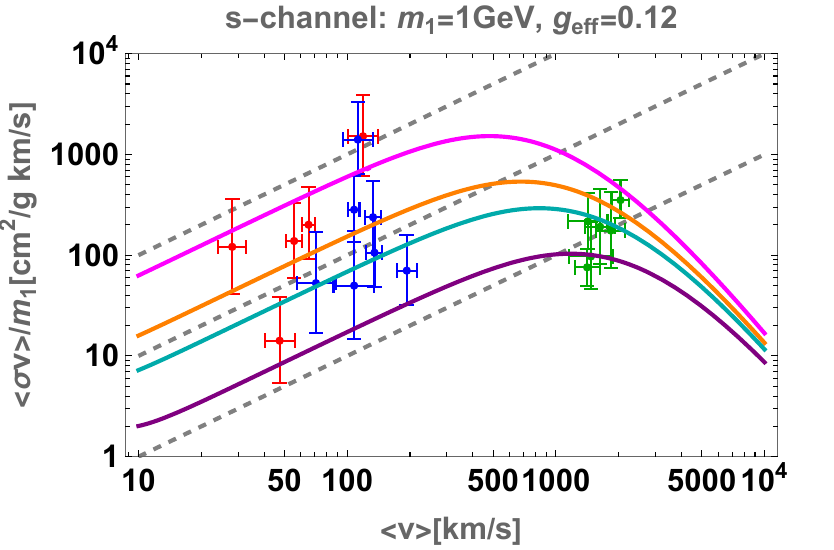} \vspace{0.3cm} \\
\includegraphics[width=0.45\textwidth,clip]{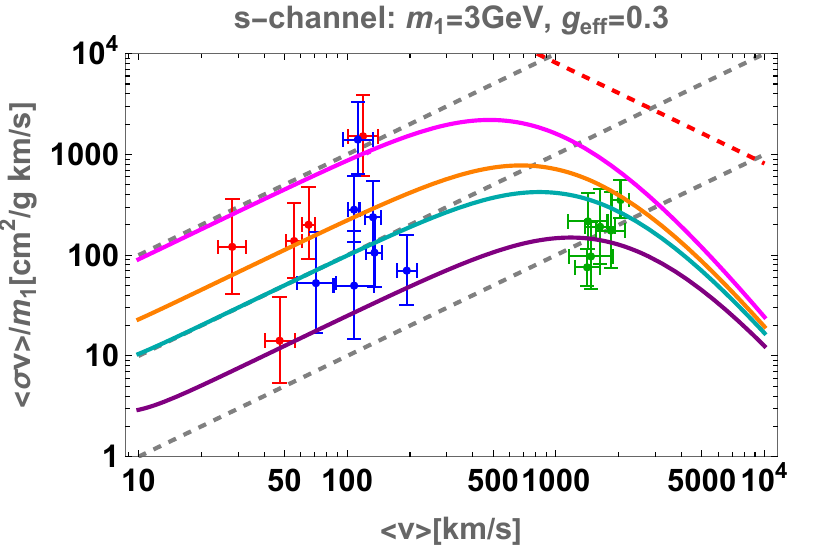} \,\,\,\,\
\includegraphics[width=0.45\textwidth,clip]{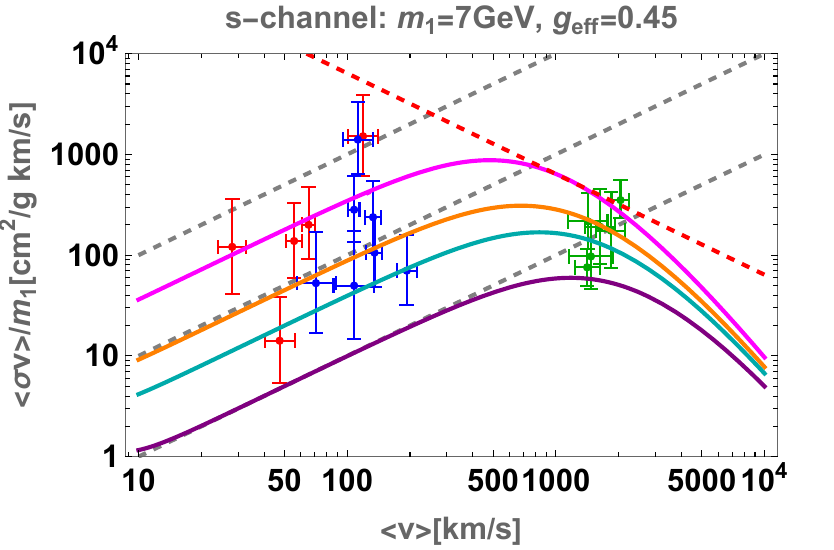}
\caption{The effective $s$-channel self-scattering cross section for $\phi_1\phi_1\to \phi^{(\dagger)}_1\phi^{(\dagger)}_1$ as a function of the averaged velocity $\langle v\rangle$.  The effective coupling is given by $g_{\rm eff}=(g^4_1+g^4_2+6g_1^2g^2_2)^{\frac{1}{4}}$. We chose $r_1=\frac{1}{2}$ for all the plots. The black dashed lines correspond to $\sigma_{\rm self}/m_1=10, 1, 0.1\,{\rm cm^2/g}$ from top to bottom, and the regions above the red dashed lines exceed the unitarity bound.
} 
\label{Fig:sch}
\end{figure}

We comment on the limits of the velocity-averaged cross section for the $s$-channel self-scattering. 
Using $\Gamma(0,a)\simeq -\gamma-\ln a$ for $a\ll 1$, we obtain the velocity integral for $v_c\ll v_0$ as
\bea
F\Big(\frac{v^2_c}{v^2_0}\Big)\simeq \frac{1}{2}\ln \frac{v^2_0}{v^2_c}\simeq \frac{1}{2}\ln \frac{v^2_0}{8\Delta}
\eea
where we approximated $v^2_c\simeq 8\Delta$ for $\Delta\ll 1$. Therefore, there is a logarithmic enhancement for the DM self-scattering cross section due to the $s$-channel resonance, but it is bounded by unitarity as $\langle \sigma v\rangle\leq 4\pi \langle\frac{v}{k^2}\rangle\equiv \langle\sigma v\rangle_{\rm max}$. For $k=\frac{1}{2}m_1 v$, we obtain $\langle\sigma v\rangle_{\rm max}=32\sqrt{\pi}/(m^2_1 v_0)$.
On the other hand, for $v_c\gg v_0$, the velocity integral becomes
\bea
F\Big(\frac{v^2_c}{v^2_0}\Big)\simeq  \frac{v^2_0}{2v^2_c}=\frac{v^2_0}{16\Delta(2-\Delta)}.
\eea
In this case, there is no enhancement for the DM self-scattering.

In Fig.~\ref{Fig:sch}, we depict the effective $s$-channel self-scattering cross section for $\phi_1\phi_1\to \phi^{(\dagger)}_1\phi^{(\dagger)}_1$ as a function of the averaged velocity $\langle v\rangle$, for $r_1=\frac{1}{2}$ and various choices of $m_1$ and $g_{\rm eff}=(g^4_1+g^4_2+6g_1^2g^2_2)^{\frac{1}{4}}$. We also chose $\Delta=10^{-6}, 2\times 10^{-6}, 3\times 10^{-6}, 6\times 10^{-6}$, in the order of magenta, orange, cyan and purple lines (namely, from top to bottom).
In the plots in the lower panel of Fig.~\ref{Fig:sch}, the regions above the red dashed lines exceed the unitarity bound.
In Fig.~\ref{Fig:sch}, we also showed the simulation data used from Ref.~\cite{velocitydep} for comparison, but we don't pursue fitting the $s$-channel self-scattering cross section.

As a result, we find that the self-scattering cross section becomes velocity-dependent due to the $s$-channel resonance such that it becomes suppressed at large DM velocities for galaxy clusters. However, for perturbative DM couplings as in the upper panel in Fig.~\ref{Fig:sch}, for a sub-GeV scale DM mass, we need $\Delta\sim 10^{-6}$ to get the velocity-dependent self-scattering from the $s$-channel consistent with the observed data. For instance, for $m_1=3\,{\rm GeV}$ and  $g_{\rm eff}=0.3$, we can take a wide range of $\Delta$ such as $\Delta=1-6\times 10^{-6}$ for the $s$-channel case, while  a smaller $\Delta$ such as $\Delta=(1-3)\times 10^{-6}$ is favored for the $u$-channel case.  Therefore, the $s$-channel enhancement for the DM self-scattering is less efficient than in the case with the $u$-channels.

\section{Relic density for dark matter}

We present the Boltzmann equations governing the relic densities for two-component dark matter in our model and show some benchmark models for DM and mediator masses of weak-scale or sub-GeV-to-GeV scale, satisfying the correct relic density.

\subsection{Dark matter annihilations}

Multi-component dark matter in our model can annihilate through the self-interactions in the dark sector in eq.~(\ref{scalars1}) such as a $U(1)'$ invariant term, $\phi^\dagger_2 \phi^2_1$, a $Z_4$ invariant term, $\phi_2 \phi^2_1$, and the interactions involving the dark Higgs in eq.~(\ref{scalars2}) and the extra gauge boson in eq.~(\ref{gauge}). 

We consider the benchmark models with $m_s\simeq 2m_1$ or $m_a\simeq 2m_1$ for the $u$-channel resonances in our model.
Thus, the self-interactions for dark matter give rise to the self-annihilations, $ss(aa)\to \phi_1\phi^\dagger_1$; the dark Higgs and the extra gauge boson lead to the extra  self-annihilations, $ss(aa), \phi_1\phi^\dagger_1\to f\,{\bar f}, VV, h_1 h_1, h_2 h_2, h_1 h_2, XX$, with $f$ being the SM fermions and $V=W,Z$, and the co-annihilations, $sa\to f\,{\bar f}$, $\phi^{(\dagger)}_1 s(a)\to \phi^{(\dagger)}_1 h_1 (h_2)$, $\phi^{(\dagger)}_1 s(a)\to \phi^{(\dagger)}_1 X$, $\phi_1 s, \phi_1 a\to \phi^\dagger_1 h_1(h_2)$, $\phi^{(\dagger)}_1 s(a)\to \phi^{(\dagger)}_1 a(s)$.

In particular, for $m_s=m_a=m_2$,  the self-interactions for dark matter give rise to the self-annihilations, $\phi_2\phi^\dagger_2\to \phi_1\phi^\dagger_1$, $\phi_2\phi_2\to \phi_1\phi^\dagger_1$, $\phi^\dagger_2 \phi^\dagger_2\to \phi_1\phi^\dagger_1$; the dark Higgs and the extra gauge boson lead to the extra  self-annihilations, $\phi_2\phi^\dagger_2, \phi_1\phi^\dagger_1\to f\,{\bar f}, VV,  h_1 h_1, h_2 h_2,h_1 h_2, XX$,  and the co-annihilations, $\phi_2\phi^\dagger_1\to \phi_1 X(h_1, h_2)$, $\phi^\dagger_2\phi^\dagger_1\to \phi_1 X(h_1,h_2)$, and their complex conjugates.

\subsection{Boltzmann equations}

For simplicity, we consider the case with $m_s=m_a=m_2$.
We also assume that two-component dark matter keeps in kinetic equilibrium during the freeze-out process \footnote{The co-annihilations, $\phi_2\phi^\dagger_1\to \phi_1 X(h_1, h_2)$, $\phi^\dagger_2\phi^\dagger_1\to \phi_1 X(h_1,h_2)$, could make the dark matter $\phi_1$ heated  due to the mass due to the mass difference between $\phi_2$ and the mediator particles. But, the Sommerfeld effects for the co-annihilation processes during the freeze-out are small and the elastic scattering between $\phi_1$ and the SM particles can maintain the kinetic equilibrium for dark matter, so the heating effect is small.}. For instance, relatively small couplings from dark photon or Higgs portals in our model are sufficient to maintain the two-component dark matter in kinetic equilibrium \cite{gmix1,gmix2}.  
Then, the evolution of the number densities are governed by the Boltzmann equations,
\bea
{\dot n}_{\phi_1}+3H n_{\phi_1} &=& -\langle\sigma v\rangle_{\phi_1\phi^\dagger_1\to f\,{\bar f}, VV,h_1 h_1, h_2 h_2,h_1 h_2,XX}\Big(n_{\phi_1}n_{\phi^\dagger_1}-n^{\rm eq}_{\phi_1}n^{\rm eq}_{\phi^\dagger_1}\Big) \nonumber  \\
&&+\langle\sigma v\rangle_{\phi_2\phi^\dagger_2\to \phi_1\phi^\dagger_1} \Big(n_{\phi_2}n_{\phi^\dagger_2}- n_{\phi_1}n_{\phi^\dagger_1}\Big) \nonumber \\
&&+\langle\sigma v\rangle_{\phi_2\phi_2\to  \phi_1\phi^\dagger_1} \Big(n_{\phi_2}^2 - n_{\phi_1}n_{\phi^\dagger_1}\Big) \nonumber \\
&&+\langle\sigma v\rangle_{\phi^\dagger_2\phi^\dagger_2\to  \phi_1\phi^\dagger_1} \Big(n_{\phi^\dagger_2}^2 - n_{\phi_1}n_{\phi^\dagger_1}\Big) \nonumber \\
&&-2\langle\sigma v\rangle_{\phi_1\phi_1\to  \phi^\dagger_1\phi^\dagger_1} \Big(n_{\phi_1}^2 - (n_{\phi^\dagger_1})^2\Big) \nonumber \\
&&+\sum_{i=X,h_1,h_2}\bigg[\langle\sigma v\rangle_{\phi_2\phi^\dagger_1\to  \phi_1i} \Big(n_{\phi_2}n_{\phi^\dagger_1}  - n_{\phi_1}n_i\Big)  \nonumber \\
&&\qquad+\langle\sigma v\rangle_{\phi^\dagger_2\phi^\dagger_1\to  \phi_1i} \Big(n_{\phi^\dagger_2}n_{\phi^\dagger_1}  - n_{\phi_1}n_i\Big)\bigg] \nonumber \\
&&-\sum_{i=X,h_1,h_2}\bigg[\langle\sigma v\rangle_{\phi^\dagger_2\phi_1\to  \phi^\dagger_1i} \Big(n_{\phi^\dagger_2}n_{\phi_1}  - n_{\phi^\dagger_1}n_i\Big)  \nonumber \\
&&\qquad+\langle\sigma v\rangle_{\phi_2\phi_1\to  \phi^\dagger_1i} \Big(n_{\phi_2}n_{\phi_1}  - n_{\phi^\dagger_1}n_i\Big)\bigg],  \label{Boltz1}
\eea
\bea
{\dot n}_{\phi_2}+3H n_{\phi_2} &=&-\langle\sigma v\rangle_{\phi_2\phi^\dagger_2\to f\,{\bar f}, VV, h_1 h_1, h_2 h_2,h_1 h_2, XX}\Big(n_{\phi_2}n_{\phi^\dagger_2}-n^{\rm eq}_{\phi_2}n^{\rm eq}_{\phi^\dagger_2}\Big) \nonumber  \\
&&-\langle\sigma v\rangle_{\phi_2\phi^\dagger_2\to \phi_1\phi^\dagger_1} \Big(n_{\phi_2}n_{\phi^\dagger_2}- n_{\phi_1}n_{\phi^\dagger_1}\Big) \nonumber \\
&&-\langle\sigma v\rangle_{\phi_1\phi_2\to \phi_1\phi^\dagger_2} n_{\phi_1}\Big(n_{\phi_2}- n_{\phi^\dagger_2}\Big) \nonumber \\
&&-2\langle\sigma v\rangle_{\phi_2\phi_2\to  \phi_1\phi^\dagger_1} \Big(n_{\phi_2}^2 - n_{\phi_1}n_{\phi^\dagger_1}\Big) \nonumber \\
&&-2\langle\sigma v\rangle_{\phi_2\phi_2\to  \phi^\dagger_2\phi^\dagger_2} \Big(n_{\phi_2}^2 - (n_{\phi^\dagger_2})^2\Big) \nonumber \\
&&-\langle\sigma v\rangle_{\phi_2\phi_2\to  \phi^\dagger_2\phi_2} n_{\phi_2}\Big(n_{\phi_2} - n_{\phi^\dagger_2}\Big) \nonumber \\
&&-\sum_{i=X,h_1,h_2}\bigg[\langle\sigma v\rangle_{\phi_2\phi^\dagger_1\to  \phi_1i} \Big(n_{\phi_2}n_{\phi^\dagger_1}  - n_{\phi_1}n_i\Big) \nonumber \\
&&\qquad+\langle\sigma v\rangle_{\phi_2\phi_1\to  \phi^\dagger_1i} \Big(n_{\phi_2}n_{\phi_1}  - n_{\phi^\dagger_1}n_i\Big)\bigg], \label{Boltz2}
\eea
and the corresponding Boltzmann equations for $n_{\phi^\dagger_1}$ and  $n_{\phi^\dagger_2}$.
Here, we assumed that $h_X, X$ are in thermal equilibrium with the SM plasma.

Assuming no CP violation in the dark sector, we can take $n_{\phi_1}=n_{\phi^\dagger_1}=\frac{1}{2}n_1$ and $n_{\phi_2}=n_{\phi^\dagger_2}=\frac{1}{2}n_2$ and $n_i=n^{\rm eq}_2$ for the co-annihilation channels, so the Boltzmann equations in eqs.~(\ref{Boltz1}) and (\ref{Boltz2}) become
\bea
{\dot n}_{1}+3H n_{1} &=& -\frac{1}{2}\langle\sigma v\rangle_{\phi_1\phi^\dagger_1\to f\,{\bar f}, VV, h_1 h_1, h_2 h_2,h_1 h_2, XX}\Big(n_1^2-(n^{\rm eq}_1)^2\Big) \nonumber  \\
&&+\frac{1}{2}\langle\sigma v\rangle_{\phi_2\phi^\dagger_2,\phi_2\phi_2,\phi^\dagger_2\phi^\dagger_2\to \phi_1\phi^\dagger_1} \Big(n_2^2 - n_1^2 \Big),
\eea
\bea
{\dot n}_2+3H n_2 &=&-\frac{1}{2}\langle\sigma v\rangle_{\phi_2\phi^\dagger_2\to f\,{\bar f}, VV,  h_1 h_1, h_2 h_2,h_1 h_2, XX}\Big(n_2^2-(n^{\rm eq}_2)^2\Big) \nonumber  \\
&&-\frac{1}{2}(\langle\sigma v\rangle_{\phi_2\phi^\dagger_2\to \phi_1\phi^\dagger_1}+2\langle\sigma v\rangle_{\phi_2\phi_2\to  \phi_1\phi^\dagger_1}) \Big(n_2^2- n_1^2\Big) \nonumber \\
&&-\frac{1}{2}\sum_{i=X,h_1,h_2}(\langle\sigma v\rangle_{\phi_2\phi^\dagger_1\to  \phi_1i} +\langle\sigma v\rangle_{\phi_2\phi_1\to  \phi^\dagger_1i} )\,n_1\Big(n_2  -n^{\rm eq}_2\Big).
\eea

 \begin{figure}[tbp]
  \begin{center}
    \includegraphics[height=0.45\textwidth]{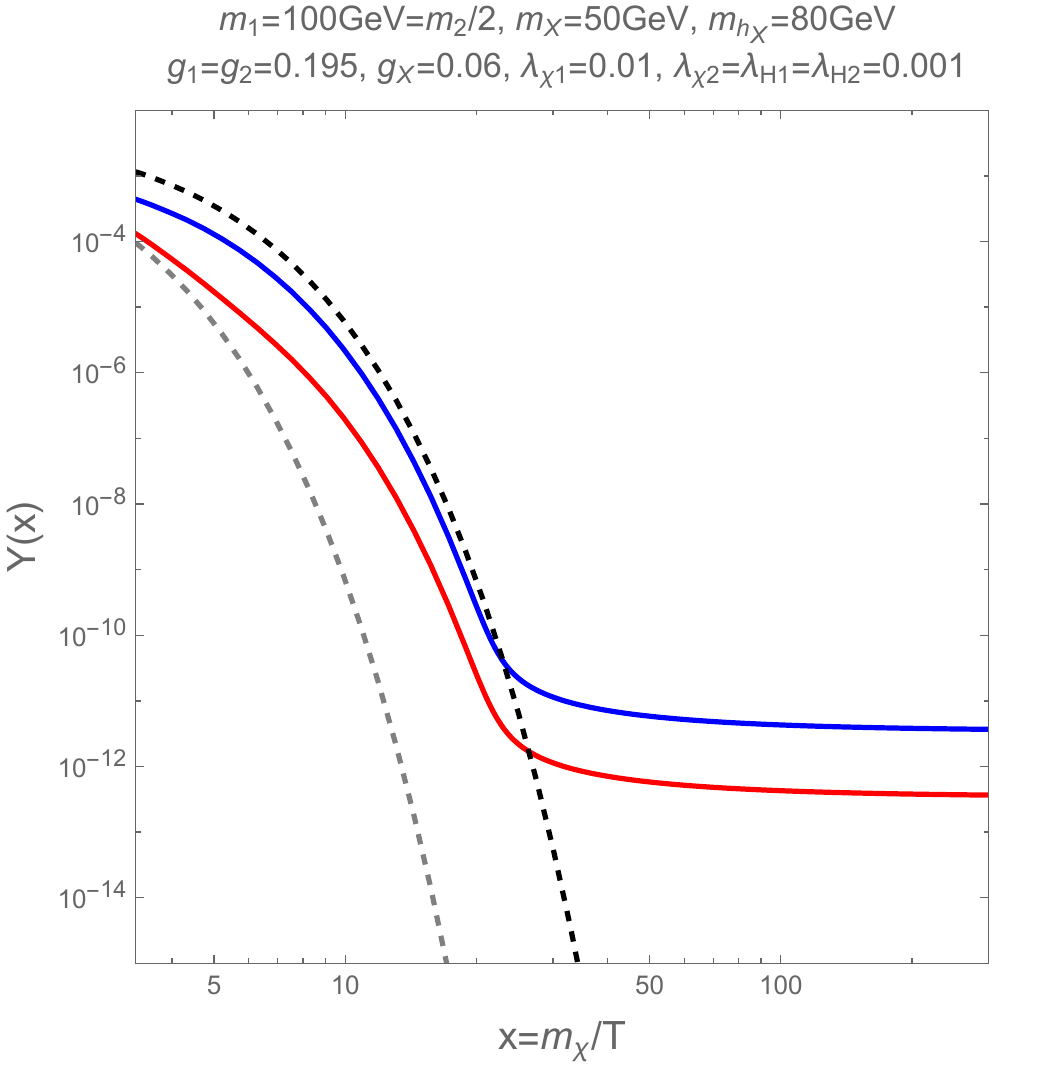}\,\,\,
        \includegraphics[height=0.45\textwidth]{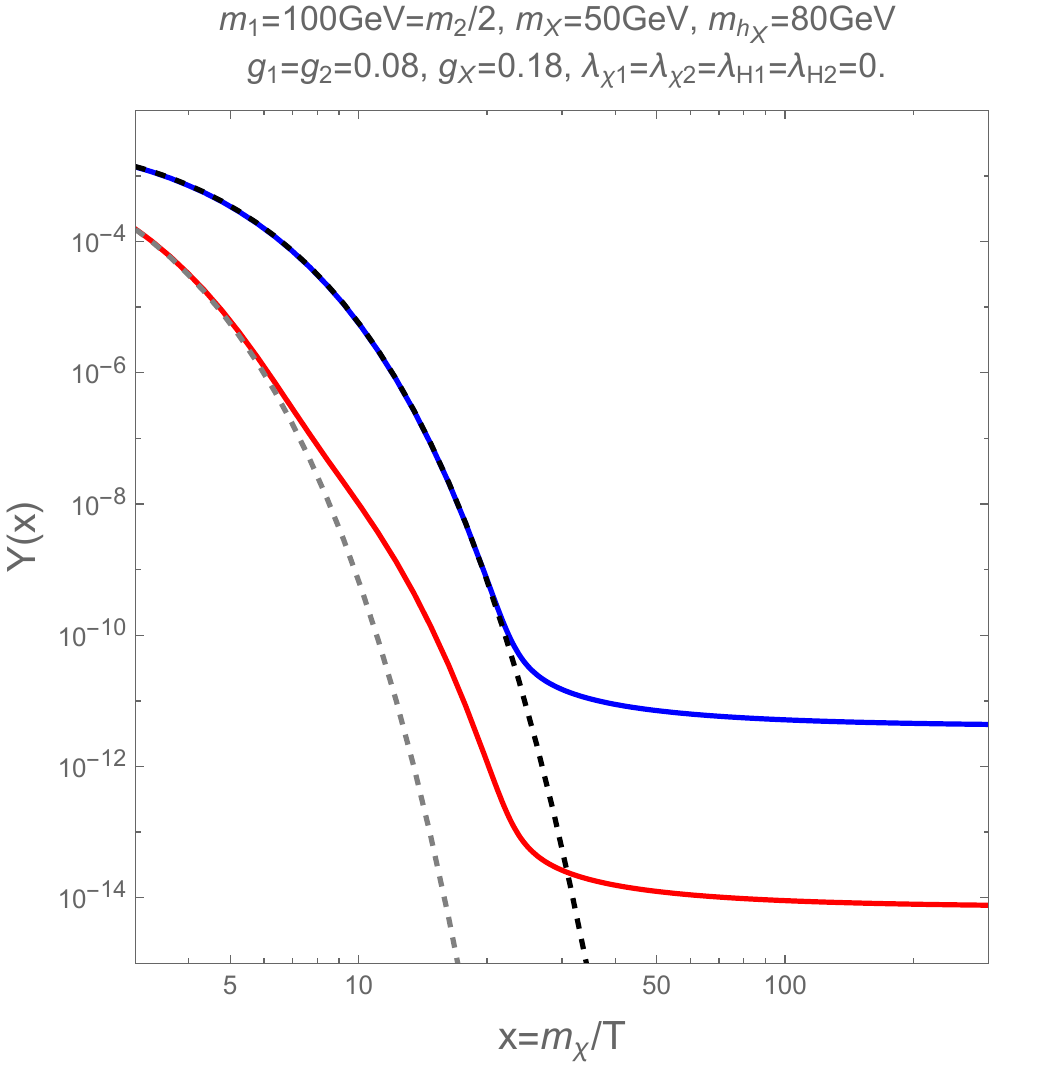}
  \end{center}
  \caption{The relic abundances for two-component dark matter, $\phi_1$ and $\phi_2$, in blue and red lines, respectively.  We  took  $m_1=m_2/2=100\,{\rm GeV}=2m_X$, $m_{h_1}=80\,{\rm GeV}$, and $g_1=g_2=0.195$, $g_X=0.06$, and the nonzero quartic couplings are $\lambda_{\chi 1}=0.01,  \lambda_{\chi 2}=0.001$, and $\lambda_{H1}=\lambda_{H2}=0.001$, on left (BM1). We took  $m_1=m_2/2=100\,{\rm GeV}=2m_X=m_{h_1}$, and  $g_1=g_2=0.08$, $g_X=0.18$, and the other dark sector couplings in the potential are set to zero, on right (BM2). The dashed lines correspond to the cases where $\phi_1, \phi_2$ are in thermal equilibrium. }
  \label{relic1}
\end{figure}

 \begin{figure}[tbp]
  \begin{center}
    \includegraphics[height=0.45\textwidth]{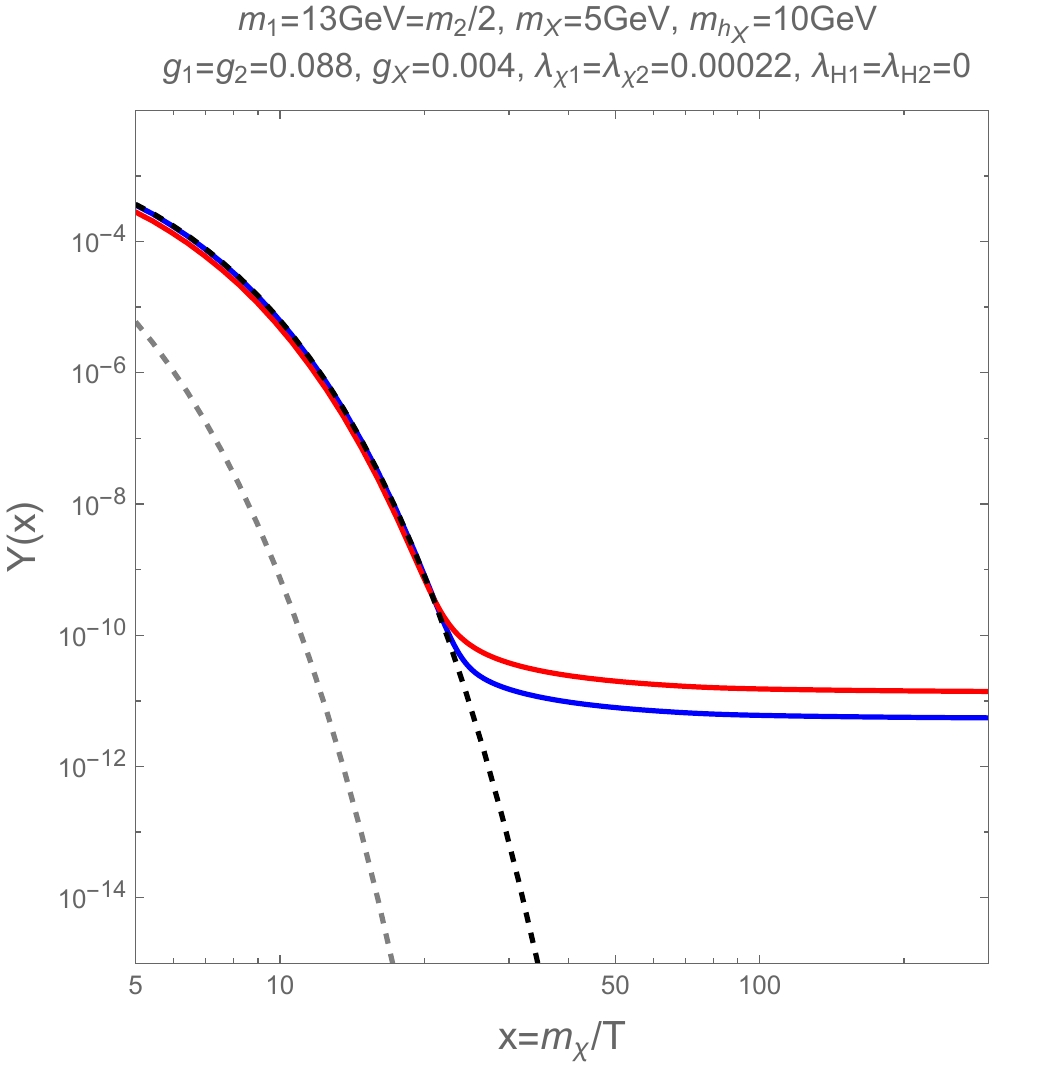}\,\,\,
        \includegraphics[height=0.45\textwidth]{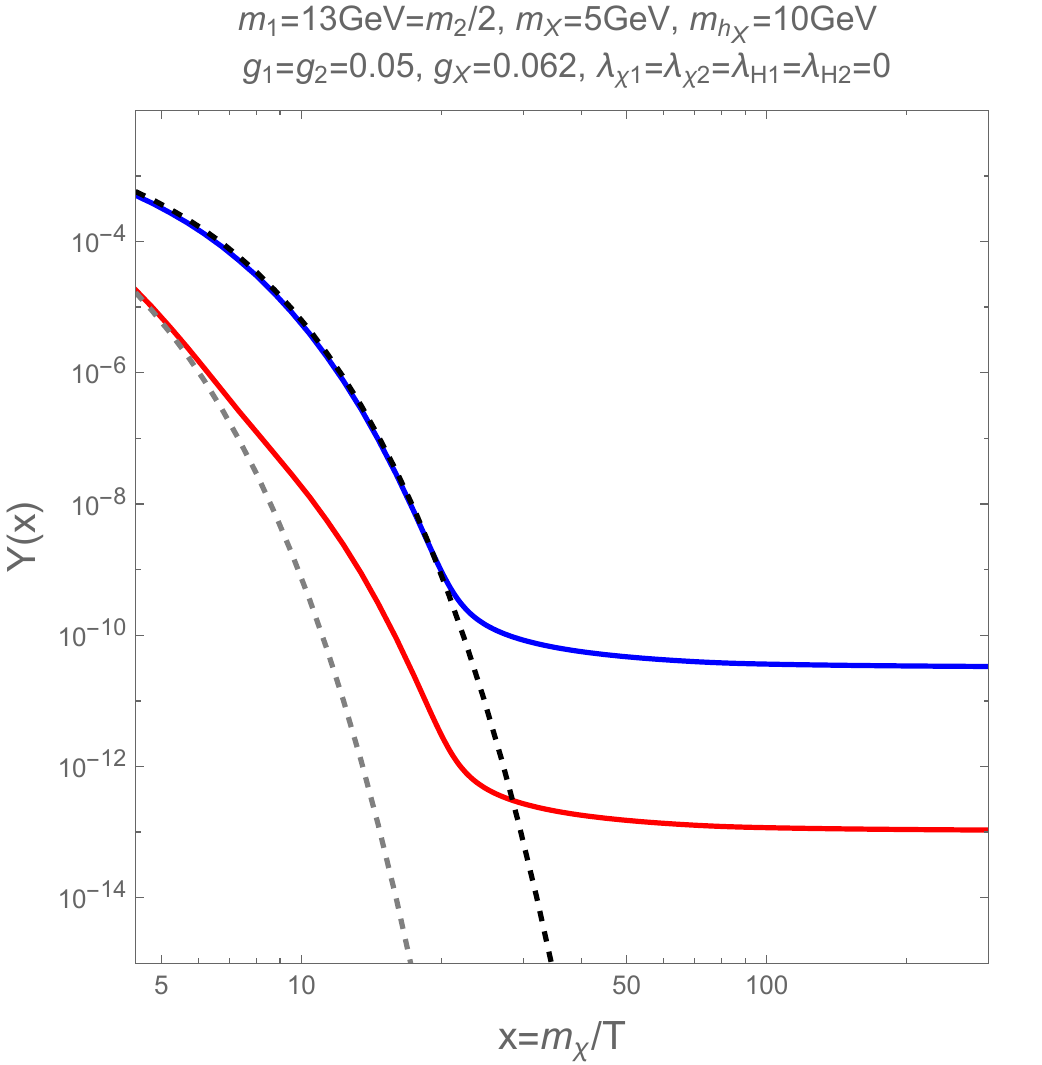}
  \end{center}
  \caption{The relic abundances for two-component dark matter, $\phi_1$ and $\phi_2$, in blue and red lines, respectively.  We  took  $m_1=m_2/2=13\,{\rm GeV}$, $m_X=5\,{\rm GeV}$, $m_{h_1}=10\,{\rm GeV}$, for both plots. We chose  $g_1=g_2=0.088$, $g_X=0.004$,  and the nonzero quartic couplings are $\lambda_{\chi 1}=\lambda_{\chi 2}=0.00022$, on left (BM3). We took  $g_1=g_2=0.05$, $g_X=0.062$, and the other dark sector couplings in the potential are set to zero, on right (BM4). The dashed lines correspond to the cases where $\phi_1, \phi_2$ are in thermal equilibrium. }
  \label{relic2}
\end{figure}

As a result, the abundances for both $\phi_1$ and $\phi_2$ components depend on the annihilations of the heavier pair into the lighter pair and the self-annihilation channels into the SM and dark sector particles. But, only the abundance for $\phi_2$ is affected directly by the semi-annihilation channels between two components of dark matter. When the gauge kinetic mixing and the dark Higgs-portal couplings are small, the self-annihilation channels of dark matter into the SM and dark sectors are suppressed, so the self-interactions between two components of dark matter and the associated dark Higgs couplings are dominant in determining the relic density. In this case, the faction of $\phi_2$ in the total relic abundance gets highly suppressed, due to the strong annihilations of the heavier component $\phi_2$.

In Figs.~\ref{relic1} and \ref{relic2}, we depict the evolutions of the relic abundances for two-component dark matter by varying the DM and mediator masses as well as the DM couplings. 
In the left of Fig.~\ref{relic1} (BM1), we took a relatively small dark gauge coupling but turned on nonzero couplings between dark matter and dark Higgs (or SM Higgs). So, in this case, the relic abundance for $\phi_2$ is relatively large. On the other hand, in the right of Fig.~\ref{relic1} (BM2), we took a sizable dark gauge coupling and turned off  the couplings between dark matter and dark Higgs (or SM Higgs), so the dark matter can annihilate dominantly into a pair of dark photons. Thus, in this case, the dark matter abundance is dominantly determined by $\phi_1$. Similarly, in Fig.~\ref{relic2} (BM3 and BM4), we considered the case with smaller masses for DM and mediators, finding that relatively small couplings for dark matter are necessary for the correct relic density.  

\begin{table}[hbt!]
  \begin{center}
  \scalebox{0.75}{
 \begin{tabular}{c|c|c|c|c|c|c|c|c|c|c|c}
      \hline\hline
      &&&&&&&&&&&\\[-2mm]
      &   $m_2$ &  $m_X$  & $m_{h_1}$ & $g_{1,2}$ & $g_X$ & $\langle\sigma v\rangle^0_{X} $  & $\langle\sigma v\rangle^0_{ h_1} $ & $r_1$ & $S_0$ & $\sigma_{\rm u,eff}/m_1$ & $\sigma_{\rm s,eff}/m_1$    \\
           & [GeV]  &  [GeV] &  [GeV] &  &   &  $[{\rm cm^3/s}]$  & $[{\rm cm^3/s}]$  & $=\frac{\Omega_1}{\Omega_{\rm DM}}$  &   &   $[{\rm cm^2/g}]$    &  $[{\rm cm^2/g}]$  
           \\[2mm]
      \hline
      &&&&&&&&&&&\\[-2mm]
      $\rm BM1$ & $200$ & 50& 80 & 0.195  & 0.06 & $1.7\times 10^{-25}$  &  $1.6\times 10^{-25}$  &  $0.834$ &  $1.42\times10^5$   & 0.347& $0.00108$ 
                     \\[2mm]
      \hline
      &&&&&&&&&&&\\[-2mm]
 $\rm BM2$ & $200$  & 50   & 80 & 0.08  &  0.18 & $2.5\times 10^{-25}$   &  $2.5\times 10^{-25}$  & $0.996$ &  $181$  & $0.00844$  & $4.41\times 10^{-5}$
                     \\[2mm]
                     \hline
      &&&&&&&&&&&\\[-2mm]
 $\rm BM3$ & $ 26$ & 5   & 10  & 0.088  & 0.004 &  $1.6\times 10^{-26}$   & $1.3\times 10^{-26}$  &  $0.165$ &  $243$   & $1.52$ & $3.92\times 10^{-4}$
                     \\[2mm]
                     \hline
      &&&&&&&&&&&\\[-2mm]
 $\rm BM4$ & $ 26$ &  5 &  10  & 0.05  & 0.062 & $1.2\times 10^{-24}$  & $1.1\times 10^{-24}$  &  $0.994$ & $28.8$    & $6.00$ & $0.00305$
                     \\[2mm]
                              \hline
      &&&&&&&&&&&\\[-2mm]
 $\rm BM5$ & $ 1$ &  1.5 &  0.8  & 0.014  & 1.8 & $-$  & $5.1\times 10^{-22}$  &  $0.9998$ & $1.019$    & $37.8$ & $0.333$
                     \\[2mm]
                         \hline
      &&&&&&&&&&&\\[-2mm]
 $\rm BM6$ & $ 0.2$ &  0.6 &  0.04  & 0.0039  & 0.5 & $-$  & $3.4\times 10^{-23}$  &  $0.9996$ & $1.016$    & $0.0231$ & $0.252$
                     \\[2mm]
      \hline\hline
    \end{tabular}}
  \end{center}
    \caption{The Sommerfeld factor and the effective self-scattering cross sections in benchmark models. We took $m_2\simeq 2m_1$ with $\Delta=10^{-6}$ and $v=20\,{\rm km/s}$ for the effective self-scattering cross sections, in all the benchmark models.  The other parameters for BM1-BM4 are chosen as in Figs.~\ref{relic1} and \ref{relic2}.  For BM5 and BM6, we took $\varepsilon= 10^{-3}$, and other nonzero parameters
to $ \lambda_{\chi 1}= \lambda_{\chi 2} = 10^{-4}$ in BM5, and $ \lambda_{\chi 1} = \lambda_{\chi 2} = 3\times 10^{-7}$ in BM6. We note that $\langle\sigma v\rangle^0_X\equiv \langle\sigma v\rangle^0_{\phi^\dagger_1\phi^\dagger_2\to \phi_1 X}$, $\langle\sigma v\rangle^0_{h_1}\equiv \langle\sigma v\rangle^0_{\phi^\dagger_1\phi^\dagger_2\to \phi_1 h_1} $, and $\sigma_{\rm u,eff}$ and $\sigma_{\rm s,eff}$ are the effective self-scattering cross sections for $u$-channel and $s$-channel, respectively.}
      \label{table2}
\end{table}

In Table~\ref{table2}, including the benchmark models in  Figs.~\ref{relic1} and \ref{relic2}, we make a summary of the predictions for the semi-annihilation cross sections for multi-component dark matter at tree level,  the DM fraction of $\phi_1$,  the $u$-channel Sommerfeld factor, as well as the effective self-scattering cross sections for $u$-channel or $s$-channel. 
We note that the correct DM relic density and other direct and indirect DM constraints are satisfied for the benchmark models.

The effective self-scattering cross sections for $u$-channel and $s$-channel and the Sommerfeld factor with the $u$-channel resonance depend on the detuning parameter from the resonance mass condition, $\Delta=1-m_2/(2m_1)$, which is positive and small for the $u$-channel resonance. In Table~\ref{table2}, we chose $\Delta=10^{-6}$ for all the benchmark models and the choice of the other parameters is explained in Table~\ref{table2}. 
Then, we computed numerically the effective self-scattering cross sections for  $u$-channel or $s$-channel at the scales of dwarf galaxies for the DM relative velocity chosen to $v=20\,{\rm km/s}$. As a result, we find that the effective self-scattering cross section for $u$-channel is much larger than the one for  the $s$-channel in most of the benchmark models except BM6. For a larger value of $\Delta$, both $u$-channel and $s$-channel enhancements become less significant, because the effective mass of the $u$-channel mediator ($\phi_1$) gets heavier and the $s$-channel gets off the resonance.

On the other hand, we also obtained the Sommerfeld factor in each benchmark model in Table~\ref{table2} by setting the DM velocity to $v=10^{-3}$ at galaxies. Then, we can set the CMB bound on  the decay branching ratios of the dark photon or dark Higgs into $e^+e^-$, as will be discussed in Section 6.2.  We note that at the scales of dwarf galaxies with the DM velocity $v=10^{-4}$, the Sommerfeld factor reads in the order of benchmark models in Table~\ref{table2}: $S_0=1.20\times 10^5, 224, 267, 88.8, 1.34, 1.08$. Thus, the Sommerfeld factor increases slightly in dwarf galaxies as compared to the case at galaxies, except BM1.

We note that the Sommerfeld factor is calculated numerically for the $u$-channel case, but  we can refer to the $t$-channel case in Sections 3.3 and 3.4 about  the qualitative and quantitative discussion on  a large Sommerfeld factor. As a result, we found that a sizable Sommerfeld factor can be obtained on or off the resonances for the benchmark models (BM1-BM4) in Table 2. On the other hand, the Sommerfeld factor is $S_0\simeq 1$ for the other benchmark models, BM5-BM6, in Table 2, because of small self-interactions for dark matter, i.e. $\alpha'=(g^2_1+g^2_2)/(4\pi)$, which is much smaller than the DM velocity, $v\sim 10^{-4}-10^{-3}$, at dwarf galaxies and galaxies.

\section{Probes of self-resonant dark matter}

We propose ways of testing the self-resonant dark matter models, such as  direct detection for halo dark matter, CMB recombination, direct detection for  boosted dark matter, etc, and show how the proposed benchmark models can be constrained.

\subsection{Direct detection for halo dark matter}

The DM-nucleus elastic scattering, $\phi^{(\dagger)}_1 N\to \phi^{(\dagger)}_1 N$, is possible through the exchanges of the dark Higgs or the extra gauge boson \cite{DD}. On the other hand, the dark Higgs is responsible for $s N\to s N$ or $a N\to a N$, but $s N\to a N$ or $a N \to s N$ with the extra gauge boson exchange is kinematically forbidden in the standard halo model if $\delta \equiv |m_s-m_a|\gtrsim 100\,{\rm keV}$ \cite{massplitting}. But, for $\delta \equiv |m_s-m_a|\lesssim 100\,{\rm keV}$, the inelastic processes, $s N\to a N$ or $a N \to s N$, are kinematically allowed, so the gauge kinetic mixing can be constrained by direct detection experiments for the inelastic processes.

For $m_s=m_a=m_2$,  both  $\phi^{(\dagger)}_1 N\to \phi^{(\dagger)}_1 N$ and $\phi^{(\dagger)}_2 N\to \phi^{(\dagger)}_2 N$ are relevant for direct detection through the dark Higgs and the extra gauge boson exchanges \cite{DD}.

We also note that the DM-electron elastic scatterings, $\phi^{(\dagger)}_1 e\to \phi^{(\dagger)}_1 e$,  $s(a) e\to s(a) e$ or $s(a)e\to a(s)e$,  are also relevant for light dark matter with sub-GeV mass \cite{Xenon1,Xenon2} or boosted dark matter coming from the DM co-annihilations \cite{DD}.
For $m_s=m_a=m_2$, both  $\phi^{(\dagger)}_1 e\to \phi^{(\dagger)}_1 e$ and $\phi^{(\dagger)}_2 e\to \phi^{(\dagger)}_2 e$ are present due to the dark Higgs and the extra gauge boson exchanges \cite{DD}. 

For $m_s=m_a=m_2$, from the results in the appendix A, the averaged cross section for the $\phi_i$-nucleon coherent scattering is given by
\bea
\sigma^{\rm coh}_{\phi_i,\phi^\dagger_i-A}=  \frac{\mu^2_{A,i}}{4\pi m^2_i}\,r_i\bigg[\Big(Z c^{(i)}_p f_p+(A-Z)c^{(i)}_n f_n \Big)^2+Z^2(g^{(i)}_p)^2\bigg]
\eea
where 
$r_i=\Omega_i/\Omega_{\rm DM}$, $\mu_{A,i}=m_i m_A/(m_A+m_i)$ are the reduced masses for the system with $\phi_i$ and a target nucleus $A$, and $c^{(i)}_{p,n}, g^{(i)}_p$ are the effective couplings to nucleons, and $f_{p,n}$ are the nucleon form factors.
For instance, for $^{131}{\rm Xe}$ nucleus, $A=131$ and $N=54$.

Similarly, the elastic scattering cross section between dark matter $\phi_i$ and electron is given by
\begin{align}
\sigma_{\phi_i,\phi^\dagger_i -e}&=\frac{\mu^2_{e,i}}{4\pi m_i^2}\,r_i\bigg[\bigg(\frac{\lambda_{e2}y_{h_2 \phi^\dagger_i\phi_i}}{m_{h_2}^2}+\frac{\lambda_{e1}y_{h_1 \phi^\dagger_i\phi_i}}{m_{h_1}^2}\bigg)^2+\frac{4e^2 q^2_i g^2_X \varepsilon^2 m^2_i}{m_X^4} \bigg]
\end{align}
where $\mu_{e,i}=m_i m_e/(m_e+m_i)$ is the reduced mass of the $\phi_i$-electron system.

\subsection{CMB recombination}

The dark matter annihilation, $\phi^\dagger_1\phi^{(\dagger)}_2\to \phi_1 Y$ with $Y=X, h_1$, can be constrained by the CMB bound \cite{planck15}. We define the effective efficiency parameter for CMB recombination \cite{slatyer3} as
\bea
f_{\rm eff}(m_2)= \frac{\int^{m_2/2}_0 dE_e \, E_e\, 2f^{e^+e^-}_{\rm eff}\, \frac{dN_e}{dE_e}}{m_2}
\eea
where $f^{e^+e^-}_{\rm eff}$ is the efficiency factor\footnote{Efficiency factors are computed for $e^+e^-$ or photons produced from the DM annihilations in Ref.~\cite{slatyer3}, but we used the results only for $e^+e^-$ because they are produced from the direct decays of the mediator particles in our work.} for $Y\to e^+e^-$, and we took the range of the lepton energy to $m_2/2$ when two leptons coming from the $Y$ decay are back-to-back. As compared to the DM self-annihilation, ${\rm DM}+{\rm DM}\to e^+e^-$, where the rate for the energy deposition is given by $\langle \sigma v\rangle m_{\rm DM} n^2_{\rm DM}$, the total deposit energy, $2m_{\rm DM}$, is replaced by $m_2$ and the interaction rate, $\langle \sigma v\rangle n^2_{\rm DM}/2=\langle \sigma v\rangle \rho^2_{\rm DM}/(2m^2_{\rm DM})$, is replaced by $\langle \sigma v\rangle n_1 n_2=\langle \sigma v\rangle r_1(1-r_1) \rho^2_{\rm DM}/(4m_1 m_2)$.
Thus, at the CMB recombination, the relative velocity for dark matter gets small \cite{SRDM}, so the Sommerfeld factors for the enhanced cross sections for the semi-annihilation channels in our model can be  constrained by the Planck data \cite{planck15, SRDM} as in the following\footnote{We have corrected a typo in the formula in Ref.~\cite{SRDM}, from $m_2$ to $4m_1$.},
\bea
\langle\sigma v\rangle_{\phi^\dagger_1\phi^{(\dagger)}_2\to \phi_1 Y} \cdot {\rm BR}(Y\to e^+e^-)< 4.1\times 10^{-25}\,{\rm cm^3/s}\,\bigg(\frac{f_{\rm eff}}{0.1}\bigg)^{-1} \cdot\frac{4}{r_1(1-r_1)} \cdot\bigg(\frac{m_1}{100\,{\rm GeV}} \bigg), \label{CMB}
\eea
with $r_1=\Omega_1/\Omega_{\rm DM}$.

\begin{table}[hbt!]
  \begin{center}
  \scalebox{0.75}{
 \begin{tabular}{c|c|c|c|c|c|c}
      \hline\hline
      &&&&&&\\[-2mm]
      &   $m_2[{\rm GeV}]$ &  $m_X[{\rm GeV}]$  & $m_{h_1}[{\rm GeV}]$ &  $S_0$ & ${\rm BR}(X\to e^+e^-)$ & ${\rm BR}(h_1\to e^+e^-)$    
           \\[2mm]
      \hline
      &&&&&&\\[-2mm]
      $\rm BM1$ & $200$ & 50& 80 &  $1.42\times10^5$   & $2.3\times 10^{-4}$ & $2.5\times 10^{-4}$ 
                     \\[2mm]
      \hline
      &&&&&&\\[-2mm]
 $\rm BM2$ & $200$  & 50   & 80  & $181$  & $0.18$  & $0.18$
                     \\[2mm]
                     \hline
      &&&&&&\\[-2mm]
 $\rm BM3$ & $ 26$ & 5   & 10  & $243$   & $0.14$ & $0.16$
                     \\[2mm]
                     \hline
      &&&&&&\\[-2mm]
 $\rm BM4$ & $ 26$ &  5 &  10  & $28.8$    & $0.35$ & $0.38$
                     \\[2mm]
                              \hline
      &&&&&&\\[-2mm]
 $\rm BM5$ & $ 1$ &  1.5 &  0.8   & $1.019$ &  $-$ & $0.017$
                     \\[2mm]
                         \hline
      &&&&&&\\[-2mm]
 $\rm BM6$ & $ 0.2$ &  0.6 &  0.04 & $1.016$    & $-$ & $0.014$
                     \\[2mm]
      \hline\hline
    \end{tabular}}
  \end{center}
    \caption{CMB bounds on the  branching ratios of the mediator decays, $X/h_1\to e^+e^-$.}
      \label{table3}
\end{table}

In Table~\ref{table3}, taking the same benchmark models as in Table~\ref{table2}, we show the CMB bounds on the branching ratios of the mediators, $X/h_1\to e^+ e^-$. 
In most of the benchmark models, the DM abundances are highly asymmetric abundances for two-component dark matter, the tree-level semi-annihilation rates are already small due to the lack of one of the components of dark matter. Thus, the CMB limits on the decay branching ratios of the mediator particles into electron and positron become weak in most of the benchmark models, except BM1, for which a very large Sommerfeld factor is achieved.

\subsection{Direct detection for boosted dark matter}

The co-annihilation channels, $\phi^{(\dagger)}_1 s(a)\to \phi^{(\dagger)}_1 h_1$, $\phi^{(\dagger)}_1 s(a)\to \phi^{(\dagger)}_1 X$, $\phi_1 s, \phi_1 a\to \phi^\dagger_1 h_1$,  can make the resultant dark matter $\phi^{(\dagger)}_1$ boosted for $m_{h_1}, m_X\lesssim m_{s(a)}$. So, it leads to an interesting consequence for direct detection \cite{boosted,semi-ann,boosted2}. 
Moreover, the corresponding annihilation cross sections can be enhanced by the $u$-channel resonances at present, so the DM flux as well as the indirect signals from the decay products of $h_1$ or $X$ can be increased accordingly \cite{indirect,box,SRDM}. 

Focusing on the case with $m_s=m_a=m_2$, we consider the DM co-annihilation channels, $\phi_2\phi^\dagger_1\to \phi_1 X(h_1)$, $\phi^\dagger_2\phi^\dagger_1\to \phi_1 X(h_1)$ and their complex conjugates, so the resultant dark matter $\phi^{(\dagger)}_1$ gets boosted for $m_{h_1}, m_X\lesssim m_2$. In this case, as discussed in Section 2.2, only if the initial states for the co-annihilation channels is a mixture of two DM pairs, $\psi_1\sim \psi_A+\frac{g_2}{g_1}\psi_B$ or its conjugate, with $A=(\phi_1\phi_2)$ and $B=(\phi_1\phi^\dagger_2)$, there is a Sommerfeld enhancement due to the DM co-scattering processes, $(\phi_2^\dagger,\phi_2)\to \phi_2\phi^\dagger_1$ and $(\phi_2^\dagger,\phi_2)\to \phi^\dagger_2\phi^\dagger_1$. As a result, the DM co-annihilation cross sections get enhanced by an universal Sommerfeld factor $S_0$, as follows, 
\bea
\langle\sigma v\rangle_{\phi_2\phi^\dagger_1\to \phi_1 X(h_1)}&=&S_0\, \langle\sigma v\rangle^0_{\phi_2\phi^\dagger_1\to \phi_1 X(h_1)}, \\
\langle\sigma v\rangle_{\phi^\dagger_2\phi^\dagger_1\to \phi_1 X(h_1)}&=&S_0\, \langle\sigma v\rangle^0_{\phi^\dagger_2\phi^\dagger_1\to \phi_1 X(h_1)}, 
\eea
with $\langle\sigma v\rangle^0_{\phi^{(\dagger)}_2\phi^\dagger_1\to \phi_1 X(h_1)}$ being the tree-level DM co-annihilation cross sections.

Moreover, we consider the differential flux for boosted dark matter (BDM) $\phi^{(\dagger)}_1$ at the location of the Solar System, coming from  the semi-annihilation processes for dark matter in our galaxy, $\phi_2\phi^\dagger_1\to \phi_1 Y$ with $Y=h_1, X$, or their CP conjugate processes,  is given by
\bea
\frac{d\Phi_{\phi_1}}{dE_1} = \frac{1}{8\pi m_1 m_2}\, \langle\sigma v\rangle_{\psi^\dagger_1\to \phi_1 Y} \, \cdot\frac{dN_{\phi_1}}{d E_1} \, \frac{1}{2}r_1(1-r_1)\, \int d\Omega \int_{\rm l.o.s.} ds\,  \rho^2_{\rm DM}(r)=\frac{d\Phi_{\phi^\dagger_1}}{dE_1} 
\eea
where $r(s,\theta)=\sqrt{r^2_{\odot}+s^2-2r_\odot s \cos\theta}$ with $r_\odot=8.33\,{\rm kpc}$ being the distance of the Sun to the galactic center and  $r_1=\Omega_1/\Omega_{\rm DM}$. The differential number of the boosted dark matter $\phi_1$ is given \cite{SRDM} by
\bea
\frac{d N_{\phi_1}}{dE_1} =\delta(E_1-E'_1),
\eea
where
\bea
E'_1 = \frac{E_{\rm cm}^2+m^2_1-m^2_Y}{2E_{\rm cm}}, \quad E_{\rm cm}\equiv  m_1+m_2 +\frac{1}{2}\mu v^2_{\rm rel},
\eea
with $\mu=m_1 m_2/(m_1+m_2)$ and $v_{\rm rel}$ being the relative velocity between $\phi^\dagger_1$ and $\phi_2$,
so the kinetic energy of the boosted dark matter, $T'_1=E'_1-m_1$, is also given by
\bea
T'_1=\frac{(E_{\rm cm}-m_1)^2-m^2_Y}{2E_{\rm cm}}. \label{DMkin}
\eea
Here, since there is no directional information in the current direct detection experiments, we need to integrate the solid angle over the whole sky in the computation of the flux.

We note that for $K\ll m_1, m_2$ and $K\ll (m^2_2-m^2_Y)/(2m_2)$, with $K\equiv \frac{1}{2}\mu v^2_{\rm rel}$, we get the energy of the boosted dark matter  as
\bea
E'_1\simeq \frac{2m^2_1+2m_1m_2+m^2_2-m^2_Y}{2(m_1+m_2)}
\eea
or the kinetic energy  as
\bea
T'_1\simeq \frac{m^2_2-m^2_Y}{2(m_1+m_2)}. \label{DMkin}
\eea
This is a good approximation, as far as $(m_2-m_Y)/m_2\gtrsim v^2_{\rm rel}\sim 10^{-6}$ for $v_{\rm rel}\sim 10^{-3}$ in our galaxy. 

As a result, we can rewrite the differential flux in the following form,
\bea
\frac{d\Phi_{\phi_1}}{dT_1} =\Phi_{\rm BSM} \,\delta(T_1-T'_1)
\eea
with
\bea
\Phi_{\rm BSM} =\frac{m_1}{2m_2}\,r_1(1-r_1)\cdot \Big(3.2\times 10^{-3}\,{\rm cm}^{-2}{\rm s}^{-1}\Big)\bigg(\frac{m_1}{100\,{\rm MeV}}\bigg)^{-2}\bigg(\frac{ \langle\sigma v\rangle_{\phi_2\phi^\dagger_1\to \phi_1 Y} }{10^{-26}\,{\rm cm}^{3}\,{\rm s}^{-1}}\bigg).
\eea

The differential rate for the boosted DM, $\phi_1$ or $\phi^\dagger_1$, per target mass for direct detection experiments is given by
\bea
\frac{dR_A}{dT_A}=\frac{1}{m_A}\int^\infty_{T_{1,{\rm min}}} dT_1  \, \frac{d\sigma_{\rm SI}}{dT_A}\,\frac{d\Phi_{\phi_1}}{dT_1}, 
\eea
with
\bea
\frac{d\sigma_{\rm SI}}{dT_A}\equiv \frac{1}{2} \bigg(\frac{d\sigma_{\phi_1-A}}{dT_A}+\frac{d\sigma_{\phi^{\dagger}_1-A}}{dT_A} \bigg).
\eea
First, we note that $T_{1,{\rm min}}$ is the minimal kinetic energy for the boosted dark matter for the  kinetic energy of the nucleus $T_A$, given by
\bea
T_{1,{\rm min}}=\bigg(\frac{T_A}{2}-m_1\bigg) \bigg[1\pm \sqrt{1+\frac{2T_A}{m_A} \frac{(m_1+m_A)^2}{(T_A-2m_1)^2} } \bigg],
\eea
in which either $(+)$ or $(-)$ signs are taken for $T_A>2m_1$ or $T_A<2m_1$.
Moreover, the total differential cross section is composed of coherent and incoherent parts \cite{semi-ann}, as follows,
\bea
\frac{d\sigma_{\rm SI}}{dT_A}=\bigg(\frac{d\sigma_{\rm SI}}{dT_A}\bigg)_{\rm coh}+\bigg(\frac{d\sigma_{\rm SI}}{dT_A}\bigg)_{\rm inc}
\eea 
with
\bea
\bigg(\frac{d\sigma_{\rm SI}}{dT_A}\bigg)_{\rm coh} &=&\frac{\sigma^{\rm coh}_{\rm SI}}{T_{A,{\rm max}}}\,|F_{\rm SI}(q)|^2 ,\\
\bigg(\frac{d\sigma_{\rm SI}}{dT_A}\bigg)_{\rm inc} &=&\frac{\sigma^{\rm inc}_{\rm SI}}{T_{A,{\rm max}}}\,(1-|F_{\rm SI}(q)|^2).
\eea
Here, the maximal kinetic energy of the target nucleus after scattering is given by
\bea
T_{A,{\rm max}} = \frac{T_1(T_1+2m_1)}{(m_1+m_A)^2/(2m_A)+T_1}, \label{recoilmax}
\eea
and the form factor for the target nucleus is, in the dipole nucleon approximation \cite{Perdrisat:2006hj},
\bea
F_{\rm SI}(q)=\bigg(1+\frac{q^2}{\Lambda_A}\bigg)^{-2},
\eea
with $\Lambda_A\simeq 0.843\,{\rm GeV} (0.8791\,{\rm fm}/R_A)$ and $R_A$ being the charge radius of the target nucleus. 
Moreover, the spin-independent cross sections for coherent and incoherent scatterings are given by
\bea
\sigma^{\rm coh}_{\rm SI}&\equiv &\frac{2}{r_1}\,\sigma^{\rm coh}_{\phi_1,\phi^\dagger_1-A} \nonumber \\
&=&\frac{\mu^2_{A,1}}{2\pi m^2_1}\bigg[\Big(Z c^{(1)}_p f_p+(A-Z) c^{(1)}_n f_n\Big)^2+Z^2(g^{(1)}_p)^2\bigg], \\
\sigma^{\rm inc}_{\rm SI}&\equiv & 2\,\sigma^{\rm inc}_{\phi_1,\phi^\dagger_1-A} \nonumber \\
&=& \frac{\mu^2_{A,1}}{2\pi m^2_1} \bigg[Z\Big(c^{(1)}_p f_p\Big)^2+Z (g^{(1)}_p)^2+(A-Z) \Big(c^{(1)}_n f_n\Big)^2 \bigg],
\eea
respectively.
Then, the total event rate in units of the number events per day and kg is given by 
\bea
R=\sum_A\int dT_A\,\frac{dR_A}{dT_A}.
\eea

\begin{figure}[t]
\centering
\includegraphics[width=0.55\textwidth,clip]{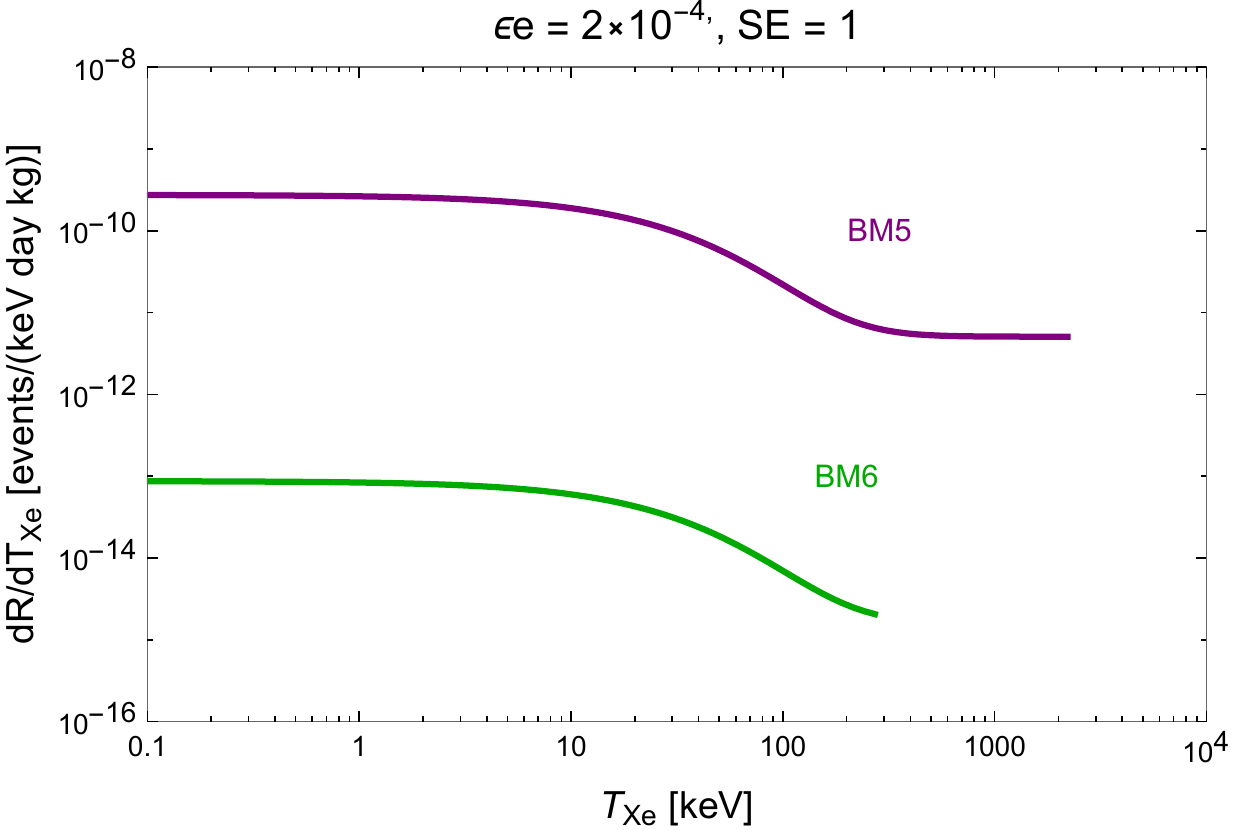} 
\caption{The differential event rate for the spin-independent elastic scattering of boosted dark matter as a function of the kinetic energy of the Xenon target nucleus. We took benchmark models for light dark matter, BM5 and BM6, in Table~\ref{table2}. 
} 
\label{Fig:event}
\end{figure}

In Fig.~{\ref{Fig:event}}, we show the differential event rate for the spin-independent elastic scattering of boosted dark matter per day per kg in units of ${\rm keV}^{-1}$ in the case of the Xenon target nucleus. We fixed the semi-annihilation cross section for $\phi^\dagger_1 \phi^{(\dagger)}_2\to \phi_1 Y$ {\color{blue}with $Y=X, h_1$} for the benchmark models, BM5, BM6, respectively, and we also took the gauge kinetic mixing to $\varepsilon e=2\times 10^{-4}$ and the $u$-channel Sommerfeld factor to $S_0=1$. If we increase the Sommerfeld factor, the differential event rate in Fig.~{\ref{Fig:event}} increases by a factor of $S_0$.

We note that there is a maximal kinetic energy for the target nucleus given in eq.~(\ref{recoilmax}). As the nuclear recoil becomes undetectable below the detector threshold $T_{\rm rh}$, there is a lower bound on the DM kinetic energy $T_1$, from $T_{A,{\rm max}}\geq T_{\rm th}$.  For instance, taking the kinetic energy of the boosted dark matter from eq.~(\ref{DMkin}) with $m_2\simeq 2m_1$ and $m_Y\ll m_1$, we obtain the lower bound on the DM mass for direct detection \cite{semi-ann}, as follows,
\bea
m_{1,{\rm min}}\simeq \frac{15m_A}{\frac{32 m_A}{T_{\rm th}}-9}\,\bigg(1+\frac{4}{5}\sqrt{1+\frac{2m_A}{T_{\rm th}}} \bigg).
\eea
Therefore, we need to take the DM mass to be greater than $m_{1,\,{\rm min}}$ for a direct detection of boosted dark matter at XENONnT. For instance, for the case of the XENONnT experiment, the detector threshold  is given by $T_{\rm th}=3.3\,{\rm keV}$ \cite{xenonnt}, so we obtain $m_{1,{\rm min}}\simeq 10.7\,{\rm MeV}$. For $m_Y\lesssim m_2$, we would need a larger DM mass to pass the detector threshold. However, we increase $m_1$, the DM flux coming from our galaxy gets suppressed. Thus, we can constrain our model by the direct detection of boosted dark matter with relatively small masses.

We also remark that the semi-annihilation processes, $\phi_2\phi^\dagger_1\to \phi_1 X(h_1)$, $\phi^\dagger_2\phi^\dagger_1\to \phi_1 X(h_1)$, and their conjugates, can be enhanced by the Sommerfeld factor with the $u$-channel resonance in our model as discussed in Section 3, so we could take a larger cross section for the semi-annihilation processes at present than the thermal cross section. 
In this case, we could take a larger DM mass well to go above the detector threshold, as the DM flux can be sizable even for  a larger DM mass due to the enhanced annihilation cross sections.

\begin{figure}[t]
\centering
\includegraphics[width=0.45\textwidth,clip]{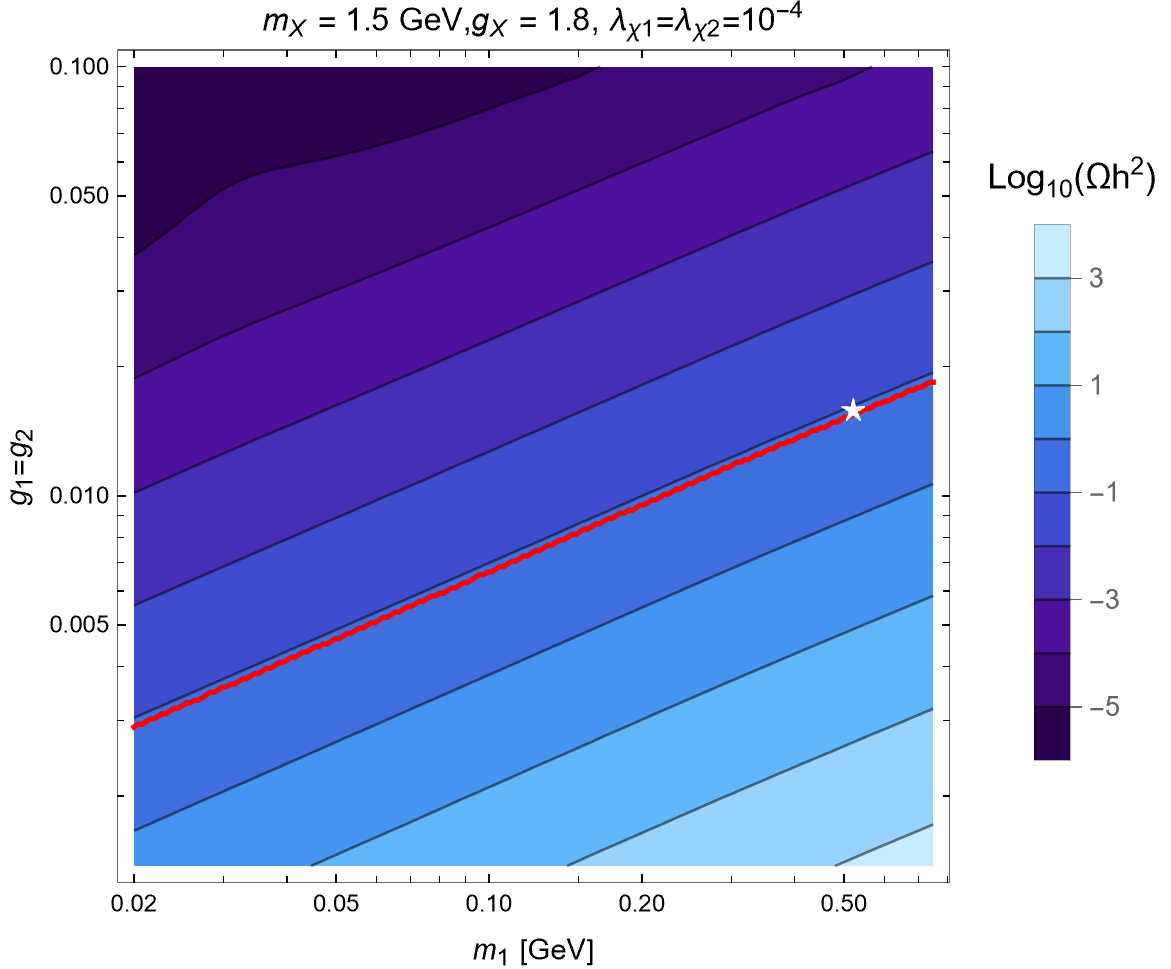}  \,\,\,\,
\includegraphics[width=0.50\textwidth,clip]{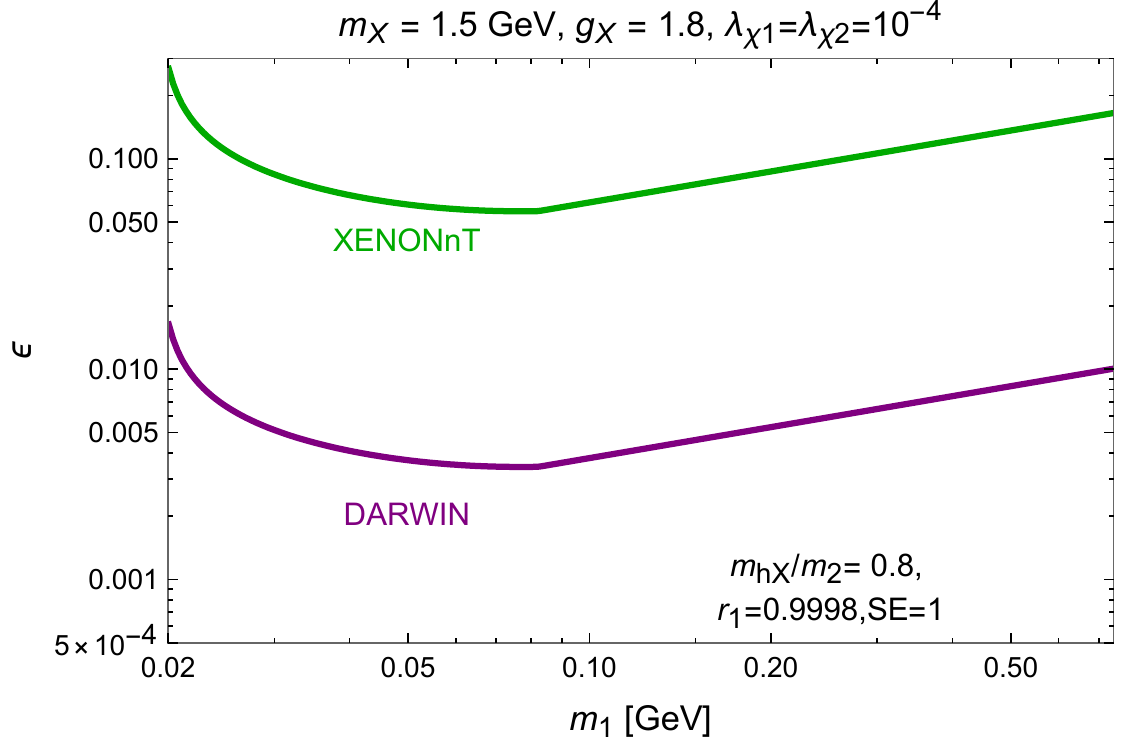} 
\caption{(Left) Parameter scan for the relic density for self-resonant light dark matter in the DM self-couplings ($g_1=g_2$) vs $m_1\simeq m_2/2$.  The benchmark model, BM5, (marked in white star), lies on the red line which indicates the contour line satisfying the observed relic density. (Right) XENONnT and DARWIN (projected) bounds for boosted dark matter in the gauge kinetic mixing ($\varepsilon$) vs  $m_1$. We consider the $X$ gauge portal for BDM-nucleus elastic scattering.
} 
\label{Fig:bdm}
\end{figure}

In the left of Fig.~\ref{Fig:bdm}, we show the parameter scan for the relic density for self-resonant light dark matter in the parameter space for DM self-couplings ($g_1=g_2$) vs $m_1$.  We took $m_2\simeq 2m_1$ for the self-resonant condition and $m_{h_1}=1.6m_1<m_2$ and $m_X>m_2$, for which the semi-annihilation channels with dark Higgs, $\phi_2\phi^\dagger_1\to \phi_1 h_1$ and $\phi^\dagger_2\phi^\dagger_1\to \phi_1 h_1$, are kinematically open. In this case, there is more freedom to make the decay branching fraction of the dark Higgs into $e^+e^-$ small enough in the presence of new light particles in the dark sector, being consistent with the CMB bound. We also chose the other parameters to $g_X=1.8$ and $\lambda_{\chi1}=\lambda_{\chi 2}=10^{-4}$ and set the Higgs-portal couplings and Higgs mixing angle to zero. The parameter scan results contain the benchmark model BM6 on the red line along which the observed relic density is obtained.  The color gradation in  the left of Fig.~\ref{Fig:bdm}  indicates the total abundance of self-resonant dark matter in the parameter space.

On the other hand, in the right of Fig.~\ref{Fig:bdm}, we also show the limits on the gauge kinetic mixing  ($\varepsilon$) vs the DM mass $m_1$ from the direct detection for boosted dark matter at XENONnT \cite{xenonnt} and DARWIN (in prospect) \cite{darwin}. The flux for boosted dark matter relies on the semi-annihilations into the dark Higgs while the direct detection for BDM-nucleus elastic scattering is dominated by the gauge kinetic mixing of the $X$ gauge boson. But, we ignored the Higgs portal couplings for the dark Higgs.
Here, we took  $m_2\simeq 2m_1$, $m_X=1.5\,{\rm GeV}$ and  $\lambda_{\chi1}=\lambda_{\chi 2}=10^{-4}$, as in the left plot, and set $g_1=g_2=0.014$, $m_{h_X}/m_2=0.8$, $r_1=0.9998$, and the $u$-channel Sommerfeld factor to $S_0=1$.  
Thus, the direct detection limit can constrain the parameter space for boosted dark matter extended from the benchmark model BM7. 

We remark that for a large Sommerfeld factor, the bounds on $\varepsilon$ from XENONnT and DARWIN get more stringent by the factor of $\sqrt{S_0}$. For instance, if we choose larger values for $g_1, g_2$ for sub-GeV scale dark matter, the relic density would be smaller due to larger DM annihilation cross sections. However, if there is a way to reproduce dark matter from the decays of a heavy particle that dominates at late times, larger DM annihilation cross sections would be helpful for the correct relic density \cite{latedecay}. 

As a result, we find that the gauge kinetic mxing $\varepsilon$ is bounded to be below $10^{-4}-10^{-3}$ in the range, $m_1=20-750\,{\rm MeV}$. For $\varepsilon\gtrsim 0.1$, a large BDM-nucleus scattering cross section would attenuate the kinetic energy of boosted dark matter during the propagation throughout the Earth  below the detection threshold \cite{semi-ann}.  We also note that for sub-GeV scale dark photon,  the parameter space up to  $\varepsilon=10^{-5} - 10^{-3}$ is  already excluded by the direct detection for halo dark matter (non-boosted dark matter) and collider data such as Belle-II \cite{belle2,belle2-summary,belle2-limit}, BaBar \cite{babar}, NA64 \cite{NA64}, etc, depending on the decay branching ratios of the dark photon.

We also remark that the couplings of the mediator particles with sub-GeV scale masses can be constrained by Big Bang Nucleosynthesis (BBN) or CMB ($N_{\rm eff}$), depending on their lifetimes and the fraction of the energy density  \cite{BBN-review,BBN1,BBN2}. For instance, if the dark photon of masses below GeV scale decay dominantly into the SM particles, the BBN bound on the lifetime of the dark photon would be $\tau_X\lesssim 100\,{\rm s}$ \cite{BBN1}. Thus, the dark photon must decay into the dark sector such that the energy injection into the SM during BBN gets suppressed.  This is consistent with the limit from CMB discussed in Section 6.2 where a small decay branching fraction of the dark photon into $e^+e^-$ is required for a sizable $u$-channel Sommerfeld factor.

\section{Conclusions}

We presented a new model for two-component scalar dark matter, consisting of $\phi_1$ and $\phi_2$,  where both the DM components are stable due to the remaining $Z_4$ gauge symmetry, as far as the DM masses satisfy $m_2<2m_1$.  The $Z_4$ symmetry in the model ensures the stability of two-component dark matter. This is in contrast to the case for the $Z_2$ model \cite{SRDM} where $\phi_1$ is a complex singlet scalar field charged under the $U(1)'$ and  $\phi_2$ is a neutral singlet scalar under the $U(1)'$, being subject to loop-induced decays due to $Z'$ interactions. 

For DM masses with $m_2$ being slightly smaller than $2m_1$, we obtained the resonance condition for the elastic co-scattering processes between $\phi_1$ and $\phi_2^{(\dagger)}$ ($u$-channel processes) for which the corresponding cross sections at tree-level become divergent at a zero momentum transfer for dark matter. Thus, applying the Bethe-Salpeter formalism for the resummation of the ladder diagrams for the $u$-channel with two dark matter scalars, we derived a nonzero effective Yukawa potential with an effective mass given by the amount of detuning $m_2$ from $2m_1$, only for one of the linear combinations of the initial states for dark matter.

We showed that the effective co-scattering processes for dark matter become enhanced and velocity-dependent to solve the small-scale problems at galaxies and they are more important than the DM self-scattering processes for $\phi_1$, which are similarly enhanced by the $s$-channel mechanism.  The same DM couplings responsible for the co-scattering processes give rise to the the semi-annihilation processes for $\phi_1$ and $\phi_2^{(\dagger)}$ annihilating into $\phi^\dagger_1$ and dark photon/Higgs, which are enhanced by the $u$-channel Sommerfeld factor. Thus, there is a nontrivial upper bound on the Sommerfeld factor from CMB recombination, depending the decay branching fractions of dark photon/Higgs. Focusing on the benchmark models for two-component dark matter satisfying the observed relic density, we considered the possibility of producing boosted dark matter with sub-GeV scale mass from the semi-annihilation processes at the galactic center and showed the parameter space for the gauge kinetic mixing for dark photon versus the DM mass, which is consistent with the XENONnT and DARWIN experiments.

\section*{Acknowledgments}

The work is supported in part by Basic Science Research Program through the National
Research Foundation of Korea (NRF) funded by the Ministry of Education, Science and
Technology (NRF-2022R1A2C2003567). 
This research (LRJ) was supported by the Chung-Ang University Young Scientist Scholarship in 2024.
This research (SSK) was supported by Global - Learning and Academic research institution for Master’s·PhD students, and Postdocs(G-LAMP) Program of the National Research Foundation of Korea(NRF) grant funded by the Ministry of Education(No. RS-2025-25442707).

\def\theequation{A.\arabic{equation}}

\setcounter{equation}{0}

\vskip0.8cm
\noindent
{\Large \bf Appendix A: Dark matter annihilation cross sections}

We list the annihilation cross sections for multi-component dark matter composed of $\phi_1, \phi_2$, which is the case with $\kappa_1=0$ in our model.

The annihilation cross section of $\phi^\dagger_i\phi_i(i=1,2)$ into a pair of SM fermions is given by
\begin{eqnarray}
\langle\sigma v\rangle_{\phi_i\phi^\dagger_i\rightarrow {\bar f}f} &=& \frac{\varepsilon^2e^2 q^2_i g^2_X N_c}{\pi x_i}\, \frac{m^2_i+ \frac{1}{2} m^2_f}{(4m^2_i-m^2_X)^2+m^2_X \Gamma^2_X}\, \sqrt{1-\frac{m^2_f}{m^2_i}} \nonumber \\
&& + \frac{N_c}{4\pi} \Big(1-\frac{m^2_f}{m^2_i}\Big)^{3/2}\bigg|\frac{\lambda_{f1}y_{h_1 \phi^\dagger_i\phi_i}}{4m^2_i-m^2_{h_1}} +\frac{\lambda_{f2} y_{h_2 \phi_i^\dagger \phi_i}}{4m^2_i-m^2_{h_2}} \bigg|^2
\,,
\end{eqnarray}
with $x_i\equiv m_i/T$ and $q_i=1,2$  being the $U(1)'$ charges for $\phi_i$ with $i=1,2$.

The cross section for $\phi^\dagger_i\phi_i(i=1,2)$ annihilating into $VV$  with $V=Z, W$ are given by
\begin{equation}
\begin{aligned}
\langle \sigma v\rangle_{\phi_i\phi^\dagger_i\rightarrow VV}\ =\ &\frac{\delta_V e^4}{256\pi \sin^4{\theta_W}\cos^4{\theta_W}}\frac{v^2m_i^2}{m_V^4}\bigg(4-4\frac{m_V^2}{m_i^2}+3\frac{m_V^4}{m_i^4}\bigg)\\
&\times \sqrt{1-\frac{m_V^2}{m_i^2}}\bigg(\frac{v\lambda_{f1}}{m_f}\frac{y_{h_1\phi^\dagger_i\phi_i}}{4m_i^2-m_{h_1}^2}+\frac{v\lambda_{f2}}{m_f}\frac{y_{h_2\phi^\dagger_i\phi_i}}{4m_i^2-m_{h_2}^2}\bigg)^2\,,
\end{aligned}
\end{equation}
with $\delta_V=1$ $(2\cos^4\theta_W)$ for $V=Z$ ($W$).
The cross-section for dark matter annihilating into $\gamma\gamma$ is loop-induced, given by
\begin{equation}
\begin{aligned}
\langle\sigma v\rangle_{\phi_i\phi^\dagger_i\rightarrow \gamma\gamma}\ =\ &\frac{e^4 m^2_i}{64\pi^5 v^2}\bigg|\sum_f N_c Q_f^2A_{1/2}(x_f)+A_1(x_w)\bigg|^2\bigg(\frac{v\lambda_{f1}}{m_f}\frac{y_{h_1 \phi^\dagger_i\phi_i}}{4m^2_i-m^2_{h_1}} +\frac{v\lambda_{f2}}{m_f} \frac{y_{h_2 \phi^\dagger_i\phi_i}}{4m^2_i-m^2_{h_2}} \bigg)^2
\,.
\end{aligned}
\end{equation}
We note that for a vanishing Higgs mixing angle, the DM couplings to the Higgs-like scalars become $y_{h_1\phi^\dagger_1\phi_1}=\lambda_{\chi 1} v_\chi$,  $y_{h_2\phi^\dagger_1\phi_1}=\lambda_{H 1} v$, $y_{h_1\phi^\dagger_2\phi_2}=\lambda_{\chi 2} v_\chi$, and $y_{h_2\phi^\dagger_2\phi_2}=\lambda_{H2} v$; $\lambda_{f1}=0$ and $\lambda_{f2}=\frac{m_f}{v}$.

We also list the cross sections for the additional DM annihilations, in the limit of vanishing Higgs mixing (i.e. $\lambda_{\chi H}=0$) and gauge kinetic mixing (i.e. $\xi=0$), as follows,
\bea
\langle\sigma v\rangle_{ij\to kl}=\frac{|{\cal M}_{ij\to kl}|^2}{8\pi (m_i+m_j)^2}\, \sqrt{1-\frac{(m_k-m_l)^2}{(m_i+m_j)^2}} \,\sqrt{1-\frac{(m_k+m_l)^2}{(m_i+m_j)^2}} \bigg(1- \frac{(m_i-m_j)^2}{(m_i+m_j)^2}\bigg)^{-1},
\eea
with
\bea
|{\cal M}_{\phi_1\phi^\dagger_1\to h_1 h_1}|^2&=& \lambda_{\chi 1}^2 \bigg(\frac{\left(2\lambda_{\chi 1} v_\chi^2+m_{h_1}^2-2
m_1^2\right)^2}{2\left(m_{h_1}^2-2  m_1^2\right)^2}-\frac{3 \sqrt{2}
   m_{h_2} v_\chi \left(2 \lambda_{\chi 1} v_\chi^2+m_{h_1}^2-2
   m_1^2\right)}{m_{h_1}^4-6 m_{h_1}^2 m_{1}^2+8
 m_1^4}  \nonumber  \\
   &&+\frac{9m_{h_2}^2 v_\chi^2}{\left(m_{h_1}^2-4
  m_1^2\right)^2}\bigg), \\
  |{\cal M}_{\phi_2\phi^\dagger_2\to h_1 h_1}|^2&=& \lambda_{\chi 2}^2 \bigg(\frac{\left(2  \lambda_{\chi 2} v_\chi^2 +m_{h_1}^2-2
  m_2^2\right)^2}{2 \left(m_{h_1}^2-2m_2^2\right)^2}-\frac{3
   \sqrt{2} m_{h_2} v_\chi \left(2 \lambda_{\chi 2} v_\chi^2+m_{h_1}^2-2
  m_2^2\right)}{m_{h_1}^4-6 m_{h_1}^2 m_2^2+8
 m_2^4} \nonumber \\
   &&+\frac{9 m_{h_2}^2 v_\chi^2}{\left(m_{h_1}^2-4
   m_2^2\right)^2}\bigg), \\
   |{\cal M}_{\phi_1\phi^\dagger_1\to h_1 h_2}|^2&=& \frac{16 \lambda_{H1}^2 \lambda_{\chi 1}^2 v^2 v_\chi^2}{\left(m_{h_1}^2+m_{h_2}^2-4
  m_1^2\right)^2}, \\
     |{\cal M}_{\phi_2\phi^\dagger_2\to h_1 h_2}|^2&=& \frac{16 \lambda_{H2}^2 \lambda_{\chi 2}^2 v^2 v_\chi^2}{\left(m_{h_1}^2+m_{h_2}^2-4
  m_2^2\right)^2},
\eea
{\small
\bea
  |{\cal M}_{\phi_1\phi^\dagger_1\to h_2 h_2}|^2&=& \frac{\lambda_{H1}^2 \bigg(2m_{h_2}^2 \Big(v (\lambda_{H1}v+m_{h_1})-3
  m_1^2\Big)-4 m_1^2 v (2 \lambda_{H1} v+m_{h_1})+m_{h_2}^4+8
  m_1^4\bigg)^2}{2(m^2_{h_2}-2m^2_1)^2(m^2_{h_2}-4m^2_1)^2},  \nonumber \\ \\
     |{\cal M}_{\phi_2\phi^\dagger_2\to h_2 h_2}|^2&=& \frac{\lambda_{H2}^2 \bigg(2m_{h_2}^2 \Big(v (\lambda_{H2}v+m_{h_1})-3
  m_2^2\Big)-4 m_2^2 v (2 \lambda_{H2} v+m_{h_1})+m_{h_2}^4+8
  m_2^4\bigg)^2}{2(m^2_{h_2}-2m^2_2)^2(m^2_{h_2}-4m^2_2)^2}, \nonumber \\
\eea
\bea
  |{\cal M}_{\phi^\dagger_2\phi^\dagger_1\to h_1 \phi_1}|^2&=& \bigg[\sqrt{2} \kappa_2 m_2(2 m_1+m_2) \left(m_1
   (2 m_1+m_2)- m_{h_1}^2\right) \left(m_2^2 (2m_1+m_2) -m_{h_1}^2 m_1\right) \nonumber \\
   &&-2 g_2 m_1
  v_\chi \bigg(\lambda_{\chi 2} m_2 (m_1+m_2) (2
   m_1+m_2) \left(m_1 (2 m_1+m_2)-m_{h_1}^2\right) \nonumber \\
  && -\lambda_{\chi 1} \left(m_{h_1}^2+m_2
   (2 m_1+m_2)\right) \left(m_{h_1}^2 m_1-m_2^2 (2
m_1+m_2)\right)\bigg)\bigg]^2 \\
  &\times& \bigg(m_2^2 (2
  m_1+m_2)^2 \left(m_{h_1}^2-m_1 (2
   m_1+m_2)\right)^2   \left(m_{h_1}^2 m_1-m_2^2 (2
   m_1+m_2)\right)^2\bigg)^{-1}, \nonumber \\
  |{\cal M}_{\phi_2\phi^\dagger_1\to h_1 \phi_1}|^2&=& 
  4 g^2_1 m^2_1
  v^2_\chi \bigg(\lambda_{\chi 2} m_2 (m_1+m_2) (2
   m_1+m_2) \left(m_1 (2 m_1+m_2)-m_{h_1}^2\right) \nonumber \\
  && -\lambda_{\chi 1} \left(m_{h_1}^2+m_2
   (2 m_1+m_2)\right) \left(m_{h_1}^2 m_1-m_2^2 (2
m_1+m_2)\right)\bigg)^2 \\
  &\times & \bigg(m_2^2 (2
  m_1+m_2)^2 \left(m_{h_1}^2-m_1 (2
   m_1+m_2)\right)^2   \left(m_{h_1}^2 m_1-m_2^2 (2
   m_1+m_2)\right)^2\bigg)^{-1}, \nonumber 
\eea
\bea
  |{\cal M}_{\phi^\dagger_2\phi^\dagger_1\to h_2 \phi_1}|^2&=&4 g_2^2 m_1^2 v^2 \bigg( \lambda_{H2} m_2
   (m_1+m_2) (2 m_1+m_2)  \left(m_1 (2
   m_1+m_2)-m_{h_2}^2\right) \nonumber \\
   &&-\lambda_{H1} \left(m_{h_2}^2+m_2
   (2 m_1+ m_2)\right) \left(m_{h_2}^2 m_1-m_2^2 (2
   m_1+m_2)\right)\bigg)^2  \\
   &\times& \bigg( m_2^2 (2 m_1+m_2)^2
   \left(m_{h_2}^2-m_1 (2 m_1+m_2)\right)^2 \left(m_{h_2}^2
   m_1- m_2^2 (2 m_1+m_2)\right)^2\bigg)^{-1},  \nonumber \\
    |{\cal M}_{\phi_2\phi^\dagger_1\to h_2 \phi_1}|^2&=&4 g_1^2 m_1^2 v^2 \bigg( \lambda_{H2} m_2
   (m_1+m_2) (2 m_1+m_2)  \left(m_1 (2
   m_1+m_2)-m_{h_2}^2\right) \nonumber \\
   &&-\lambda_{H1} \left(m_{h_2}^2+m_2
   (2 m_1+ m_2)\right) \left(m_{h_2}^2 m_1-m_2^2 (2
   m_1+m_2)\right)\bigg)^2  \\
   &\times& \bigg( m_2^2 (2 m_1+m_2)^2
   \left(m_{h_2}^2-m_1 (2 m_1+m_2)\right)^2 \left(m_{h_2}^2
   m_1- m_2^2 (2 m_1+m_2)\right)^2\bigg)^{-1}, \nonumber 
\eea
}
\bea
  |{\cal M}_{\phi_1\phi^\dagger_1\to X X}|^2&=&\frac{2g_X^4 q_1^4 \left(8 m_1^4-8 m_1^2 m_X^2+3
   m_X^4\right)}{\left(m_X^2-2 m_1^2\right)^2}, \\
    |{\cal M}_{\phi_2\phi^\dagger_2\to X X}|^2&=&\frac{2 g_X^4 q_2^4 \left(8 m_2^4-8 m_2^2 m_X^2+3
   m_X^4\right)}{\left(m_X^2-2 m_2^2\right)^2}, 
\eea
{\small
\bea
  |{\cal M}_{\phi^\dagger_2\phi^\dagger_1\to X\phi_1}|^2&=&4 g_2^2 g_X^2 m_1^2 (m^2_2-m^2_X)
\left((2m_1+m_2)^2-m_X^2\right) \nonumber \\
   &\times& \bigg(q_2 m_2^2  (2 m_1+m_2) \left(m_1(2
   m_1+m_2)- m_X^2\right) \nonumber \\
   &\quad& + q_1 \left(2 m_1 (2
   m_1+m_2)- m_X^2\right) \left(2 m_1
   m_2^2-m_1 m_X^2+m_2^3\right)\bigg)^2  \\
   &\times& \bigg( m_2^2
   m_X^2 (2 m_1+m_2)^2 \left(m_X^2-m_1 (2
   m_1+m_2)\right)^2 \left(2 m_1  m_2^2-m_1
   m_X^2+m_2^3\right)^2\bigg)^{-1}, \nonumber  \\
    |{\cal M}_{\phi_2\phi^\dagger_1\to X\phi_1}|^2&=&4 g_1^2 g_X^2  m_1^2 (m^2_2-m^2_X)
  \left((2 m_1+m_2)^2-m_X^2\right) \nonumber \\
   &\times& \bigg( q_2m_2^2  (2 m_1+ m_2)
   \left(m_1 (2m_1+m_2)-m_X^2\right) \nonumber \\
   &\quad&-q_1 \left(2 m_1 (2 m_1+m_2)-m_X^2\right)
   \left(2 m_1 m_2^2- m_1 m_X^2+m_2^3\right) \bigg)^2  \\
   &\times& \bigg( m_2^2
   m_X^2 (2 m_1+m_2)^2 \left(m_X^2-m_1 (2
   m_1+m_2)\right)^2 \left(2 m_1  m_2^2-m_1
   m_X^2+m_2^3\right)^2\bigg)^{-1}, \nonumber 
\eea
}
\bea
  |{\cal M}_{\phi_2\phi^\dagger_2\to \phi_1 \phi^\dagger_1}|^2&=&\bigg(\lambda_{12}-\frac{4 m_1^2
   \left(g_1^2+g_2^2\right)}{m_2^2}-\frac{ \lambda_{H1} \lambda_{H2}
   v^2}{m_{h_2}^2-4m_2^2}-\frac{ \lambda_{\chi 1} \lambda_{\chi 2}
   v_\chi^2}{m_{h_1}^2-4 m_2^2}\bigg)^2,  \\
    |{\cal M}_{\phi_2\phi_2\to \phi_1 \phi^\dagger_1}|^2&=& \frac{64 g_1^2 g_2^2 m_1^4}{m_2^4}.
\eea

\def\theequation{B.\arabic{equation}}

\setcounter{equation}{0}

\vskip0.8cm
\noindent
{\Large \bf Appendix B: Direct detection cross sections}

We list the cross section formulas for direct detection \cite{DD}.

For $m_s=m_s=m_2$, we consider the direct direction of multi-component dark matter with $\phi_1$ and $\phi_2$ in our model. The effective Lagrangian for dark matter-quark elastic scattering is given by
\begin{align}
{\cal L}_{q,{\rm eff}}&= \sum_{i=1,2}\bigg(\frac{\lambda_{q2}y_{h_2 \phi^\dagger_i\phi_i}}{m^2_{h_2}}+\frac{\lambda_{q1}y_{h_1 \phi^\dagger_i\phi_i}}{m^2_{h_1}}\bigg) |\phi_i|^2 {\bar q} q + \sum_{i=1,2}\frac{g_X q_i e\,\varepsilon Q_q}{m^2_{Z'}}\, i(\phi_i\partial_\mu\phi^\dagger_i-\phi^\dagger_i\partial_\mu \phi_i) {\bar q}\gamma^\mu q,
\end{align}
with $q_1=1, q_2=2$.
Then, the relevant matching conditions between quark and nucleon operators are given by
\begin{align}
\langle N|{\bar q}q |N\rangle = \frac{m_N}{m_q}\,f^{(N)}_{Tq}
\,, 
\end{align}
for light quarks ($q=u,d,s$), and 
\begin{align}
\langle N|{\bar q}q |N\rangle &= \frac{2}{27}\,\frac{m_N}{m_q}\, f^{(N)}_{TG}, \quad f^{(N)}_{TG}=1-\sum_{q=u,d,s} f^{(N)}_{Tq}
\,,
\end{align}
for heavy quarks ($q=c,b,t$), and those for vector operators are given by
\begin{equation}\begin{aligned}
\langle N|{\bar u}\gamma^\mu u |N\rangle &= 2 {\bar N}\gamma^\mu N,
&
N&=p\,, \\
\langle N|{\bar d}\gamma^\mu d |N\rangle &= {\bar N}\gamma^\mu N, 
&
N&=p\,, \\
\langle N|{\bar u}\gamma^\mu u |N\rangle &= {\bar N}\gamma^\mu N, 
&
N&=n\,, \\
\langle N|{\bar d}\gamma^\mu d |N\rangle &= 2{\bar N}\gamma^\mu N, 
&
N&=n\,.
\end{aligned}\end{equation}
As a result, we get the effective Lagrangian for dark matter-nucleon elastic scattering as follows,
\begin{align}
{\cal L}_{N,{\rm eff}} &=\sum_{i=1,2}\frac{m_N}{m_q}\bigg(\frac{\lambda_{q2}y_{h_2 \phi^\dagger_i\phi_i}}{m^2_{h_2}}+\frac{\lambda_{q1}y_{h_1 \phi^\dagger_i \phi_i}}{m^2_{h_1}}\bigg) \Big(\sum_{q=u,d,s}f^{(N)}_{Tq}+ \frac{2}{27}f^{(N)}_{TG}\times 3  \Big)  |\phi_i|^2 {\bar N} N \nonumber \\
&\quad+\sum_{i=1,2}\frac{g_X q_i e\,\varepsilon}{m^2_X}\, i(\phi_i\partial_\mu\phi^\dagger_i-\phi^\dagger_i\partial_\mu \phi_i)\bigg((2Q_u+Q_d){\bar p}\gamma^\mu p+(Q_u+2Q_d) {\bar n}\gamma^\mu n\bigg)
\,.
\end{align}
Thus, from $2Q_u+Q_d=+1$ and $Q_u+2Q_d=0$, there is a nonzero interaction only for dark matter-proton scattering with $X$-portal.
Consequently, we obtain the total spin-independent cross sections for the $\phi_i$-nucleus coherent scattering\cite{DD}, as follows,
\begin{equation}
\begin{aligned}
& \sigma^{\rm coh}_{\phi_i -A}\ =\ \frac{\mu^2_{A,i}}{4\pi m^2_i}\Big[Z \Big(c^{(i)}_p f_p+g^{(i)}_p\Big)+(A-Z) \Big(c^{(i)}_n f_n+g^{(i)}_n\Big) \Big]^2\,, \label{coh1}
\end{aligned}
\end{equation}
where $\mu_{A,i}=m_i m_A/(m_A+m_i)$ is the reduced mass of the $\phi_i$-nucleus system with $m_A$ being the target nucleus mass, $Z, A$  are the number of protons and the atomic number, respectively, and the effective couplings and form factors are given by
\bea
c^{(i)}_N &\equiv& \frac{m_N}{m_q}\bigg(\frac{\lambda_{q2} y_{h_2 \phi_i^\dagger\phi_i}}{m^2_{h_2}}+\frac{\lambda_{q1}y_{h_1 \phi^\dagger_i\phi_i}}{m^2_{h_1}}\bigg)
\,,\\
 f_p & \equiv& \sum_{q=u,d,s}f^{(p)}_{Tq}+ \frac{2}{9}f^{(p)}_{TG} \simeq 
0.28
\,,\\
f_n & \equiv& \sum_{q=u,d,s}f^{(n)}_{Tq}+ \frac{2}{9}f^{(n)}_{TG}  \simeq  
0.28
\,,\\
 g^{(i)}_p &=& -\frac{2e q_i g_X \varepsilon m_i}{m_X^2},\\
g^{(i)}_n &\approx& 0
\,.
\eea
Similarly, the $\phi^\dagger_i$-nucleus scattering cross sections have the $X$ contributions flipped in sign, given by
\begin{equation}
\begin{aligned}
& \sigma^{\rm coh}_{\phi^\dagger_i -A}\ =\ \frac{\mu^2_{A,i}}{4\pi m^2_i}\Big[Z \Big(c^{(i)}_p f_p-g^{(i)}_p\Big)+(A-Z) \Big(c^{(i)}_n f_n-g^{(i)}_n\Big) \Big]^2 \label{coh2}
\,.
\end{aligned}
\end{equation}
Then, the averaged spin-independent cross section for dark matter-nucleus coherent scattering is given
\bea
\sigma^{\rm coh}_{\phi_i,\phi^\dagger_i-A}&=&\frac{r_i}{2} \Big( \sigma^{\rm coh}_{\phi_i -A}+ \sigma^{\rm coh}_{\phi^\dagger_i -A} \Big) \nonumber  \\
&=& \frac{\mu^2_{A,i}}{4\pi m^2_i}\,r_i\bigg[\Big(Z c^{(i)}_p f_p+(A-Z)c^{(i)}_n f_n \Big)^2+Z^2(g^{(i)}_p)^2\bigg],
\eea
with $r_i=\Omega_i/\Omega_{\rm DM}$.
As a result, the above dark matter-nucleus scattering cross section is related to the normalized-to-proton scattering cross section by
\begin{align}
\sigma^{\rm coh}_{\phi_i,\phi^\dagger_i-p}= \Big(\frac{\mu_{N,i}}{\mu_{A,i}}\Big)^2\, \frac{\sigma_{\phi_i,\phi^\dagger_i-A}}{A^2}\,,
\end{align}
with  $\mu_{N,i}=m_i m_N/(m_N+m_i)$.

On the other hand, if the dark matter-nucleus scattering is not coherent, such as for the boosted dark matter $\phi_1$, the total spin-independent cross sections in eqs.~(\ref{coh1}) and (\ref{coh2}) should be replaced by
\bea
\sigma^{\rm inc}_{\phi_1-A}&=& \frac{\mu^2_{A,i}}{4\pi m^2_1}\, \bigg[Z\Big(c^{(1)}_p f_p+g^{(1)}_p\Big)^2 +(A-Z)  \Big(c^{(1)}_n f_n+g^{(1)}_n\Big)^2 \bigg], \\
\sigma^{\rm inc}_{\phi^\dagger_1-A}&=& \frac{\mu^2_{A,i}}{4\pi m^2_i}\, \bigg[Z\Big(c^{(1)}_p f_p-g^{(1)}_p\Big)^2 +(A-Z)  \Big(c^{(1)}_n f_n-g^{(1)}_n\Big)^2 \bigg],
\eea
respectively. Then, the averaged spin-independent cross sections for dark matter-nucleus incoherent scattering are given by
\bea
\sigma^{\rm inc}_{\phi_1,\phi^\dagger_1-A}&=&\frac{1}{2} \Big( \sigma^{\rm inc}_{\phi_1 -A}+ \sigma^{\rm inc}_{\phi^\dagger_1 -A} \Big) \nonumber  \\
&=& \frac{\mu^2_{A,1}}{4\pi m^2_1} \bigg[Z\Big(c^{(1)}_p f_p\Big)^2+Z (g^{(1)}_p)^2+(A-Z) \Big(c^{(1)}_n f_n\Big)^2 \bigg].
\eea
Then, the above dark matter-nucleus scattering cross section is related to the normalized-to-proton scattering cross section by
\begin{align}
\sigma^{\rm inc}_{\phi_1,\phi^\dagger_1-p}= \Big(\frac{\mu_{N,1}}{\mu_{A,1}}\Big)^2\, \frac{\sigma^{\rm inc}_{\phi_1,\phi^\dagger_1-A}}{A}.
\end{align}

The elastic scattering between dark matter and electron is mediated by CP-even scalars and the $X$ gauge boson, so the corresponding cross sections are similarly given \cite{DD} by 
\begin{align}
\sigma_{\phi_i,\phi^\dagger_i -e}&= \frac{\mu^2_{e,i}}{8\pi m_i^2}\,r_i \bigg[\Big(\frac{\lambda_{e2}y_{h_2 \phi^\dagger_i\phi_i}}{m_{h_2}^2}+\frac{\lambda_{e1}y_{h_1 \phi^\dagger_i\phi_i}}{m_{h_1}^2}\Big)+\frac{2e q_i g_X \varepsilon m_i}{m_X^2}\bigg]^2 \nonumber \\
&\quad\quad+ \frac{\mu^2_{e,i}}{8\pi m_i^2}\, r_i\bigg[\Big(\frac{\lambda_{e2}y_{h_2 \phi^\dagger_i\phi_i}}{m_{h_2}^2}+\frac{\lambda_{e1}y_{h_1 \phi^\dagger_i\phi_i}}{m_{h_1}^2}\Big)-\frac{2e q_i g_X \varepsilon m_i}{m_X^2}\bigg]^2 \nonumber \\
&=\frac{\mu^2_{e,i}}{4\pi m_i^2}\,r_i\bigg[\Big(\frac{\lambda_{e2}y_{h_2 \phi^\dagger_i\phi_i}}{m_{h_2}^2}+\frac{\lambda_{e1}y_{h_1 \phi^\dagger_i\phi_i}}{m_{h_1}^2}\Big)^2+\frac{4e^2 q^2_i g^2_X \varepsilon^2 m^2_i}{m_X^4} \bigg]
\end{align}
where $\mu_{e,i}=m_i m_e/(m_e+m_i)$ is the reduced mass of the $\phi_i$-electron system.

\def\theequation{C.\arabic{equation}}

\setcounter{equation}{0}

\vskip0.8cm
\noindent
{\Large \bf Appendix C: Kinematics for boosted dark matter}

Consider the DM co-annihilation processes, $\phi^\dagger_1\phi^{(\dagger)}_2\to \phi_1 Y$ with $Y=h_1, X$, or its conjugate process.

In the galactic center of mass frame, from the momentum and the energy for $Y$,
\bea
p_Y&=& \frac{m_2(2m_1+m_2)}{2(m_1+m_2)}\,\bigg(1-\frac{m^2_Y}{m^2_2}\bigg)^{1/2} \Big(1-\frac{m^2_Y}{(2m_1+m_2)^2} \Big)^{1/2},  \label{Ymom} \\
E_Y &=& \frac{m^2_2+2m_1 m_2 +m^2_Y}{2(m_1+m_2)},
\eea
the velocity for the $Y$ particle \cite{SRDM} is given by
\bea
v_Y= \frac{p_Y}{E_Y}= \frac{m_2(2m_1+m_2)}{m^2_2+2m_1 m_2 +m^2_Y}\,\bigg(1-\frac{m^2_Y}{m^2_2}\bigg)^{1/2} \Big(1-\frac{m^2_Y}{(2m_1+m_2)^2} \Big)^{1/2}.
\eea
Similarly, with the momentum, $p'_1=p_Y$, and  the energy for $\phi_1$ appearing in the final states,  given by
\bea
E'_1=\frac{2m^2_1+2m_1m_2+m^2_2-m^2_Y}{2(m_1+m_2)},
\eea
the velocity for the boosted dark matter $\phi_1$ from the DM co-annihilations is given by
\bea
v'_1= \frac{p_Y}{E'_1}=\frac{m_2(2m_1+m_2)}{2m^2_1+2m_1m_2+m^2_2-m^2_Y}\,\bigg(1-\frac{m^2_Y}{m^2_2}\bigg)^{1/2} \Big(1-\frac{m^2_Y}{(2m_1+m_2)^2} \Big)^{1/2}.
\eea

Moreover, when the $Y$ particle decays into a pair of leptons, ignoring the lepton masses, we get the energy of the leptons in the rest frame of the $Y$ particle as ${\bar E}_f=\frac{m_Y}{2}$. However, the energy of the leptons in the galactic center of mass frame is boosted \cite{SRDM} to
\bea
E_f=\frac{1}{\gamma_Y}\, {\bar E}_f \,(1-v_Y\,\cos\theta)^{-1}
\eea
where $\theta$ is the angle between the velocity of the $Y$ particle and the lepton velocity, and 
\bea
\gamma_Y=\frac{1}{\sqrt{1-v^2_ Y}}=\frac{E_Y}{m_Y}. 
\eea
Therefore, the leptons can carry the energies between $E^-_f$ and $E^+_f$ with
\bea
E^\mp_f= \sqrt{\frac{1\mp v_Y}{1\pm v_Y}}\, \frac{m_Y}{2}, \label{leptonE}
\eea
and  the energy spectrum for the leptons is box-shaped  with the width being given by $\Delta E_f= E^+_f- E^-_f= v_Y \gamma_Y\, m_Y$.

We also comment on the cosmic ray signatures from boosted dark matter in our model.

Due to the same semi-annihilation processes discussed for boosted dark matter in the previous section,  namely, $\phi^\dagger_1\phi^{(\dagger)}_2\to \phi_1 Y$ with $Y=X, h_1$ and their complex conjugate processes, the subsequent decays of the boosted particle $Y$,  $Y\to f{\bar f}$, can produce leptons at galaxies, so they give rise to cosmic ray signals too \footnote{We also remark that the energetic leptons produced from the semi-annihilation processes can be also constrained by the CMB recombination \cite{SRDM}.}.
Then,  we can consider the corresponding differential flux for the leptons as 
\bea
\frac{d\Phi_f}{dE_f} = \frac{1}{4\pi m_1 m_2}\, \langle\sigma v\rangle_{\phi^\dagger_1\phi^{(\dagger)}_2\to \phi_1 Y,\, {\rm c.c.}} \, \cdot\frac{dN_f}{d E_f} \, r_1(1-r_1)\, J
\eea
where the differential number of leptons is given by
\bea
\frac{d N_f}{dE_f} = \frac{1}{E^+_f -E^-_f}\,\Theta(E_f-E^-_f) \Theta(E^+_f -E_f)\, {\rm BR}(Y\to f{\bar f}),
\eea
where
\bea
E^\mp_f= \sqrt{\frac{1\mp v_Y}{1\pm v_Y}}\, \frac{m_Y}{2},
\eea
with $v_Y$ being the velocity of the $Y$ particle in the galactic frame, given by
\bea
v_Y= \frac{m_2(2m_1+m_2)}{m^2_2+2m_1 m_2 +m^2_Y}\,\bigg(1-\frac{m^2_Y}{m^2_2}\bigg)^{1/2} \Big(1-\frac{m^2_Y}{(2m_1+m_2)^2} \Big)^{1/2}.
\eea
We note that we need to consider the $J$-factor for indirect detection, in the following way, 
\bea
J=\frac{1}{\Delta \Omega} \int_{\Delta\Omega} d\Omega \int_{\rm l.o.s.} ds\,  \rho^2_{\rm DM}.
\eea


\begin{thebibliography}{999}


\bibitem{Planck:2018vyg}
N.~Aghanim \textit{et al.} [Planck],
Astron. Astrophys. \textbf{641} (2020), A6
[erratum: Astron. Astrophys. \textbf{652} (2021), C4]
doi:10.1051/0004-6361/201833910
[arXiv:1807.06209 [astro-ph.CO]].



\bibitem{xenonnt}
E.~Aprile \textit{et al.} [XENON],
Phys. Rev. Lett. \textbf{131} (2023) no.4, 041003
doi:10.1103/PhysRevLett.131.041003
[arXiv:2303.14729 [hep-ex]].



\bibitem{PandaX:2024qfu}
Z.~Bo \textit{et al.} [PandaX],
``Dark Matter Search Results from 1.54 Tonne$\cdot$Year Exposure of PandaX-4T,''
[arXiv:2408.00664 [hep-ex]].



\bibitem{LZCollaboration:2024lux}
J.~Aalbers \textit{et al.} [LZ Collaboration],
``Dark Matter Search Results from 4.2 Tonne-Years of Exposure of the LUX-ZEPLIN (LZ) Experiment,''
[arXiv:2410.17036 [hep-ex]].



\bibitem{SIDM}
S.~Tulin and H.~B.~Yu,
Phys. Rept. \textbf{730} (2018), 1-57
doi:10.1016/j.physrep.2017.11.004
[arXiv:1705.02358 [hep-ph]].



\bibitem{SRDM}
S.~S.~Kim, H.~M.~Lee and B.~Zhu,
JHEP \textbf{05} (2022), 148
doi:10.1007/JHEP05(2022)148
[arXiv:2202.13717 [hep-ph]];
S.~S.~Kim, H.~M.~Lee and B.~Zhu,
JHEP \textbf{10} (2021), 239
doi:10.1007/JHEP10(2021)239
[arXiv:2108.06278 [hep-ph]].




\bibitem{review}
H.~M.~Lee,
[arXiv:2304.05942 [hep-ph]].


\bibitem{velocitydep}
M.~Kaplinghat, S.~Tulin and H.~B.~Yu,
Phys. Rev. Lett. \textbf{116} (2016) no.4, 041302
doi:10.1103/PhysRevLett.116.041302
[arXiv:1508.03339 [astro-ph.CO]].


\bibitem{smallscale}
  D.~N.~Spergel and P.~J.~Steinhardt,
  Phys.\ Rev.\ Lett.\  {\bf 84}, 3760 (2000)
  [astro-ph/9909386];
  W.~J.~G.~de Blok,
  Adv.\ Astron.\  {\bf 2010}, 789293 (2010)
  [arXiv:0910.3538 [astro-ph.CO]];
  M.~Boylan-Kolchin, J.~S.~Bullock and M.~Kaplinghat,
  Mon.\ Not.\ Roy.\ Astron.\ Soc.\  {\bf 415}, L40 (2011)
  [arXiv:1103.0007 [astro-ph.CO]];
  D.~H.~Weinberg, J.~S.~Bullock, F.~Governato, R.~K.~de Naray and A.~H.~G.~Peter,
  arXiv:1306.0913 [astro-ph.CO];  
  M.~Rocha, A.~H.~G.~Peter, J.~S.~Bullock, M.~Kaplinghat, S.~Garrison-Kimmel, J.~Onorbe and L.~A.~Moustakas,
  Mon.\ Not.\ Roy.\ Astron.\ Soc.\  {\bf 430}, 81 (2013)  [arXiv:1208.3025 [astro-ph.CO]].  



\bibitem{baryon}
F.~Governato, A.~Zolotov, A.~Pontzen, C.~Christensen, S.~H.~Oh, A.~M.~Brooks, T.~Quinn, S.~Shen and J.~Wadsley,
Mon. Not. Roy. Astron. Soc. \textbf{422} (2012), 1231-1240
doi:10.1111/j.1365-2966.2012.20696.x
[arXiv:1202.0554 [astro-ph.CO]];
A.~M.~Brooks and A.~Zolotov,
Astrophys. J. \textbf{786} (2014), 87
doi:10.1088/0004-637X/786/2/87
[arXiv:1207.2468 [astro-ph.CO]].


\bibitem{Z4IDM}
J.~Kim, S.~S.~Kim, H.~M.~Lee and R.~Padhan,
JHEP \textbf{08} (2025), 105
doi:10.1007/JHEP08(2025)105
[arXiv:2505.00121 [hep-ph]].


\bibitem{U1RG}
C.~Branchina, H.~M.~Lee and K.~Yamashita,
JHEP \textbf{03} (2025), 122
doi:10.1007/JHEP03(2025)122
[arXiv:2407.14826 [hep-ph]].



\bibitem{self-scattering}
Y.~J.~Kang and H.~M.~Lee,
Eur. Phys. J. C \textbf{81} (2021) no.10, 868
doi:10.1140/epjc/s10052-021-09610-x
[arXiv:2002.12779 [hep-ph]];
Y.~J.~Kang and H.~M.~Lee,
J. Phys. G \textbf{48} (2021) no.4, 045002
doi:10.1088/1361-6471/abe529
[arXiv:2003.09290 [hep-ph]].


\bibitem{boundstate1}
S.~Tulin, H.~B.~Yu and K.~M.~Zurek,
Phys. Rev. D \textbf{87} (2013) no.11, 115007
doi:10.1103/PhysRevD.87.115007
[arXiv:1302.3898 [hep-ph]].


\bibitem{boundstate2}
S.~Cassel,
J. Phys. G \textbf{37} (2010), 105009
doi:10.1088/0954-3899/37/10/105009
[arXiv:0903.5307 [hep-ph]].



\bibitem{Colquhoun:2020adl}
B.~Colquhoun, S.~Heeba, F.~Kahlhoefer, L.~Sagunski and S.~Tulin,
Phys. Rev. D \textbf{103} (2021) no.3, 035006
doi:10.1103/PhysRevD.103.035006
[arXiv:2011.04679 [hep-ph]].


\bibitem{Landau}
L.~D.~Landau and E.~M.~Lifshits,
Butterworth-Heinemann, 1991,
ISBN 978-0-7506-3539-4
doi:10.1016/C2013-0-02793-4


\bibitem{planck15}
P.~A.~R.~Ade \textit{et al.} [Planck],
Astron. Astrophys. \textbf{594} (2016), A13
doi:10.1051/0004-6361/201525830
[arXiv:1502.01589 [astro-ph.CO]].
    
    
    
   
   
\bibitem{slatyer3}  
T.~R.~Slatyer,
Phys. Rev. D \textbf{93} (2016) no.2, 023527
doi:10.1103/PhysRevD.93.023527
[arXiv:1506.03811 [hep-ph]].




\bibitem{gmix1}
S.~M.~Choi and H.~M.~Lee,
JHEP \textbf{09} (2015), 063
doi:10.1007/JHEP09(2015)063
[arXiv:1505.00960 [hep-ph]].


\bibitem{gmix2}
S.~M.~Choi, H.~M.~Lee, Y.~Mambrini and M.~Pierre,
JHEP \textbf{07} (2019), 049
doi:10.1007/JHEP07(2019)049
[arXiv:1904.04109 [hep-ph]].


\bibitem{DD}
S.~M.~Choi, J.~Kim, H.~M.~Lee and B.~Zhu,
JHEP \textbf{06} (2020), 135
doi:10.1007/JHEP06(2020)135
[arXiv:2003.11823 [hep-ph]].


\bibitem{massplitting}
J.~Kim, S.~S.~Kim, H.~M.~Lee and R.~Padhan,
Phys. Lett. B \textbf{861} (2025), 139243
doi:10.1016/j.physletb.2025.139243
[arXiv:2407.13595 [hep-ph]];
J.~Kim, S.~S.~Kim, H.~M.~Lee and R.~Padhan,
JHEP \textbf{08} (2025), 105
doi:10.1007/JHEP08(2025)105
[arXiv:2505.00121 [hep-ph]].


\bibitem{Xenon1}
H.~M.~Lee,
JHEP \textbf{01} (2021), 019
doi:10.1007/JHEP01(2021)019
[arXiv:2006.13183 [hep-ph]].


\bibitem{Xenon2}
S.~M.~Choi, H.~M.~Lee and B.~Zhu,
JHEP \textbf{04} (2021), 251
doi:10.1007/JHEP04(2021)251
[arXiv:2012.03713 [hep-ph]].


\bibitem{boosted}
K.~Agashe, Y.~Cui, L.~Necib and J.~Thaler,
JCAP \textbf{10} (2014), 062
doi:10.1088/1475-7516/2014/10/062
[arXiv:1405.7370 [hep-ph]].


\bibitem{boosted2}
B.~Fornal, P.~Sandick, J.~Shu, M.~Su and Y.~Zhao,
Phys. Rev. Lett. \textbf{125} (2020) no.16, 161804
doi:10.1103/PhysRevLett.125.161804
[arXiv:2006.11264 [hep-ph]].


\bibitem{semi-ann}
B.~Betancourt Kamenetskaia, M.~Fujiwara, A.~Ibarra and T.~Toma,
[arXiv:2501.12117 [hep-ph]].


\bibitem{indirect}
F.~D'Eramo and J.~Thaler,
JHEP \textbf{06} (2010), 109
doi:10.1007/JHEP06(2010)109
[arXiv:1003.5912 [hep-ph]].



\bibitem{box}
A.~Ibarra, H.~M.~Lee, S.~L{\'o}pez Gehler, W.~I.~Park and M.~Pato,
JCAP \textbf{05} (2013), 016
[erratum: JCAP \textbf{03} (2016), E01]
doi:10.1088/1475-7516/2013/05/016
[arXiv:1303.6632 [hep-ph]].


\bibitem{Perdrisat:2006hj}
C.~F.~Perdrisat, V.~Punjabi and M.~Vanderhaeghen,
Prog. Part. Nucl. Phys. \textbf{59} (2007), 694-764
doi:10.1016/j.ppnp.2007.05.001
[arXiv:hep-ph/0612014 [hep-ph]].




\bibitem{darwin}
J.~Aalbers \textit{et al.} [DARWIN],
JCAP \textbf{11} (2016), 017
doi:10.1088/1475-7516/2016/11/017
[arXiv:1606.07001 [astro-ph.IM]].



\bibitem{latedecay}
K.~Y.~Choi, J.~E.~Kim, H.~M.~Lee and O.~Seto,
Phys. Rev. D \textbf{77} (2008), 123501
doi:10.1103/PhysRevD.77.123501
[arXiv:0801.0491 [hep-ph]].



\bibitem{belle2}
E.~Kou \textit{et al.} [Belle-II],
PTEP \textbf{2019} (2019) no.12, 123C01
[erratum: PTEP \textbf{2020} (2020) no.2, 029201]
doi:10.1093/ptep/ptz106
[arXiv:1808.10567 [hep-ex]];


\bibitem{belle2-summary}
L.~Corona,
PHEP \textbf{2025} (2025), 4
doi:10.31526/PHEP.2025.04

\bibitem{belle2-limit}
I.~Adachi \textit{et al.} [Belle-II],
Phys. Rev. Lett. \textbf{124} (2020) no.14, 141801
doi:10.1103/PhysRevLett.124.141801
[arXiv:1912.11276 [hep-ex]];
I.~Adachi \textit{et al.} [Belle-II],
Phys. Rev. Lett. \textbf{130} (2023) no.23, 231801
doi:10.1103/PhysRevLett.130.231801
[arXiv:2212.03066 [hep-ex]].



\bibitem{babar}
J.~P.~Lees \textit{et al.} [BaBar],
Phys. Rev. Lett. \textbf{119} (2017) no.13, 131804
doi:10.1103/PhysRevLett.119.131804
[arXiv:1702.03327 [hep-ex]];
J.~P.~Lees \textit{et al.} [BaBar],
Phys. Rev. Lett. \textbf{113} (2014) no.20, 201801
doi:10.1103/PhysRevLett.113.201801
[arXiv:1406.2980 [hep-ex]].


\bibitem{NA64}
Y.~M.~Andreev \textit{et al.} [NA64],
Phys. Rev. Lett. \textbf{131} (2023) no.16, 161801
doi:10.1103/PhysRevLett.131.161801
[arXiv:2307.02404 [hep-ex]];
Y.~M.~Andreev \textit{et al.} [NA64],
Phys. Rev. Lett. \textbf{132} (2024) no.21, 211803
doi:10.1103/PhysRevLett.132.211803
[arXiv:2401.01708 [hep-ex]];
Y.~M.~Andreev \textit{et al.} [NA64],
[arXiv:2505.14291 [hep-ex]].


\bibitem{BBN-review}
R.~H.~Cyburt, B.~D.~Fields, K.~A.~Olive and T.~H.~Yeh,
Rev. Mod. Phys. \textbf{88} (2016), 015004
doi:10.1103/RevModPhys.88.015004
[arXiv:1505.01076 [astro-ph.CO]].



\bibitem{BBN1}
M.~Hufnagel, K.~Schmidt-Hoberg and S.~Wild,
JCAP \textbf{02} (2018), 044
doi:10.1088/1475-7516/2018/02/044
[arXiv:1712.03972 [hep-ph]];
M.~Hufnagel, K.~Schmidt-Hoberg and S.~Wild,
JCAP \textbf{11} (2018), 032
doi:10.1088/1475-7516/2018/11/032
[arXiv:1808.09324 [hep-ph]].

\bibitem{BBN2}
L.~Forestell, D.~E.~Morrissey and G.~White,
JHEP \textbf{01} (2019), 074
doi:10.1007/JHEP01(2019)074
[arXiv:1809.01179 [hep-ph]].




\end{thebibliography}
\end{document}